\documentclass[pra,twocolumn,superscriptaddress,floatfix,nofootinbib]{revtex4-2}
\pdfoutput=1 

\usepackage[T1]{fontenc}
\usepackage{amsmath}
\usepackage{amssymb}
\usepackage{amsfonts}
\usepackage{graphicx}
\usepackage{enumitem}
\usepackage{bm}
\usepackage{rotating}
\usepackage{hyperref}
\usepackage{pbox}
\usepackage[dvipsnames]{xcolor}
\usepackage{array}
\usepackage{physics}
\usepackage{dsfont}
\usepackage{mathtools}
\usepackage{leftidx}
\usepackage{lmodern}
\usepackage[normalem]{ulem}
\usepackage{braket}
\usepackage{tensor}
\usepackage{mathrsfs}
\usepackage{float}
\usepackage{dsfont}
\usepackage[margin=2cm]{geometry}
\usepackage[title]{appendix}
\usepackage{tcolorbox}

\newcommand{\bk}{ {\mathbf{k}} }

\newcommand{\exd}{{\hbox{d}}}

\def\smath#1{\text{\scalebox{.85}{$#1$}}}
\def\sfrac#1#2{\smath{\frac{#1}{#2}}}

\newcommand{\Ws}{ {\mathcal{W}_{\s}} }
\newcommand{\Wc}{ {\mathcal{W}_{\times} } }

\def\ssI{{\scriptscriptstyle I}}

\newcommand{\bx}{\mathbf{x}}
\newcommand{\rr}[1]{\left(#1\right)}
\renewcommand{\Re}{\text{Re}}
\renewcommand{\Im}{\text{Im}}

\newcommand{\sy}{\mathsf{y}}
\newcommand{\s}{\textsc{s}}
\newcommand{\ab}{\textsc{ab}}

\newcommand*{\Scale}[2][4]{\scalebox{#1}{$#2$}}

\begin{document}

\title{Effective master equations for two accelerated qubits}

\author{Greg Kaplanek}
\email{g.kaplanek@imperial.ac.uk}
\affiliation{Department of Physics and Astronomy, McMaster University, Hamilton, Ontario, L8S 4M1, Canada}
\affiliation{Perimeter Institute for Theoretical Physics, Waterloo, Ontario N2L 2Y5, Canada}
\affiliation{Theoretical Physics, Blackett Laboratory, Imperial College, London, SW7 2AZ, UK}

\author{Erickson Tjoa}
\email{e2tjoa@uwaterloo.ca}
\affiliation{Department of Physics and Astronomy, University of Waterloo, Waterloo, Ontario, N2L 3G1, Canada}
\affiliation{Institute for Quantum Computing, University of Waterloo, Waterloo, Ontario, N2L 3G1, Canada}

\begin{abstract}    
We revisit the problem involving two constantly accelerating Unruh-DeWitt detectors using Open Effective Field Theory methods. We study the time evolution of the joint detector state using a Markovian approximation which differs from the standard one taken in the literature. 
 {We show that this Markovian limit already implies the complete positivity of the dynamical evolution map without invoking the rotating wave approximation (RWA), in contrast to standard derivations of open system master equations. 
By calculating explicitly the domain of validity of this Markovian approximation, we argue that the lack of complete positivity in the usual microscopic derivation stems from the (subtle) fact that the Redfield equation is used outside its domain of validity. We give two well-known cases studied in the literature that violate the validity of the Markovian approximation: (i) the ``{stacked trajectory}'' limit (when detector trajectories are taken to be on top of one another), and (ii) large gap-to-acceleration ratio. Since Markovian dynamics with or without RWA can lead to different qualitative predictions for entanglement dynamics, our work emphasizes the need to properly track the regime of validity of all approximations.
}

\end{abstract}

\maketitle

\section{Motivation}

It is by now well-known that vacuum states in quantum field theory (QFT) contain correlations between disjoint spacetime regions \cite{summers1985bell,summers1987bell,Higuchi2017wedges}. In particular, vacuum entanglement is responsible for various physical phenomena such as the Unruh and Hawking effects. However, due to the ultraviolet properties of field-theoretic entanglement and the lack of direct measurement theory in QFT, one often resorts to the use of an external probe as a way to study the entanglement structure of the quantum field. In relativistic quantum information, the paradigmatic example is the Unruh-DeWitt (UDW) detector model \cite{Unruh1979evaporation,DeWitt1979}, where a two-level quantum system (qubit) is coupled locally to a quantum scalar field.

From the perspective of relativistic quantum information, the study of entanglement dynamics between two UDW detectors typically falls into two classes: (i) via the \textit{entanglement harvesting protocol} (see, e.g., \cite{reznik2003entanglement,Valentini1991nonlocalcorr,VerSteeg2009entangling,pozas2015harvesting,pozas2016entanglement,Grimmer2021algebraic,tjoa2021entanglement,henderson2018harvestingBH,smith2016topology,Caroline2022harvesting,Tjoa2020vaidya,Gallock-Yoshimura2021freefall,Caroline2022harvesting,maeso2022entanglement,salton2015acceleration,cong2019entanglement,Ng2018newtechniques,cong2020horizon,henderson2019AdSharvesting,Nadine2021delocharvesting,simidzija2018harvesting,Smith2020harvestingGW}), or (ii) through \textit{open quantum systems} methods (see, e.g., \cite{Benatti2004Unruh,BENATTI2005review,Hu2015twodetectorent,Zhou2021massiveentanglement2,Menezes2018entanglementKerr,huang2017dynamics,Soares2022open,benatti2022local,Zhang2007reflect,Zhang2007reflect2,Hu2011blackhole,Hu2013desitter,huang2019boundary,Zhou2021massiveentanglement,Zhou2021massiveentanglement2,Hu2022loss-anti,benatti2022local}). Although the objectives may vary, at the technical level the main difference between these two approaches is the interaction timescale. On the one hand, the entanglement harvesting protocol is by construction restricted to relatively short timescales, since one wants to extract entanglement from the field (and not through signalling between the two detectors \cite{tjoa2021entanglement}). On the other hand, the open quantum systems framework often aims to obtain \textit{late-time dynamics}, which is useful when one wishes to understand long-time processes such as {thermalization}. For example, open system methods allows one to directly determine if (and when) a single detector approaches a Gibbs state (see, e.g., \cite{Benito2019asymptotic,Moustos2017nonmarkov}), instead of stopping at the detailed balance condition or Planckian transition rates\footnote{These are necessary but not sufficient conditions for thermalization as they do not capture what happens to density matrix coherence.} (see, e.g., \cite{Aubry2014derivative,Satz2007transitionrate}).

In this work we revisit the open system framework between two UDW detectors undergoing uniform accelerations. Our motivation is based on two important considerations:
\begin{enumerate}[leftmargin=*,label=(\arabic*)]
    \item Making late-time predictions in perturbation theory \textit{reliably} is notoriously difficult (see, e.g., \cite{Kaplanek2018rindler,kaplanek2020hot,kaplanek2020hot2,kaplanek2021qubits}). The issue is that the strength of the detector-field coupling imposes a natural timescale for which the perturbation series at any given order is valid. In essence, perturbative expansions of quantities like $e^{- i  g H_{\mathrm{int}} t} \simeq 1 - i  g H_{\mathrm{int}} t + \ldots $ generically become suspect at late times when $g H_{\mathrm{int}} t$ becomes too large. This is also true for detectors that are switched on and off adiabatically and smoothly (``carefully'' \cite{Satz2006howoften,Satz2007transitionrate,Fewster2016wait}). In particular, one needs to have control over how long is ``long times'' when studying weakly-coupled systems --- one virtue of the open quantum system approach is that it provides a way in which to resum late-time breakdowns of perturbation theory so as to more reliably gain access to late-time behaviour.
    
    \item The use of \textit{semigroup master equations}, which captures the reduced open system dynamics in the presence of a large environment, is often done without careful or explicit analysis of the validity of the approximations that go into it. One of the most common examples is the invocation of the \textit{Gorini-Kossakowski-Sudarshan-Lindblad} (GKSL) master equation, among many others (e.g.  \cite{Benatti2004Unruh,BENATTI2005review,Hu2015twodetectorent,huang2017dynamics,Menezes2018entanglementKerr,Zhou2021massiveentanglement2,Soares2022open,benatti2022local,Zhang2007reflect,Zhang2007reflect2,Hu2011blackhole,Hu2013desitter,huang2019boundary,Zhou2021massiveentanglement,Zhou2021massiveentanglement2,Hu2022loss-anti,benatti2022local}).
\end{enumerate}

In more detail, the usual approach to a microscopic derivation of a GKSL master equation for the reduced density matrix $\rho_{\textsc{sys}}(\tau)$ of the detector (``system'') at time $\tau$ typically involves three distinct approximations:
\begin{enumerate}[leftmargin=*,label=(\arabic*)]
\item First, one perturbs the underlying Liouville-von Neumann equation for the full density matrix using the \textit{Born approximation}. This is justified when the environment is large compared to the system. However, the resulting master equation for $\rho_{\textsc{sys}}(\tau)$ is intractable because it depends on its entire history of evolution (i.e. the evolution has \textit{memory}).
\item This is where the second approximation --- the \textit{Markovian approximation} --- is employed, by working in a regime of parameter space where the evolution is ``memoryless'' (meaning here that the evolution equation for $\rho_{\textsc{sys}}(\tau)$ is time-local). At this point, the resulting evolution equation is infamously of the ``Redfield-type'': the differential equation for $\rho_\textsc{sys}(\tau)$ induces a dynamical evolution map $\Phi_\tau:\rho_{\textsc{sys}}(0)\mapsto \rho_{\textsc{sys}}(\tau)$ that is believed to be not completely positive\footnote{That is, one can end up with predictions of negative probabilities with the computed reduced density matrix. This is sometimes known as ``slippage of initial conditions'' \cite{Gaspard1999slippage,Suarez1992memory,Benatti2010entanglement2,Anderloni2007redfield}. } (CP) \cite{breuer2002theory,lidar2020lecture}.
\item The third approximation, known as the \textit{secular approximation} or \textit{rotating wave approximation} (RWA), is then used in order to make $\Phi_\tau$ a completely positive and trace-preserving (CPTP) map, i.e., a quantum channel. The inspiration for this approximation is from quantum optics in which rapidly oscillating terms in the master equation can be neglected when the system is near resonance with an oscillating environment (like a laser tuned to a specific frequency), and formally described by Davies in \cite{davies1974markovian,davies1976markovian}.
\end{enumerate}

Notably, once the three approximations are taken, the resulting master equation can be treated like a black box and it is often used liberally, at the risk of not tracking the validity of its usage.

In this work we argue that the lack of CP-property mentioned above in Step (2) and (3) arises because the Markovian limit is not carefully taken. Our point of view is that Step (2) is valid in a specific region of parameter space where the bath correlation time (in this work, the reciprocal of the Unruh temperature $\sim 1/a$) is the shortest timescale in the problem --- including timescales associated with the system operators (such as $\sim 1/\Omega$, where $\Omega$ are the qubit energy gaps). As we show here, carefully performed Markovian limits amount to approximating \textit{both} the system state $\rho_\textsc{sys}(\tau)$ in the interaction picture and the system observables (the monopole operator of the qubit detector) as memoryless.  {The manner in which we perform this approximation is inspired by the work of \cite{Martin2018cosmologydecohere} (working in the context of cosmology), which uses a slightly different version of the Markovian approximation  compared to the usual one in the literature.}

When this is done, the so-called \textit{Born-Markov approximations} (Steps (1) \textit{and} (2) in the above) produce an evolution map that is CP-preserving according to the GKSL theorem \cite{lindblad1976generators,gorini1976completely}. This implies that the RWA is \textit{not} necessary (see also \cite{whitney2008staying}) and therefore, the standard approach in employing the RWA-based GKSL master equation (see, e.g., \cite{breuer2002theory}, for standard reference) is at best valid on much smaller parameter space than what the Born-Markov approximations allow for\footnote{This is already ignoring the possibility of other problems appearing in RWA: for example, some form of RWA known as pre-trace RWA  can lead to causality violation \cite{Funai2019rotatingwave} or even the lack of a Markovian limit \cite{Fleming2010RWA}.}. We provide explicit bounds (``validity relations'') for which the Born-Markov approximation is valid. The derived bounds show that two well-known special cases studied in the literature, (i) the ``{stacked trajectory}'' limit (when detector trajectories are taken to be equal\footnote{Sometimes compared to the two-atom Dicke model \cite{DickeModel}.} with proper separation $L=0$, see e.g., \cite{Benatti2004Unruh,Menezes2018entanglementKerr,Hu2011blackhole,Hu2013desitter,huang2017dynamics}), and (ii) large gap-to-acceleration ratio $\Omega/a\gtrsim 1$ (see, e.g., \cite{Hu2015twodetectorent,huang2017dynamics,huang2019boundary,Hu2022loss-anti,huang2017dynamics,Zhang2007reflect,Zhang2007reflect2,benatti2022local}), and (iii) inertial qubits interacting with vacuum state \cite{huang2019boundary,Zhou2021massiveentanglement,Zhou2021massiveentanglement2}, all of which violate the validity of the Markovian approximation.

In this work we also compare the late-time state and entanglement dynamics between the two detectors between the RWA-based evolution vs the one without RWA. In both cases the fixed point of the evolution is a maximally mixed state (understood as a high-temperature Gibbs state), so one cannot use the late-time state to check the validity of RWA. It is also worth noting that some of the very early time dynamics within $g^2 a\tau\ll 1$ calculated using Markovian master equation may \textit{not} be reliable, since at early times non-Markovian effects are important. The takeaway is that just because the Markovian solution exists for early and late times does not mean the full dynamics agree with Markovian one at all times. Also importantly, within the Born-Markov approximations one cannot infer the Unruh temperature ${T}_\textsc{U}=a/(2\pi)$ from its late-time density matrix: the reason is because we are in the high-temperature limit and the asymptotic state is effectively the zeroth-order expansion in $\Omega/a\propto \Omega/T_{\textsc{U}}$ of the Gibbs state. The information about the Unruh temperature must be obtained elsewhere, e.g., by computing the decay rate or Fisher information (for recent examples of the latter using the RWA see \cite{du2021fisher,feng2022quantum-fisher}).

This paper is organized as follows. In \S\ref{sec: setup}, we discuss the UDW setup for two uniformly accelerated detectors. From there we develop the Nakajima-Zwanzig master equation for the joint detector state in \S\ref{sec:master}, and then further take its Markovian limit in \S\ref{sec: markovian} obtaining the late-time asymptote for the state and compute the timescales for the approach to this fixed point. In \S\ref{sec: validity} we calculate the validity relations that constrain the parameters in order for the Born-Markov approximations to be valid. Finally, in \S\ref{sec: comparison} we compare our results without the RWA with those that are derived using the RWA. We use the mostly-plus signature for the metric and set $c=\hbar=1$.

\section{Qubits in space}
\label{sec: setup}

In this section we formulate the detector-field interaction based on the Unruh-DeWitt model, sending two qubits (two-level systems) along parallel accelerated trajectories while interacting with a free massless scalar field $\phi$ in (3+1)-dimensional Minkowski space. We work in perturbation theory to order $\mathcal{O}(g^2)$ where the coupling strength of the detector-field interaction is $g\ll 1$. 

Consider two observers Alice and Bob, each carrying a pointlike two-level Unruh-DeWitt (UDW) detector, put along parallel accelerated trajectories in flat spacetime with worldlines (in Minkowski coordinates)
\begin{equation}
    \begin{aligned}
    \sy_\textsc{a}(\tau) &= (t(\tau), x(\tau), 0, 0)\,,\\
    \sy_\textsc{b}(\tau) &= (t(\tau), x(\tau), L, 0)\,,
    \label{eq: trajectories}
    \end{aligned}
\end{equation}
with acceleration $a>0$, where $t(\tau) = \frac{1}{a}\sinh(a\tau)$, $x(\tau)=\frac{1}{a}\cosh(a\tau)$ and $L>0$ is the proper separation along the $y$-direction. As they are parallel in the transverse direction, the two observers' worldlines can be parametrized by the same proper time and we set the initial proper times along each trajectory $\tau=0$ to align with $t=0$.

The massless scalar field $\phi$ in Minkowski spacetime satisfies the Klein-Gordon equation $\partial_\mu\partial^\mu \phi = 0$ with mode decomposition in the inertial quantization frame $(t,\bx)$ given by (in the interaction picture)
\begin{align}
    \phi(t,\bx) &= \int \frac{\dd^3\bk}{\sqrt{2(2\pi)^3\omega_\bk}}\bigr[a_\bk^{\phantom{\dagger}} e^{-i\omega_\bk t+ i\bk\cdot\bx} + \text{h.c.}\bigr]\,,
    \label{eq: Fourier-decomposition}
\end{align}
where $\omega_\bk = |\bk|$ is the relativistic dispersion for massless fields. The canonical commutation relation is given by $[a^{\phantom{\dagger}}_\bk,a_{\bk'}^\dagger]=\delta^3(\bk-\bk')\openone_{\phi}$ (with $\openone_{\phi}$ the identity on the field Hilbert space) and all other commutators vanish. The free Hamiltonian for the scalar field reads
\begin{align}
    \mathfrak{h}_\phi &= \frac{1}{2}\int_{\Sigma_t}\dd^3\bx\, \left[ \pi^2(t,\bx) + \partial_i\phi(t,\bx)\partial^i\phi(t,\bx)\right]\,,
\end{align}
where $\pi = \partial_t \phi$ and $\Sigma_t$ is a constant-$t$ Cauchy surface in Minkowski spacetime. This free Hamiltonian generates time translations with respect to Minkowski time $t$.

For the $4 \times 4$ matrices acting on the joint qubit subspace of the Hilbert space we use the notation
\begin{equation}
\sigma_{\alpha}^{\textsc{a}} \coloneqq \sigma_{\alpha} \otimes \openone \quad \mathrm{and} \quad \sigma_{\alpha}^{\textsc{b}} \coloneqq \openone \otimes \sigma_{\alpha}
\end{equation}
where $\openone$ is the $2\times 2$ identity matrix, $\sigma_{\alpha}$ are the standard Pauli matrices for $\alpha \in \{1,2,3, \pm \}$, and in particular $\sigma_{\pm} = \frac{1}{2}( \sigma_{1} \pm i \sigma_{2})$.

For simplicity, we consider both detectors to be identical with energy gap $\Omega$, so that the free Hamiltonian for Alice and Bob's detectors reads
\begin{equation}
\mathfrak{h} \coloneqq \mathfrak{h}_{\textsc{a}} + \mathfrak{h}_{\textsc{b}}  \qquad \mathrm{with} \  \mathfrak{h}_{j} \coloneqq \tfrac{\Omega}{2}  \sigma_{3}^{j} \ .
\end{equation}
The ground and excited states of each detector are denoted by $\ket{\downarrow}$ and $\ket{\uparrow}$ respectively, with the action of the $\mathfrak{su}(2)$ ladder operators given by $\sigma_+\ket{\downarrow}=\ket{\uparrow}$ and $\sigma_-\ket{\uparrow}=\ket{\downarrow}$. The free Hamiltonian for the detector-field system reads
\begin{align}
    H_0 & = \frac{\dd \tau}{\dd t} \;  \mathfrak{h} \otimes \openone_{\phi} + \openone \otimes \openone \otimes \mathfrak{h}_\phi  \ .
\end{align}
In the interaction picture, two detectors interact with the field via the interaction Hamiltonian
\begin{align}
    g H^{\ssI}_{\text{int}}(t) &= g \frac{\dd \tau}{\dd t} \sum_{j = \textsc{a},\textsc{b}}  \mu_{j}^{\ssI}\big(\tau(t)\big)\otimes \phi\big[\sy_j\big(\tau(t)\big)\big]  \,,
\end{align}
with the superscript $I$ denoting the interaction picture. The monopole operator of each detector appearing in $H^{\ssI}_{\text{int}}$ is given by
\begin{equation}
    \mu^\ssI_j(\tau) : = e^{+i\mathfrak{h} \tau} \mu_{j} e^{- i\mathfrak{h}\tau} =  \sigma^j_+ e^{i\Omega \tau}+\sigma^j_- e^{-i\Omega \tau} \,,
\end{equation}
where $\mu_{j} \coloneqq \sigma^{j}_{1} = \sigma^j_+ + \sigma^j_-$ is the monopole operator in the Schr\"odinger picture. For simplicity we have chosen each qubit to interact with the field with the same coupling strength $g \ll 1$.

With this setup, the interaction picture density matrix $\rho^{\ssI}(t)$ of the detector-field system evolves according to the Liouville-von Neumann equation
\begin{align} \label{liouville1}
    \frac{\dd \rho^\ssI(t)}{\dd t} = -i g \left[ H^\ssI_\text{int}(t), \rho^\ssI(t)\right] \ .
\end{align}
The aim of this work is to determine the evolution of the joint qubit state i.e.\ the reduced density matrix obtained by tracing over the field's degrees of freedom where
\begin{align}
    \rho^\ssI_{\textsc{ab}}(t) & \coloneqq \tr_\phi[\rho^\ssI(t)]\,.
\end{align}
In this work we assume, as is standard in the literature, that the initial state of the detector-field system at $t=0$ is uncorrelated such that
\begin{align}
    \rho^{\ssI}(0) = \rho_{\textsc{ab}}^{\ssI}(0)\otimes\ketbra{0}{0}\,,
    \label{eq: initial-state}
\end{align}
where $\ket{0}$ is the Minkowski vacuum defined by $a_\bk^{\phantom{\dagger}}\ket{0}=0$ for all $\bk$, and take the initial joint detector state $\rho^{\ssI}_\ab(0)$ to be arbitrary\footnote{An often unstated fact is that pointlike UDW detector model is \textit{incompatible} with arbitrary initial state of the detector, in that it will lead to ultraviolet (UV) divergences. One has to impose (a) UV cutoff or (b) spatial smearing to regulate the UV divergence. In open quantum systems problems, a hard UV cutoff is usually imposed instead of giving a finite size to the detector due to its mathematical tractability (see, e.g., \cite{Satz2006howoften}).}.

For later use, the joint qubit state in the Schr\"odinger picture, denoted by $\rho_\textsc{ab}(t)$, is related to the interaction picture version by
\begin{align}
    \rho_{\textsc{ab}}^\ssI(\tau) = e^{+i \mathfrak{h} \tau} \rho_\ab(\tau) e^{-i \mathfrak{h} \tau} \ .
    \label{eq: Schrodinger-convert}
\end{align}

\section{Late times and master equations}
\label{sec:master}

In this section we begin with the Liouville-von Neumann equation \eqref{liouville1}  and connect its standard perturbative expansion to master equations which are better suited for studying late times. Using the Nakajima-Zwanzig master equation (equivalent to the Born approximation at the order $\mathcal{O}(g^2)$ considered in this work), we then develop explicit integro-differential equations that can be used later to study the Markovian limit. 
 
\subsection{From perturbation theory to master equations}
\label{sec:pert_master}

We begin by noting an equivalent formulation of the Liouville-von Neumann equation \eqref{liouville1},
\begin{eqnarray} \label{liouville2}
    \rho^\ssI(t) &= & \rho^{\ssI}(0) -i g \int_0^t\dd t'\,[ {H^\ssI_\text{int}(t')},\rho^\ssI(0)]
    \\
    &&\hspace{-0.5cm}- g^2 \int_0^t\!\!\dd t' \int_0^{t'}\!\!\!\dd t''\,\bigr[H^\ssI_
    \text{int}(t'),[H^\ssI_\text{int}(t''),\rho^\ssI(t'')]\bigr] \ , \notag
\end{eqnarray}
which lends itself useful to perturbative calculations.  {This equation is derived by inserting the integral version of (\ref{liouville1}),
\begin{eqnarray} \label{liouville2}
    \rho^\ssI(t) &= & \rho^{\ssI}(0) -i g \int_0^t\dd t'\,[H^\ssI_\text{int}(t'),\rho^\ssI(t')]
\end{eqnarray}
into itself iteratively.}  Invoking the standard perturbative (Dyson) series expansion on (\ref{liouville2}) yields to second-order in the qubit-field coupling
\begin{eqnarray} \label{pert_g2}
   && \rho^\ssI(t) = \rho^{\ssI}(0) -i g \int_0^t\dd t'\,[ {H^\ssI_\text{int}(t')},\rho^\ssI(0)]
    \\
    && \  - g^2 \int_0^t\!\!\dd t' \int_0^{t'}\!\!\!\dd t''\,\bigr[H^\ssI_
    \text{int}(t'),[H^\ssI_\text{int}(t''),\rho^\ssI(0)]\bigr] +\mathcal{O}(g^3)\,. \notag
\end{eqnarray}
After a partial trace over the field, the second term vanishes due to the vanishing of the one-point function $\braket{0|\phi(t,\bx)|0}=0$, and the joint state of the detectors up to ${O}(g^2)$ reads
\begin{eqnarray}
   \rho^\ssI_\textsc{ab}(t) & \simeq &  \rho_\textsc{ab}(0)  \label{eq: truncated-dyson} \\
    && -g^2 \!\!\int_0^t\!\!\dd t' \!\!\int_0^{t'}\!\!\!\dd t''\,\tr_\phi\bigr[H^\ssI_
    \text{int}(t'),[H^\ssI_\text{int}(t''),\rho^\ssI(0)]\bigr]\,. \notag
\end{eqnarray}
In many contexts such as entanglement harvesting protocol \cite{reznik2003entanglement,Valentini1991nonlocalcorr,VerSteeg2009entangling,pozas2015harvesting,pozas2016entanglement,Grimmer2021algebraic,tjoa2021entanglement,henderson2018harvestingBH,smith2016topology,Caroline2022harvesting,Tjoa2020vaidya,Gallock-Yoshimura2021freefall,Caroline2022harvesting}, this is the order in which the final state of the detectors are often worked out. 

In this context, the standard perturbative approach applies in a regime where $1 \ll a \tau \ll 1/g^2$, and begins to breakdown\footnote{In this case, temporal smooth switching functions multiplying the interaction Hamiltonian are sometimes used to turn off the interaction before perturbative breakdown occurs.} when $g^2 a \tau \sim \mathcal{O}(1)$  {as outlined in \cite{kaplanek2020hot,kaplanek2020hot2,kaplanek2021qubits}}. The utility of the open system approach is that there exists a Markovian regime  truncated at the same order as \eqref{pert_g2} that allows us to study the same problem in the late-time regime $g^2 a \tau \sim \mathcal{O}(1)$ by ``resummation'' of terms to all orders in $g^2 a\tau$. The way this works is to note that perturbative series like (\ref{pert_g2}) gives time evolution of the system from  $\tau_0$ to $\tau$ so long as $g^2 a (\tau - \tau_{0}) \ll 1$ holds. Within this window, one can differentiate the perturbative expression to yield a differential equation for $\rho(\tau)$. \textit{If} this differential equation is {\it time-local} in $\rho(\tau)$ i.e.~not depending on its entire integrated history of evolution from $\tau_0$ to $\tau$, then the {\it same} differential equation applies in any other perturbatively small window from any $\tau_j$ to $\tau$, so long as $g^2 a (\tau - \tau_{j}) \ll 1$. The master equation then applies over much larger timescales, since it can be trusted over the union of such perturbatively small but overlapping time domains allowing for integration out to late times where $g^2 a\tau \sim \mathcal{O}(1)$ (but $g^4a\tau\ll 1$).

Indeed, the time-local nature of Markovian master equations is the essential property that we need to resum the late-time breakdown of (\ref{pert_g2}) to all orders in $g^2a\tau$ while neglecting $\mathcal{O}(g^4 a \tau)$ effects. This resummation argument is, in essence, a renormalization group argument familiar from particle physics (e.g. see \cite{burgess2020introduction} as well as \cite{Kaplanek2018rindler,kaplanek2020hot,kaplanek2022some} for a late-time analogy using particle decays).

Our task is now clear --- what remains is to:
\begin{enumerate}[label=(\arabic*),leftmargin=*]

    \item turn Eq.~\eqref{pert_g2} into a time-local Markovian equation that is valid up to late times as specified above using suitable approximations;
    
    \item find an explicit, late-time resummed solution to the Markovian regime;
    
    \item find the domain of validity of the approximations that go into (1) and (2), and show that the resulting equation defines a completely-positive evolution.
\end{enumerate}
\noindent Note that Step (3) is often neglected, which may lead to unphysical results. We discuss this further in \S\ref{sec: validity}.

To gain access to late times, we apply the \textit{Born approximation}\footnote{The Born approximation is a weaker requirement compared to of the standard Dyson series truncation \eqref{eq: truncated-dyson} because we can have $\rho_\textsc{ab}^\ssI(t'')\not\simeq\rho^\ssI_\textsc{ab}(0)$ for long interactions without significantly changing the bath state.} to Eq.\ (\ref{liouville2}),
\begin{align}
    \rho^\ssI(t) \simeq \rho^{\ssI}_{\textsc{ab}}(t)\otimes \ket{0}\!\bra{0} \ , 
\end{align}
which neglects correlations between the joint qubit state and the field  \footnote{These correlations can be shown to be $\mathcal{O}(g^2)$ \cite{Martin2018cosmologydecohere} so it only contributes as $\mathcal{O}(g^4)$ effect. For any significant backreaction onto the the field state, the Born approximation is not valid.}. The resulting state at $\mathcal{O}(g^2)$ reads
\begin{eqnarray}
    &&\rho^\ssI_\textsc{ab}(t) \simeq \rho_\textsc{ab}(0) 
    \label{eq: truncated-born} \\
    &&-g^2 \!\!\int_0^t\!\!\dd t' \!\!\int_0^{t'}\!\!\!\dd t''\,\tr_\phi\Scale[0.90]{\Bigr[H^\ssI_
    \text{int}(t'),\bigr[H^\ssI_\text{int}(t''),\rho^\ssI_\textsc{ab}(t'')\otimes \ket{0}\!\bra{0}\bigr]\Bigr]}\,. \notag
\end{eqnarray}
By taking the derivative with respect to $t$, we obtain the integro-differential equation
\begin{align}
    \frac{\dd\rho^\ssI_\textsc{ab}}{\dd t} & \simeq  -g^2 \!\!\int_0^{t}\!\!\!\dd t'\tr_\phi\bigr[H^\ssI_
    \text{int}(t),[H^\ssI_\text{int}(t'),\rho^\ssI_\textsc{ab}(t')\otimes \ket{0}\!\bra{0}]\bigr]\,.
    \label{eq: NZ-equation}
\end{align}
This equation is useful in that it only depends on $\rho^\ssI_\textsc{ab}$, unlike \eqref{pert_g2}. Note however that Eq.~\eqref{eq: NZ-equation} is not time-local because it depends on the entire history (``memory'') of its evolution.

It is worth emphasizing that the equation of motion of the form Eq.~\eqref{eq: NZ-equation} is precisely the $\mathcal{O}(g^2)$-truncation of the \textit{Nakajima-Zwanzig equation} \cite{Nakajima1958,Zwanzig1960} (see e.g.\cite{lidar2020lecture,breuer2002theory} for introductory material). The basic idea behind the Nakajima-Zwanzig equation is that we can define a projection $\mathcal{P}$ such that
\begin{align}
    \mathcal{P}[\rho^\ssI(t)]\coloneqq \rho_{\textsc{ab}}^\ssI(t) \otimes\ket{0}\!\bra{0} \ .
\end{align}
This splits the total state into ``relevant'' part projected by $\mathcal{P}$ (the system) and ``irrelevant'' part projected by its complement $\openone-\mathcal{P}$ (the environment). Since the Liouville-von Neumann equation (\ref{liouville1}) is linear one can use it to derive an exact equation of motion for $\mathcal{P}[\rho^\ssI(t)]$ alone which is here of the form
\begin{align}
    \frac{\partial\mathcal{P}[\rho^\ssI(t)]}{\partial t} = \int_0^t\dd t'\mathcal{K}(t,t')\mathcal{P}[\rho^\ssI(t')]\,,
    \label{eq: NZ-feshbach}
\end{align}
where $\mathcal{K}(t,t')$ is a ``memory kernel'' that measures information backflow from the detector to the field. The nice feature of the Nakajima-Zwanzig equation is that what we called the ``Born approximation'' in Eq.~(\ref{eq: NZ-equation}) is naturally built-in as the leading-order expansion of the memory kernel $\mathcal{K}(t,t')$, thus it has the natural interpretation that indeed we are neglecting memory effect due to backreaction to the field. The Nakajima-Zwanzig formalism provides a very natural organizing principle for perturbative expansion in a way that makes clear the information flow between the ``relevant part'' (the system) and ``irrelevant part'' (the environment) of the total system. 
Following \cite{lidar2020lecture}, we refer to Eq.~(\ref{eq: NZ-equation}) as the Nakajima-Zwanzig master equation at second order (NZ-ME2).

\subsection{Nakajima-Zwanzig equation for two accelerated qubits}

After tracing out the field degrees of freedom, NZ-ME2 \eqref{eq: NZ-equation} results in 
\begin{align} \label{NZ_minkowski_t}
   & \frac{\dd \rho^\ssI_\ab}{\dd t} = g^2  \sum_{j,k \in \{\mathrm{A},\mathrm{B}\}} \; \!\!\int_0^t\!\!\dd t'\frac{\dd \tau(t)}{\dd t} \frac{\dd \tau(t')}{\dd t'} \\
   & \ \ \times \Bigr(\mathcal{W}_{jk}\big( \tau(t), \tau(t') \big) \left[ \mu^\ssI_k\big( \tau(t') \big) \rho_\textsc{ab}^\ssI(t') , \mu^\ssI_j\big( \tau(t) \big) \right] + \mathrm{H.c.} \Bigr) \notag
\end{align}
where $\mathrm{H.c.}$ denotes Hermitian conjugate and the sum runs over the labels $j,k \in \{\textsc{A,B}\}$. Here $t,t'$ are Minkowski time variables, $\tau$ is the common proper time for the detectors. The pullback of the vacuum Wightman two-point function along the trajectories $\sy_j,\sy_k$ denoted $\mathcal{W}_{jk}$, is given by
\begin{align} \label{W_general}
    \mathcal{W}_{jk}(\tau,\tau') \coloneqq \braket{0|\phi(\sy_j(\tau)) \phi(\sy_k(\tau'))|0} .
\end{align}
For the parallel accelerated trajectories \eqref{eq: trajectories}, $\mathcal{W}_{jk}$ simplifies greatly: we get the ``self-correlations'' \cite{sciama1981quantum}
\begin{eqnarray} \label{eq: Wightman-functions1}
    \mathcal{W}_{\s}(\Delta \tau ) & : = & \mathcal{W}_{\textsc{aa}}(\tau,\tau') = \mathcal{W}_{\textsc{bb}}(\tau,\tau') \notag\\
    &=& -\frac{a^2}{16\pi^2}\frac{1}{\sinh^2\left[\frac{a}{2}(\Delta \tau -i\epsilon)\right]}\,,
\end{eqnarray}
and ``cross-correlations'' 
\begin{eqnarray} \label{eq: Wightman-functions2}
    \mathcal{W}_{\times}(\Delta \tau ) & \coloneqq & \mathcal{W}_{\textsc{ab}}(\tau,\tau')=\mathcal{W}_{\textsc{ba}}(\tau,\tau')\notag\\
    &= &-\frac{a^2}{16\pi^2}\frac{1}{\sinh^2\left[\frac{a}{2}(\Delta \tau -i\epsilon)\right]-(\frac{aL}{2})^2} 
\end{eqnarray}
where $\Delta\tau=\tau-\tau'$\footnote{ {These are derived from the Wightman function (\ref{W_general}), $\langle 0| \phi(x) \phi(x') |0 \rangle = \left[ 4 \pi^2 ( - (x^0 - x^{\prime 0} - i \epsilon)^2 + |\mathbf{x} - \mathbf{x}'|  )\right]^{-1}$, evaluated for $x = \sy_j(\tau)$ and $x' = \sy_k(\tau')$. }}. 

By performing a change of variable $s=\Delta\tau$, the resulting NZ-ME2 can be re-expressed in terms of proper time $\tau$ such that
\begin{eqnarray}
    & \ & \frac{\dd \rho^\ssI_\textsc{ab}}{\dd \tau} \simeq \; g^2 \sum_{j,k \in \{\mathrm{A},\mathrm{B}\}} \; \int_0^\tau\dd s\; \label{eq: NZ-final} \\
   & \  & \quad  \times \Bigr( \; \mathcal{W}_{jk}(s) \bigr[\mu_{j}^\ssI(\tau-s)\rho^\ssI_\textsc{ab}(\tau-s), \mu^\ssI_{k}(\tau)\bigr] + \mathrm{H.c.} \; \Bigr)\notag \,. 
\end{eqnarray}
Using the uncoupled basis $\{\ket{\uparrow\uparrow},\ket{\uparrow\downarrow},\ket{\downarrow\uparrow},\ket{\downarrow\downarrow}\}$, the integro-differential equation \eqref{eq: NZ-final} for $\rho_{\textsc{ab}}^\ssI(\tau)$ has two decoupled components: that is, we can split the density matrix into
\begin{equation} \label{XO_split}
\rho^\ssI_\textsc{ab}(\tau) = \rho^\ssI_{\textsc{ab},X}(\tau) + \rho^\ssI_{\textsc{ab},O}(\tau)
\end{equation}
where
\begin{subequations}
\label{eq: blocks}
\begin{align}
    \rho^\ssI_{\textsc{ab},X}(\tau) &= 
        \left[\begin{matrix}
             \rho^\ssI_{11}(\tau) & 0  & 0 & \rho^\ssI_{14}(\tau) \\
             0 & \rho^\ssI_{22}(\tau) & \rho^\ssI_{23}(\tau) & 0 \\
             0 & \rho^{\ssI}_{32}(\tau) & \rho^\ssI_{33}(\tau) & 0 \\
             \rho^{\ssI}_{41}(\tau) & 0 & 0 & \rho_{44}^{\ssI}(\tau)
        \end{matrix}\right]\,,\\
    {\rho^\ssI_{\textsc{ab},O}(\tau)} &=     
        \left[\begin{matrix}
         0 & \rho^\ssI_{12}(\tau) & \rho^\ssI_{13}(\tau) & 0 \\
         \rho^{\ssI}_{21}(\tau) & 0 & 0 & \rho^\ssI_{24}(\tau) \\
         \rho^{\ssI}_{31}(\tau) & 0 & 0 & \rho^\ssI_{34}(\tau) \\
         0 & \rho^{\ssI}_{42}(\tau) & \rho^{\ssI}_{43}(\tau) & 0
    \end{matrix}\right]\,.
\end{align}
\end{subequations}
\noindent We call these decoupled pieces the $X$-block and $O$-block respectively (due to the positions of the nonzero matrix elements). Note that $\rho_{\textsc{ab},X}^\ssI(\tau)$ is known as an $X$-state and some of its properties have been investigated in the literature (see e.g. \cite{Ali2010discord}). The components of Eq.~\eqref{eq: blocks} are not all independent since we can use 
\begin{equation}
    \begin{aligned}
    \rho_{44}^{\ssI}(\tau) & = 1 - \rho_{11}^{\ssI}(\tau) - \rho_{22}^{\ssI}(\tau) - \rho_{33}^{\ssI}(\tau) \ , \\
    \rho_{nm}^{\ssI}(\tau) & = \rho_{mn}^{\ssI\ast}(\tau) \quad \mathrm{for\ }n \neq m \in \{1,2,3,4\}\,.
    \end{aligned}
\end{equation}
Therefore, for the $X$-block we have a system of seven coupled ordinary differential equations (ODE), while for $O$-block we have eight coupled ODEs with eight variables. The full explicit expressions for the ODEs for $X$-block and $O$-block are given in Appendix~\ref{sec: NZ-ME2_X} and Appendix~\ref{sec: NZ-ME2_O} respectively.

Before we solve these equations, let us remark on the choice of initial state $\rho_\textsc{ab}(0)$. Interestingly, many existing studies involving two UDW detectors in open system framework restricts their attention to an $X$-state as the initial state \cite{Benatti2004Unruh,Zhou2021massiveentanglement,Zhou2021massiveentanglement2,Hu2015twodetectorent,huang2017dynamics,benatti2022local,Soares2022open,Zhang2007reflect} because the time evolution preserves the $X$-block \cite{Tanas2013}. Here we see that this restriction is unnecessary since $X$-block completely decouples from $O$-block, so one can evolve the $O$-block independently anyway. We see in the next section (for nonzero energy gap $\Omega>0$) that the $O$-block tends towards zero at late-times, so only the $X$-block survives in the long time limit.

\section{Two-qubit asymptotic state in the Markovian regime}
\label{sec: markovian}

As mentioned earlier, the main obstruction to solving Eq.~\eqref{eq: NZ-final} (or equivalently the ODEs given in Appendices \ref{sec: NZ-ME2_X} and \ref{sec: NZ-ME2_O}) and obtaining late-time results is that it is not time-local: the integrals on the RHS have memory over their entire history of the evolution integrating over functions of $\tau-s$. To gain access to late time, we must enter a \textit{Markovian} regime where the dynamics is memoryless. As we will shall see, there are subtleties involved in taking Markovian limit of Eq.~\eqref{eq: NZ-final}.

\subsection{A different Markovian approximation}
\label{subsec: correct-markov}

Let us start with a Markovian approximation of Eq.~(\ref{eq: NZ-final}) that is \textit{different} from the one usually taken in the literature. We see how this compares with the standard approach in \S\ref{subsec: markovian-common}.

The physical essence of the Markovian limit is the observation that the environment correlators $\mathcal{W}_{\textsc{s}, \times}(s)$ are sharply peaked about $s=0$: this implies that there exists a regime in which the timescales associated with the system evolve much slower than the timescale set by the environment. This timescale is  {set by $1/a$} as can be seen from Eqs.~(\ref{eq: Wightman-functions1}) and (\ref{eq: Wightman-functions2}) where the environment correlators fall off exponentially fast:
\begin{subequations}
    \begin{align} 
        \label{Ws_falloff}
        \mathcal{W}_{\s}(s) &\simeq - \tfrac{a^2}{4\pi^2} e^{- a s} \quad 
        as \gg 1\,,\\
        \label{Wx_falloff}
        \mathcal{W}_{\times}(s) &\simeq - \tfrac{a^2}{4\pi^2} e^{- a s} \quad 
        as \gg \max\left\{ 1, \tfrac{1}{2} \log(\tfrac{aL}{2}) \right\}\!.
    \end{align}
\end{subequations}
This means that when the system evolves slower than the environment, the history/memory dependence in the RHS of (\ref{eq: NZ-final}) becomes negligible\footnote{It is sometimes colloquially stated that fast (Markovian) environment dynamics means that $\mathcal{W}_{\s, \times}(s)\sim \delta(s)$ underneath the integral sign in (\ref{eq: NZ-final}) --- from this point of view it also makes sense that one removes the history of integration as done in Eq.~\eqref{eq: memoryless}.} and we can perform Taylor series about $s=0$ where
\begin{eqnarray} 
\label{MarkovianTaylor}
    && \mu_{j}^{\ssI}(\tau - s) \rho_{\textsc{ab}}^{\ssI}(\tau - s)  \\
    && \ \ \simeq  \mu^{\ssI}_{j}(\tau)  \rho_{\textsc{ab}}^{\ssI}(\tau)  - s \big( \Scale[0.93]{ \mu_{j}^{\ssI}(\tau) \dot{\rho}_{\textsc{ab}}^{\ssI}(\tau) + \dot{\mu}_j^{\ssI}(\tau) \rho_{\textsc{ab}}^{\ssI}(\tau) } \big) + \mathcal{O}(s^2)\,, \notag
\end{eqnarray}
and a similar Taylor series for the opposite ordering of operators $\rho_{\textsc{ab}}^{\ssI}(\tau - s) \mu_{j}^{\ssI}(\tau - s)$ in (\ref{eq: NZ-final}). Therefore, the leading order of the series expansion is memoryless and reads
\begin{equation}
    \frac{\dd \rho^\ssI_\textsc{ab}}{\dd \tau} \simeq g^2 \sum_{j,k} \int_0^\tau\dd s\; \Scale[0.90]{ \Bigr( \mathcal{W}_{jk}(s) \bigr[\mu_{j}^\ssI(\tau)\rho^\ssI_\textsc{ab}(\tau), \mu^\ssI_{k}(\tau)\bigr] + \mathrm{H.c.}  \Bigr) }\,, \label{eq: memoryless}
\end{equation}
and only the correlators $\mathcal{W}_{\s,\times}$ depend on $s$. 

Physically, the series expansion leading to \eqref{eq: memoryless} essentially makes $1/a$ the shortest timescale of the problem, i.e. the Markovian limit is where the environment dynamics are extremely rapid as compared to the system dynamics. In this regime, the environment erases the history of integration in Eq.~(\ref{eq: NZ-final}), {\it including} the integration over the (system) monopole operators $\mu_{j}^{\ssI}$. This is a very important point that we stress from time to time in this work: since we are in the interaction picture, memoryless Markovian limit requires that the memory is neglected from \textit{both} the monopole operators \textit{and} the states, otherwise the residual memory leads to problems as we see later on.

Another noteworthy point that is often neglected is that the Taylor approximation step from (\ref{eq: NZ-final}) to (\ref{eq: memoryless}) provides a means of quantifying {\it when} the Markovian approximation applies: indeed, it is now clear that the Markovian limit begins to fail when the subleading derivative terms in (\ref{MarkovianTaylor}) become too large. Bounding the next-to-leading-order terms in (\ref{MarkovianTaylor}) to be small relative to the leading-order terms maps out the parameter space where the Markovian approximation applies. We explore such Markovian validity bounds in full detail in \S\ref{sec: validity} so as to determine the parameter ranges for which we can trust the Markovian evolution predicted by Eq.~(\ref{eq: memoryless}).

\subsubsection*{Time-dependent coefficients and late times}

Finally, while Eq.~\eqref{eq: memoryless} is already memoryless (or time-local), it turns out the upper limit of $\tau$ on the integrals prevents straightforward integration out to very late times for the problem at hand. To see why, note that Eq.~\eqref{eq: memoryless} can be recast into a system of \textit{ordinary differential equations} (ODEs) for a vector $\mathbf{u}(\tau)$ whose components are built out of the entries of the density matrix  $\rho_\textsc{ab}^\ssI(\tau)$: schematically, it takes the general form
\begin{align}
    \frac{\dd\mathbf{u}(\tau)}{\dd\tau} = \mathbb{A}(\tau)\mathbf{u}(\tau) + \mathbf{v}(\tau) \,,
    \label{eq: matrix-ODE-formal}
\end{align}
where $\mathbf{v}(\tau)$ is some possibly nonzero vector and $\mathbb{A}(\tau)$ is a square matrix, both of whose entries contain the integral transforms of environment correlators $\mathcal{W}_{\s,\times}(s)$ appearing in Eq.~\eqref{eq: memoryless} (with upper limits $\tau$ on the integrals). Even for the case when $\mathbf{v}(\tau)=\mathbf{0}$, the matrix ODE above cannot be solved in closed form unless $\mathbb{A}(\tau)$ obeys very specific properties\footnote{Such as when $[\mathbb{A}(s),\mathbb{A}(s')]=0$ for all $s,s'\in (0,\tau)$.}. Formal solutions to \eqref{eq: matrix-ODE-formal} generally involve time-ordered exponentials which in turn requires perturbative treatments that the late-time resummation was meant to avoid.

Another important point that is often missed is that although Eq.~\eqref{eq: matrix-ODE-formal} can be organized into a Lindblad-like form (see Eq.~\ref{eq: master-equation-dissipator}), the resulting coefficients are \textit{time-dependent} and therefore this master equation does {\it not} obey the assumptions of the Gorini-Kossakowski-Sudarshan-Lindblad (GKSL) theorems stated in  \cite{lindblad1976generators,kossakowski1972quantum}. In particular, if one tries to put Eq.~\eqref{eq: matrix-ODE-formal} into ``Lindblad form'', the corresponding \textit{Kossakowski matrix} of Lindblad coefficients can be apparently negative-definite especially at early times, which is usually taken to be a violation of the complete-positivity (CP) property for the evolution. However, Lindblad's theorem only applies for master equations in Lindblad form with time-independent Kossakowski matrix (see \cite{whitney2008staying} for more details\footnote{In essence this is because the Lindblad generator as defined in \cite{gorini1976completely,lindblad1976generators} is a single time-independent (possibly unbounded) operator.}). Consequently, one cannot make claims about the CP property of the evolution (or lack thereof) by invoking Lindblad's theorem at this stage. In fact, since we cannot perform late-time resummation yet with time-dependent $\mathbb{A}(\tau)$, it is pointless to check the CP properties though the GKSL theorems at this juncture.

The upshot is that the \textit{practical} calculation of late-time resummation requires more than just removing memory effect: we need to find a regime where the time-local equation \eqref{eq: memoryless} can be approximated as a matrix ODE with constant coefficients, which is exactly solvable in closed form. That is, we need to work in the regime where Eq.~\eqref{eq: matrix-ODE-formal} reduces to
\begin{align}
    \frac{\dd\mathbf{u}(\tau)}{\dd\tau} = \mathbb{A}\mathbf{u}(\tau) + \mathbf{v} \,,
    \label{eq: matrix-ODE-markov}
\end{align}
where $\mathbb{A}$ is now a constant matrix and $\mathbf{v}$ is also a constant vector. Indeed, this is the case in later sections when we consider late-time dynamics of the detectors.

To this end, given the falloff of the environment correlators given in (\ref{Ws_falloff}) and (\ref{Wx_falloff}), we assume that 
\begin{equation} \label{infinity_approx}
a \tau \gg \max\left\{ 1, \tfrac{1}{2} \log(\tfrac{aL}{2}) \right\} 
\end{equation}
so that we can approximate $\sim \infty$ on the upper limit of the integrals in our Markovian equation of motion \eqref{eq: memoryless} giving
\begin{equation}
    \frac{\dd \rho^\ssI_\textsc{ab}}{\dd \tau} \simeq \; g^2 \sum_{j,k} \int_0^\infty\!\!\!\!\dd s\; \Scale[0.90]{ \Bigr( \mathcal{W}_{jk}(s) \bigr[\mu_{j}^\ssI(\tau)\rho^\ssI_\textsc{ab}(\tau), \mu^\ssI_{k}(\tau)\bigr] + \mathrm{H.c.}  \Bigr) } \label{eq: memoryless2} \,.
\end{equation}
This is the master equation whose late-time resummed solution is amenable to explicit computation. 

We solve this equation in \S\ref{subsec: GKLS}, but first we take a small detour to discuss why we use the stated Markovian approximation instead of the standard one usually taken in the literature.

\subsection{On other commonly-used versions of the Markovian approximation}
\label{subsec: markovian-common}

In the majority of the literature,  {most of which are based on the approach described in \cite{breuer2002theory,lidar2020lecture}}, the ``standard'' Markovian approximation is usually taken as the approximation
\begin{equation} 
\label{MarkovianTaylor_lit}
    \rho_{\textsc{ab}}^{\ssI}(\tau - s) \simeq  \rho_{\textsc{ab}}^{\ssI}(\tau)  - s \; \dot{\rho}_{\textsc{ab}}^{\ssI}(\tau) + \ldots 
\end{equation}
as opposed to Eq.~(\ref{MarkovianTaylor}) taken above.  {That is, instead of applying the series expansion about $\tau$ on both the observable and the state $\mu_{j}^{\ssI}(\tau - s) \rho_{\textsc{ab}}^{\ssI}(\tau - s)$, the usual approach only applies the series expansion to the density operator $\rho_{\textsc{ab}}^{\ssI}(\tau - s)$ while keeping the monopole operator $\mu_{j}^{\ssI}(\tau-s)$ intact. The folklore is that Markovian regime is when the state has no memory about its past history.}

We now argue that the self-consistent way of taking the standard approximation (\ref{MarkovianTaylor_lit}) for the setup at hand is exactly the one considered in this work  {(which in turn is inspired by \cite{Martin2018cosmologydecohere}).} The usual logic for keeping only the leading-order term of the state $\rho_{\textsc{ab}}^{\ssI}(\tau - s) \simeq  \rho_{\textsc{ab}}^{\ssI}(\tau)  - s \dot{\rho}_{\textsc{ab}}^{\ssI}(\tau) + \ldots$  in any integro-differential master equations, such as those that arise from Nakajima-Zwanzig formalism, is that we are guaranteed that $\dot{\rho}_{\textsc{ab}}^{\ssI}(\tau) \sim \mathcal{O}(g^2)$ from the master equation we start from. Since these derivative terms in the approximation are suppressed by two extra powers of the coupling strength, this seems to imply that all the subleading terms in the Taylor series can be safely ignored.  What this argument fails to account for is that the derivative terms can become dangerously large if energy scales associated with the system become too large (compared to the energy scales of the environment). 

In order to demonstrate where the problem lies, it is most easily illustrated for the single detector case where $\Omega / a \gtrsim 1$ alone causes the Markovian approximation to break down (for two detectors there are more conditions, as we see in \S\ref{sec: validity}). By ignoring Bob's detector entirely and only focusing on Alice's detector, one gets the analog of the Nakajima-Zwanzig equation (\ref{eq: NZ-final}) for Alice's reduced density matrix $\rho^{\ssI}_{\textsc{a}} \coloneqq \mathrm{tr}_\textsc{b}\left[ \rho^{\ssI}_{\textsc{ab}} \right]$ where (see \S3.2 of \cite{kaplanek2020hot} and also \cite{kaplanek2020hot2,kaplanek2021qubits,kaplanek2021qubit} for more details)
\begin{equation} \label{NZ_Alice}
\frac{\dd \rho^\ssI_{\textsc{a}}}{\dd \tau} \simeq g^2 \int_0^\tau\dd s\ \Bigr(\Scale[0.90]{ \mathcal{W}_{\s}(s)\bigr[\mu^\ssI(\tau-s)\rho^\ssI_\textsc{a}(\tau-s), \mu^\ssI(\tau) \bigr] + \mathrm{h.c.} } \Bigr)
\end{equation}
where $\mu^{\ssI} \coloneqq \mathrm{tr}_{\textsc{b}} [ \mu_\textsc{a}^\ssI ]$. The failure of the Markovian approximation in the case of $\Omega / a \gtrsim 1$ is most easily appreciated by studying the evolution of the off-diagonal components $\rho^{\ssI}_{\textsc{a},12}$ of Alice's reduced density matrix\footnote{In Eq.~(\ref{NZ_Alice_off}) and those that follow, one should include a renormalization of Alice's detector gap in order to get the Markovian solution (\ref{Alice_sol}) --- we omit this detail here, see Appendix \ref{App:Alice}.}
\begin{eqnarray} \label{NZ_Alice_off}
& \ & \frac{\dd \rho^\ssI_{\textsc{a},12}}{\dd \tau} \simeq  - 2g^2 \int_0^\tau \dd s\ \mathrm{Re}\left[ \mathcal{W}_{\s}(s) \right] e^{+ i \Omega s} \rho^\ssI_{\textsc{a},12}(\tau - s) \\
& \  & \qquad \ \ \  + 2g^2 e^{+ 2 i \Omega \tau} \int_0^\tau \dd s\ \mathrm{Re}\left[ \mathcal{W}_{\s}(s) \right] e^{- i \Omega s} \rho^{\ssI\ast}_{\textsc{a},12}(\tau - s) \ . \notag
\end{eqnarray}
If one uses (\ref{MarkovianTaylor_lit}), as is usually done in the literature, then the resulting Markovian equation of motion for the off-diagonal components yields
\begin{eqnarray} \label{Markov_Alice_off}
\frac{\dd \rho^\ssI_{\textsc{a},12}}{\dd \tau} & \simeq & - 2 g^2 \int_0^\infty \dd s\ \mathrm{Re}\left[ \mathcal{W}_{\s}(s) \right] e^{+ i \Omega s} \rho^\ssI_{\textsc{a},12}(\tau) \\
& \  & + 2 g^2 e^{+ 2 i \Omega \tau} \int_0^\infty \dd s\ \mathrm{Re}\left[ \mathcal{W}_{\s}(s) \right] e^{- i \Omega s} \rho^{\ssI\ast}_{\textsc{a},12}(\tau) \ . \notag
\end{eqnarray}
where we have also assumed $a \tau \gg 1$ so the upper limit on the integral can be approximated by $\simeq \infty$. The time-local differential equation (\ref{Markov_Alice_off}) turns out to have the solution in the non-degenerate regime $g^2 \mathcal{C} \ll \Omega$
\begin{equation} \label{Alice_sol}
\rho^\ssI_{\textsc{a},12}(\tau) \simeq \mathcal{A} e^{ - g^2 \mathcal{C} \tau} + \mathcal{B} e^{(- g^2 \mathcal{C} + 2 i \Omega ) \tau}
\end{equation}
where $\mathcal{C}$ is given by
\begin{equation} \label{mathcalC_def}
\mathcal{C} \coloneqq 2 \int_0^\infty \exd s \Scale[0.90]{\ \mathrm{Re}\left[ \mathcal{W}_{\s}(s) \right] \cos(\Omega s) } = \sfrac{\Omega}{4\pi} \coth\left( \sfrac{\pi\Omega}{a} \right)  \ , 
\end{equation}
and $\mathcal{A}$ and $\mathcal{B}$ are constant coefficients which can be explicitly computed but whose form is not important in what follows (see Appendix \ref{App:Alice}).

To check when the Markovian approximation is valid, one can insert the solution (\ref{Markov_Alice_off}) and see when (\ref{NZ_Alice_off}) is well-approximated by (\ref{Markov_Alice_off}). Although somewhat tedious (see Appendix \ref{App:Alice}), it can be shown that we require
\begin{equation} \label{Alice_whentrue}
\int_0^\tau \exd s\; \Scale[0.90]{ \mathrm{Re}\left[ \mathcal{W}_{\s}(s) \right] e^{+ i \Omega s} e^{ + g^2 \mathcal{C} s} } \simeq \int_0^\infty \exd s\;  \Scale[0.90]{  \mathrm{Re}\left[ \mathcal{W}_{\s}(s) \right] e^{- i \Omega s} }\,.
\end{equation}
Noting that $\mathrm{Re}\left[ \mathcal{W}_{\s}(s) \right] \propto e^{-as}$ for $as \gg 1$, the above approximation can only work when
\begin{subequations}
    \begin{align}
        g^2 \mathcal{C} &\ll a \label{g2C}\,,\\
    e^{+ i \Omega s} &\simeq e^{- i \Omega s} \simeq 1\,.
    \label{exps}
    \end{align}
\end{subequations}
The main point here is that while the first condition \eqref{g2C} can be suppressed by making $g$ sufficiently small, the second condition (\ref{exps}) can hold underneath the integral sign in (\ref{Alice_whentrue}) only when $\Omega \ll a$. In other words, for $\Omega/a \gtrsim 1$ it is impossible to satisfy the approximation (\ref{Alice_whentrue}), and so the Markovian approximation cannot be consistently applied. The argument given here is related to the diagnostic for the failure of the Markovian approximation described in \cite{kaplanek2020hot} where it was explicitly shown that derivative terms in the Taylor series (\ref{MarkovianTaylor_lit}) become too large when $\Omega/a \gtrsim 1$. This extra requirement that $\Omega/a\ll 1$ is often missed in the literature {(although stated in \cite{Moustos2017nonmarkov,kaplanek2020hot,kaplanek2020hot2,kaplanek2021qubit})}.

What we have seen above is that the correct way of taking the Markovian approximation necessarily accounts for the requirement that $\Omega/a\ll 1$. We now claim that our version of Markovian approximation in \S\ref{subsec: correct-markov} {(inspired by the work in \cite{Martin2018cosmologydecohere})} does account for this. Following \S\ref{subsec: correct-markov}, the right way of taking the approximation for Alice's qubit is to perform Taylor series expansion about $\tau$ on \textit{both} the monopole and the state:
\begin{eqnarray} \label{MarkovianTaylor2}
    & & \mu^{\ssI}(\tau - s) \rho_{\textsc{a}}^{\ssI}(\tau - s)   \\
    & & \ \ \ \simeq  \mu^{\ssI}(\tau)  \rho_{\textsc{a}}^{\ssI}(\tau)  - s \big( \Scale[0.93]{ \mu^{\ssI}(\tau) \dot{\rho}_{\textsc{a}}^{\ssI}(\tau) + \dot{\mu}^{\ssI}(\tau) \rho_{\textsc{a}}^{\ssI}(\tau) } \big) + \ldots \ , \notag
\end{eqnarray}
Note that by approximating $\mu_{j}^{\ssI}(\tau - s) \simeq \mu_{j}^{\ssI}(\tau)$ underneath the integral sign, we are automatically requiring that the energy scales associated with the system (in this case $\Omega$ appearing in the detector's monopole operator) are small compared to those associated with the environment (encoded in the falloff of the environment correlators). This amounts to erasing any history-dependence in the system operators {\it as well as} the state itself. Furthermore, by bounding the derivative terms in (\ref{MarkovianTaylor2}) relative to the leading-order term, it can be checked explicitly for a single detector that indeed the Markovian approximation is valid only when $\Omega / a \ll 1$. 

In fact, what we believe to be the correct Markovian approximation also solves other problems that otherwise require further renormalization of divergences or other sleights of hands. Using the approximation (\ref{MarkovianTaylor2}) on (\ref{NZ_Alice_off}) gives us a different Markovian equation of motion ({\it c.f.} Eq.~(\ref{Markov_Alice_off}))  
\begin{eqnarray} 
\frac{\dd \rho^\ssI_{\textsc{a},12}}{\dd \tau} & \simeq & - 2 g^2 \int_0^\infty \dd s\ \mathrm{Re}\left[ \mathcal{W}_{\s}(s) \right] \rho^\ssI_{\textsc{a},12}(\tau) \\
& \  & + 2 g^2 e^{+ 2 i \Omega \tau} \int_0^\infty \dd s\ \mathrm{Re}\left[ \mathcal{W}_{\s}(s) \right] \rho^{\ssI\ast}_{\textsc{a},12}(\tau) \ . \notag
\end{eqnarray}
which has the solution ({\it c.f.} Eq.~(\ref{Alice_sol})) 
\begin{align}
    \label{Alice_sol2}
    &\rho^\ssI_{\textsc{a},12}(\tau)  \simeq  A_{\s} e^{ - g^2 C_\s \tau} + B_{\s} e^{(- g^2 C_\s + 2 i \Omega ) \tau}\,,\\
    A_{\s} &= \rho^\ssI_{\textsc{a},12}(0) +  \frac{ig^2 C_{\s}}{2\Omega} \rho^{\ssI\ast}_{\textsc{a},12}(0)\,,\,
    B_{\s} = - \frac{ig^2 C_{\s}}{2\Omega} \rho^{\ssI\ast}_{\textsc{a},12}(0)\notag\,,
\end{align}
and the constant $C_\s$ reads
\begin{equation} \label{Cs_def}
C_{\s} \coloneqq \lim_{\Omega \to 0^{+}} \mathcal{C} = \int_0^{\infty} \exd s \; \mathrm{Re}\left[ \mathcal{W}_{\s}(s) \right] = \frac{a}{4\pi^2} \ .
\end{equation}
This solution has three crucial features: (1) it is \textit{much} simpler than Eq.~\eqref{Alice_sol}, (2) it is independent of any UV divergences, and most importantly (3) contrary to what is often said in the literature, it preserves complete-positivity (CP) property for the entirety of its evolution \textit{without further approximations} (including RWA) \cite{kaplanek2020hot}. The requirement $\Omega/a\ll1$ is encoded in the definition of $C_\s$, since $\mathcal{C}/a=C_\s/a +\mathcal{O}(\Omega^2/a^2)$. It is important to note that the Markovian solution still has dependence on $\Omega$: it just cannot appear in the ``matrix coefficients'' of the Markovian equation of motion.

In the remainder of this paper, we use this logic in applying the analog of the Markovian approximation (\ref{MarkovianTaylor2}), namely Eq.~\eqref{MarkovianTaylor}, to our more complicated two-detector problem. We see later that the ``correct'' Markovian approximation also circumvents the need for tracking any UV divergences often associated with the ``standard'' Markovian approximation and will preserve complete positivity without RWA in the regime where Markovian limit is valid. Before doing so, we first show in the next subsection that (\ref{MarkovianTaylor2}) results in a Lindblad equation which makes manifest the CP-preserving property and hence why RWA is not necessary.

\subsection{Gorini-Kossakowski-Sudarshan-Lindblad (GKSL) form}
\label{subsec: GKLS}

The best way to show why the Markovian approximation (\ref{MarkovianTaylor2}) results in CP-preserving dynamics (without further approximation) is to cast our master equation into a Schr\"odinger picture equation of the form
\begin{align}
    \frac{\partial\rho_\ab}{\partial \tau } &=  -i[\mathfrak{h}_{\mathrm{eff}},\rho_\ab(\tau)] + \mathcal{D}[\rho_\ab(\tau)]\,.
    \label{eq: master-equation-dissipator}
\end{align}
Eq.~\eqref{eq: master-equation-dissipator} takes the form analogous to the the Liouville-von Neumann equation with an effective Hamiltonian $\mathfrak{h}_{\mathrm{eff}}$ (sometimes called the Lamb shift) and with an extra term involving the dissipation superoperator $\mathcal{D}$ capturing the non-unitary open dynamics of the time evolution generated by the master equation. 
For bipartite qubits, Eq.~\eqref{eq: master-equation-dissipator} is said to be in \textit{Gorini-Kossakowski-Sudarshan-Lindblad} (GKSL) \textit{form} if the dissipator has the form \cite{gorini1976completely,lindblad1976generators}: %
\begin{align} \label{dissipator}
    \mathcal{D}[\rho_\ab ] =   \sum_{\alpha,\beta=1}^3\sum_{j,k=\textsc{a,b}} \Scale[0.95]{ \gamma^{\alpha\beta}_{jk}  \rr{\sigma_\beta^k\rho_\ab \sigma_\alpha^j - \tfrac{1}{2}\left\{\sigma_\alpha^j\sigma_\beta^k,\rho_\ab\right\}} } \,
\end{align}
where the matrix $\boldsymbol{\gamma} \coloneqq [\gamma_{jk}^{\alpha\beta}]$ is known as the \textit{Kossakowski matrix}. The time evolution of the joint qubit state $\rho_\textsc{ab}$ is said to be \textit{completely positive} (CP) if the linear superoperator $\Phi_\tau: \rho_{\textsc{ab}}(0)\mapsto \rho_\textsc{ab}(\tau)$ is a quantum channel, i.e., $\Phi_\tau$ is completely positive and trace-preserving (CPTP) map. It is known that that the dynamical evolution is CP if and only if the Kossakowski matrix $\boldsymbol{\gamma}$ is positive semidefinite,  $\mathfrak{h}_{\mathrm{eff}}$ is Hermitian, and the ``jump operators'' (here $\sigma_{\alpha}^j $) form an orthonormal basis for the Hilbert-Schmidt operators on the joint detector Hilbert space \cite{lindblad1976generators,gorini1976completely,gorini1978properties,kossakowski1972quantum,ingarden1975connection,whitney2008staying,breuer2002theory}.

Let us now recast the Markovian dynamics encoded in equation (\ref{eq: memoryless2}) into GKSL form. Converting (\ref{eq: memoryless2}) to the Schr\"odinger picture using (\ref{eq: Schrodinger-convert}) yields
\begin{eqnarray} \label{our_memoryless}
    & & \frac{\partial \rho_{\ab} }{\partial \tau} \simeq  -i [ \mathfrak{h}_\textsc{a} + \mathfrak{h}_\textsc{b}, \rho_{\ab}(\tau)] \\
    & & \hspace{0.5cm}+ g^2\sum_{j,k}\int_0^\infty\dd s\; \Scale[0.90]{ \bigg(\mathcal{W}_{jk}(s)[\mu_k(0)\rho_\ab(\tau),\mu_j(0)] + \mathrm{h.c.} \bigg) }\,, \notag
\end{eqnarray}
with the sum running over $j,k\in\{\text{A,B}\}$ as before. We stress again that a crucial difference that distinguishes the usual procedure in the literature from ours is the way the Markovian approximation is implemented\footnote{The ``standard'' Markovian approximation would have given
\begin{align}
    \frac{\partial \rho_{\ab} }{\partial \tau} &\simeq  -i [ \mathfrak{h}_\textsc{a} + \mathfrak{h}_\textsc{b}, \rho_{\ab}(\tau)]  \notag \\
    &\hspace{0.5cm}+ g^2\sum_{j,k}\int_0^\infty\dd s\; \Scale[0.90]{ \bigg(\mathcal{W}_{jk}(s)[\mu_k(-s)\rho_\ab(\tau),\mu_j(0)] + \mathrm{h.c.} \bigg) } \notag \,.
\end{align}}

In addition to the constant $C_{\s}$ defined in (\ref{Cs_def}), we also define two other constants that depend on separation $L$:
\begin{subequations}
\begin{align}
    & C_{\times}  \coloneqq  2 \int_0^\infty \!\!\!\!\exd s \; \mathrm{Re}\left[ \mathcal{W}_{\times}(s)\right] =\sfrac{\sinh^{-1}\left( aL/2 \right)}{2\pi^2 L \sqrt{1+(aL/2)^2}} \,, \label{Cx_def} \\
    & K_{\times}  \coloneqq  2 \int_0^\infty \!\!\!\!\exd s \; \mathrm{Im}\left[ \mathcal{W}_{\times}(s)\right] = - \sfrac{1}{4\pi L \sqrt{ 1 + (aL/2)^2}}  \,. \label{Kx_def}
\end{align}
\end{subequations}
The Markovian master equation (\ref{our_memoryless}) can be reorganized into GKSL form (\ref{eq: master-equation-dissipator}) with 
\begin{align}
    \mathfrak{h}_{\text{eff}} = \mathfrak{h}_{\textsc{a}} + \mathfrak{h}_{\textsc{b}} + g^2 \mathcal{K}_{\times} \sigma_{1}^\textsc{a}  \sigma_{1}^\textsc{b}\,.
    \label{heff_ours}
\end{align}
The entries of the Kossakowki matrix (\ref{dissipator}) are given by
\begin{subequations}
\begin{align}
    [\gamma_{\textsc{aa}}^{\alpha\beta}] &= [\gamma_{\textsc{bb}}^{\alpha\beta}] = 
    \left[\begin{array}{ccc}
        g^2C_\s & 0 & 0\\
        0 & 0 & 0 \\
        0 & 0 & 0
    \end{array}
    \right]\,,\label{eq: kossakowski-matrix-2}\\
    [\gamma_{\textsc{ab}}^{\alpha\beta}] &= [\gamma_{\textsc{ba}}^{\alpha\beta}] =
    \left[\begin{array}{ccc}
        g^2C_\times & 0 & 0\\
        0 & 0 & 0 \\
        0 & 0 & 0 
    \end{array}
    \right]\,.\label{eq: kossakowski-matrix-3}
\end{align}
\end{subequations}
The effective Hamiltonian operator $\mathfrak{h}_{\text{eff}}$ is here the joint free Hamiltonian of the detectors $\mathfrak{h}_\textsc{a}+\mathfrak{h}_\textsc{b}$ together with an extra degeneracy-lifting environment-induced interaction term $g^2\mathcal{K}_\times\sigma_1^\textsc{a}\otimes \sigma_1^\textsc{b}$. Notably, this so-called ``Lamb shift'' term is \textit{finite} and the degeneracy in the joint free Hamiltonian is lifted because the spectrum of $\mathfrak{h}_{\text{eff}}$ is $\{-\Omega, -g^2 K_{\times}, +g^2 K_{\times},\Omega\}$ while the spectrum of $\mathfrak{h}_\textsc{a}+\mathfrak{h}_\textsc{b}$ is $\{-\Omega, 0, 0, \Omega\}$.

We now claim that the dynamics described by $\mathfrak{h}_{\text{eff}}$ and the $6\times 6$ Kossakowski matrix,
\begin{align} \label{eq: kossakowski-matrix-1}
\boldsymbol{\gamma} = \begin{bmatrix} \gamma_{\textsc{aa}}^{\alpha\beta} &  \gamma_{\textsc{ab}}^{\alpha\beta}  \\  \gamma_{\textsc{ba}}^{\alpha\beta} &  \gamma_{\textsc{bb}}^{\alpha\beta} \end{bmatrix} 
\end{align}
defines a CP-preserving evolution map \textit{without} further approximation. This follows directly from the fact that $\mathfrak{h}_\text{eff}$ is Hermitian and the Kossakowski matrix $\boldsymbol{\gamma}$ is indeed positive-semidefinite, where the only two nonzero eigenvalues are
\begin{align}
    \lambda_1[\boldsymbol{\gamma}] = g^2(C_\s + C_\times) \; , \quad  \lambda_2[\boldsymbol{\gamma}] = g^2(C_\s - C_\times)\,,
\end{align}
which are non-negative since $C_\s \geq C_\times > 0$ for any $L \geq 0$. In other words, the Markovian master equation (\ref{our_memoryless}) studied in this work using the correct Markovian approximation \eqref{MarkovianTaylor} \textit{already} defines CP-preserving evolution without the need for any additional approximations such as the RWA.

\subsection{Markovian limit of the $X$-block}

We are now ready to perform the late-time resummation for the late-time dynamics of the detectors. Recall that the evolution decouples\footnote{The full set of integro-differential equations for the $X$-block from NZ-ME2 (before the Markovian approximation) are shown in Appendix~\ref{sec: NZ-ME2_X}.} into two sets of matrix ODE for the $X$- and $O$-blocks as stated in (\ref{XO_split}).  In what follows we find it more convenient to work in the Schr\"odinger picture.

Starting from the Markovian dynamics \eqref{our_memoryless}, the $X$-block component can be rearranged into a matrix ODE with constant coefficient of the form \eqref{eq: matrix-ODE-markov}: this reads
\begin{align}
    \frac{\dd \mathbf{x}}{\dd \tau} = (\mathbb{M}_0+g^2\mathbb{M}_2)\mathbf{x}(\tau) + g^2\mathbf{b}\,,
    \label{eq: matrix-ODE-X-block}
\end{align}
where we define the vector of $X$-block components $\mathbf{x}(\tau)$ and the constant vector $\mathbf{b}$ to be
\begin{widetext}
\begin{subequations}
\begin{align}
    \mathbf{x}(\tau) & : = \left[ \begin{matrix} \; \rho_{11}(\tau) & \rho_{22}(\tau) & \rho_{33}(\tau) & \mathrm{Re}\left[ \rho_{14}(\tau) \right] & \mathrm{Im}\left[ \rho_{14}(\tau) \right] & \mathrm{Re}\left[ \rho_{23}(\tau) \right] & \mathrm{Im}\left[ \rho_{23}(\tau) \right]  \end{matrix} \; \right]^{T} \ , \\
    \mathbf{b} & : = \left[ \begin{matrix} 0 & C_{\s} & C_{\s} & -C_{\times} & -K_{\times} & C_{\times} & 0 \end{matrix} \right]^{T} \ ,
\end{align}
\end{subequations}
and $\mathbb{M}_0,\mathbb{M}_2$ are constant matrices 
\begin{align}
    \mathbb{M}_0 &= \left[
        \begin{array}{ccccccc}
         0 & 0 & 0 & 0 & 0 & 0 & 0 \\
         0 & 0 & 0 & 0 & 0 & 0 & 0 \\
         0 & 0 & 0 & 0 & 0 & 0 & 0 \\
         0 & 0 & 0 & 0 & 2 \Omega  & 0 & 0 \\
         0 & 0 & 0 & -2 \Omega  & 0 & 0 & 0 \\
         0 & 0 & 0 & 0 & 0 & 0 & 0 \\
         0 & 0 & 0 & 0 & 0 & 0 & 0 \\
        \end{array}
        \right],\quad 
    \mathbb{M}_2 =\left[
    \begin{array}{ccccccc}
     -2 C_{\s} & C_{\s} & C_{\s} & -2 C_{\times} &  {-}2 K_{\times} & 2 C_{\times} & 0 \\
     0 & -3 C_{\s} & -C_{\s} & 2 C_{\times} & 0 & -2 C_{\times} & -2 K_{\times} \\
     0 & -C_{\s} & -3 C_{\s} & 2 C_{\times} & 0 & -2 C_{\times} & 2 K_{\times} \\
     0 & 2 C_{\times} & 2 C_{\times} & -2 C_{\s} & 0 & 2 C_{\s} & 0 \\
     2 K_{\times} & K_{\times} & K_{\times} & 0 & -2 C_{\s} & 0 & 0 \\
     0 & -2 C_{\times} & -2 C_{\times} & 2 C_{\s} & 0 & -2 C_{\s} & 0 \\
     0 & K_{\times} & -K_{\times} & 0 & 0 & 0 & -2 C_{\s} \\
    \end{array}
    \right]\,.
\end{align}
\end{widetext}
The determinant of the matrix is given by
\begin{align}
    \label{detM}
    &\det\left(\mathbb{M}_0+g^2\mathbb{M}_2 \right) \notag\\
    &\hspace{0.5cm} =  - 256 g^{10} \Omega^2 C_{\mathrm{s}} \left( C_{\mathrm{s}}^2 - C_{\times}^2 \right)\left( C_{s}^2 + K_{\times}^2 \right)
\end{align}
which is nonzero so long as $\Omega > 0$ and $L > 0$ (since $C_\times\leq C_\s$ with $C_{\times} \to C_{\mathrm{s}}$ in the limit $L \to 0^{+}$). Remarkably, this alone already implies that any inferences made for the dynamics immediately become suspect in the limit that the trajectories are ``stacked'' on top of one another with $L \to 0^{+}$ (considered, e.g., in \cite{Benatti2004Unruh}).

The eigenvalues for $\mathbb{M}_0+g^2\mathbb{M}_2 $ are given by\footnote{These are computed as a series in the coupling $g$ since the the characteristic polynomial has high degrees so we neglect $\mathcal{O}(g^4)$ contributions here --- note however that $\lambda_{4}^{X}$ and $\lambda_{5}^{X}$ are exact eigenvalues of $\mathbb{M}_0+g^2\mathbb{M}_2$).}
\begin{equation}
    \begin{aligned}
        \lambda_1^{X} \ & \simeq \ \qquad && - 2 g^2 C_{\mathrm{s}}\,, \\
        \lambda_2^{X} \ & \simeq \ && - 3 g^2 C_{\mathrm{s}} - g^2 \sqrt{ C_{\mathrm{s}}^2 + 8 C_{\times}^2 }\,, \\
        \lambda_3^{X} \ & \simeq \ && - 3 g^2 C_{\mathrm{s}} + g^2 \sqrt{ C_{\mathrm{s}}^2 + 8 C_{\times}^2 }\,,  \\
        \lambda_4^{X} \ & = \ && - 2 g^2 \left( C_{\mathrm{s}} - i K_{\times} \right)\,, \\
        \lambda_5^{X} \ & = \ && - 2 g^2 \left( C_{\mathrm{s}} + i K_{\times} \right)\,, \\
        \lambda_6^{X} \ & \simeq \ - 2 i \Omega \  && - 2 g^2 C_{\mathrm{s}}\,,  \\
        \lambda_7^{X} \ & \simeq \ + 2 i \Omega \  && - 2 g^2 C_{\mathrm{s}}\,.
    \end{aligned}
    \label{lambdaX}
\end{equation}
Since the determinant (\ref{detM}) is nonzero this means that the matrix is invertible (so long as $\Omega, L  > 0$). This is sufficient for computing the steady state solution
\begin{align}
    \frac{\dd \mathbf{x}_\star}{\dd \tau}=0 \ \Longrightarrow \ \mathbf{x}_\star = -
    (\mathbb{M}_0+g^2\mathbb{M}_2)^{-1}g^2\mathbf{b}\,.
\end{align}
Furthermore since $\mathrm{Re}[\lambda_{j}^{X}] < 0$ for all $j$  {at $\mathcal{O}(g^2)$}, we can conclude that the dynamics sink towards the above steady-state $\mathbf{x}_\star$ for any initial state $\mathbf{x}(0)$. The stationary solution $\mathbf{x}_\star$ reads
\begin{align}
    \mathbf{x}_\star 
    & \simeq  \rr{\sfrac{1}{4},\sfrac{1}{4},\sfrac{1}{4},0,0,0,0}^T \,,
    \label{eq: steady-x-vector}
\end{align}
and more generally, the solution to (\ref{eq: matrix-ODE-X-block}) for arbitrary $\tau$ is given by
\begin{align}
    \mathbf{x}(\tau) &= e^{(\mathbb{M}_0+g^2\mathbb{M}_2)\tau}(\mathbf{x}(0)-\mathbf{x}_\star) + \mathbf{x}_\star\,.
    \label{eq: general-sol-X-block}
\end{align}
This is the late-time resummed solution for the $X$-block since it is valid for $g^2a\tau\sim 1$.

We can evaluate this general solution \eqref{eq: general-sol-X-block} numerically or perturbatively along the lines of \cite{kaplanek2020hot} (e.g.~see Eq.~(\ref{14_sol_noRWA})). The numerical calculation of \eqref{eq: general-sol-X-block} will be relevant, for instance, when we want to calculate the amount of entanglement that the two detectors acquire via interaction with the quantum field.

\subsection{Markovian limit for {$O$-block}}

For the {$O$-block}, we perform the same procedure as before (see Appendix~\ref{sec: NZ-ME2_O} for the NZ-ME2 before taking the Markovian approximation) which leads to
\begin{align}
    \frac{\dd\mathbf{y}(\tau)}{\dd \tau} = (\mathbb{N}_0+g^2\mathbb{N}_2) \mathbf{y}(\tau)\,,
\end{align}
where we define
\begin{widetext}
\begin{align}
    \mathbf{y}(\tau)  \coloneqq \left[ \; \begin{matrix} \rho_{12}(\tau) & \rho_{13}(\tau) & \rho_{24}(\tau) & \rho_{34}(\tau) & \rho^*_{12}(\tau) & \rho^*_{13}(\tau) & \rho^*_{24}(\tau) & \rho^*_{34}(\tau) \end{matrix} \;  \right]^T
\end{align}
and where $\mathbb{N}_0,\mathbb{N}_2$ are constant matrices
\begin{align}
    \mathbb{N}_0 &= \Scale[0.90]{ \left[
    \begin{array}{cccccccc}
     -i \Omega  & 0 & 0 & 0 & 0 & 0 & 0 & 0 \\
     0 & -i \Omega  & 0 & 0 & 0 & 0 & 0 & 0 \\
     0 & 0 & -i \Omega  & 0 & 0 & 0 & 0 & 0 \\
     0 & 0 & 0 & -i \Omega  & 0 & 0 & 0 & 0 \\
     0 & 0 & 0 & 0 & i \Omega  & 0 & 0 & 0 \\
     0 & 0 & 0 & 0 & 0 & i \Omega  & 0 & 0 \\
     0 & 0 & 0 & 0 & 0 & 0 & i \Omega  & 0 \\
     0 & 0 & 0 & 0 & 0 & 0 & 0 & i \Omega  \\
    \end{array}
    \right] } \; , \quad   
    \mathbb{N}_2 =  \Scale[0.90]{  \left[
    \begin{array}{cccccccc}
     -2 C_\s & -\alpha & C_\times & C_\s & C_\s & C_\times & -\alpha^{\ast} & 0 \\
     -\alpha & -2 C_\s & C_\s & C_\times & C_\times & C_\s & 0 & -\alpha^{\ast} \\
     C_\times & C_\s & -2 C_\s & -\alpha^{\ast} & -\alpha & 0 & C_\s & C_\times \\
     C_\s & C_\times & -\alpha^{\ast} & -2 C_\s & 0 & -\alpha & C_\times & C_\s \\
     C_\s & C_\times & -\alpha & 0 & -2 C_\s & -\alpha^{\ast} & C_\times & C_\s \\
     C_\times & C_\s & 0 & -\alpha & -\alpha^{\ast} & -2 C_\s & C_\s & C_\times \\
     -\alpha^{\ast} & 0 & C_\s & {C_{\times}} & C_\times & C_\s & -2 C_\s & -\alpha \\
     0 & -\alpha^{\ast} & {C_{\times}} & C_\s & C_\s & C_\times & -\alpha & -2 C_\s \\
    \end{array}
    \right] \ , }
    \label{eq: matrix-ODE-O-block}
\end{align}
\end{widetext}
with the shorthand $\alpha \coloneqq C_\times - i K_\times$. Furthermore we find the determinant
\begin{align}
    &\det(\mathbb{N}_0+g^2\mathbb{N}_2) \notag \\
    &\hspace{0.5cm} =  \Omega^8 + 16 g^4 \Omega^6 C_{\mathrm{s}}^2 + 64 g^8 \Omega^4 C_{\mathrm{s}}^2 \left( C_{\mathrm{s}}^2 - C_{\times}^2 \right)
\end{align}
which is always nonzero in the perturbative limit for nonzero gap $(\Omega>0)$. The matrix eigenvalues at $\mathcal{O}(g^2)$ are given by
\begin{subequations}
\begin{equation}
    \begin{aligned}
    \lambda_1^{O} \ & \simeq \  - i \Omega + g^2 \big( \Scale[0.8]{ - 2 C_{\mathrm{s}} + C_{\times} -  \sqrt{ (C_{\mathrm{s}} - C_{\times})^2 - K_{\times}^2 } } \big) \\
    \lambda_2^{O} \ & \simeq \  - i \Omega + g^2 \big( \Scale[0.8]{ - 2 C_{\mathrm{s}} + C_{\times} +  \sqrt{ (C_{\mathrm{s}} - C_{\times})^2 - K_{\times}^2 } } \big) \\
    \lambda_3^{O} \ & \simeq \  - i \Omega + g^2 \big( \Scale[0.8]{ - 2 C_{\mathrm{s}} - C_{\times} -  \sqrt{ (C_{\mathrm{s}} + C_{\times})^2 - K_{\times}^2 } } \big) \\
    \lambda_4^{O} \ & \simeq \  - i \Omega + g^2 \big( \Scale[0.8]{ - 2 C_{\mathrm{s}} - C_{\times} +  \sqrt{ (C_{\mathrm{s}} + C_{\times})^2 - K_{\times}^2 } } \big) 
    \end{aligned}
\end{equation}
with the corresponding conjugate pairs
\begin{align}
 \lambda_5^{O} = \lambda_1^{O\ast}\,, \; 
 \lambda_6^{O} = \lambda_2^{O\ast}\,, \; 
 \lambda_7^{O} = \lambda_3^{O \ast}\,, \; 
 \lambda_8^{O} = \lambda_4^{O \ast} \ .
\end{align}
\label{lambdaO}
\end{subequations}
Since the matrix $\mathbb{N}_0+g^2\mathbb{N}_2$ is invertible (for $\Omega > 0$), we have the general late-time resummed solution
\begin{align}
    \mathbf{y}(\tau) = e^{(\mathbb{N}_0+g^2\mathbb{N}_2)\tau}\mathbf{y}(0)\,.
    \label{eq: general-sol-O-block}
\end{align}
At large $\tau\to \infty$, we have $\mathbf{y}(\tau)\to \mathbf{0}$, i.e. the $O$-block decays to zero. This follows from the fact that the stationary solution $\mathbf{y}_\star$ is given by
\begin{align}
    \frac{\dd\mathbf{y}_\star}{\dd\tau}=0 \ \Longrightarrow \  \mathbf{y}_\star = 0\,,
\end{align}
because all the eigenvalues of $\mathbb{N}_0+g^2\mathbb{N}_2$ have negative real parts (so long as $\Omega>0$). This result implies that regardless of the initial state chosen for the two detectors, only the $X$-block contribution survives in the late-time limit.

\subsection{Thermalization and (lack of) entanglement}

The late-time steady-state solution for the NZ-ME2 equation follows by combining the steady-state solutions $\mathbf{x}_\star$ and $\mathbf{y}_\star$ for both $X$-block and $O$-block we obtained earlier. The result is that the late-time stationary state is maximally mixed for any initial joint state of both qubits:
\begin{align}
    \rho_{\textsc{ab}}(\infty)
    &= \frac{\openone}{2} \otimes \frac{\openone}{2}  + \mathcal{O}(g^4)\,,
    \label{eq: separable-final-state}
\end{align}
which is clearly separable. Observe that the steady state solution \eqref{eq: separable-final-state} is independent of $\Omega, a, L$ and any UV cutoffs (like $\epsilon$ used to regulate coincident limit divergences). In other words, calculations in the Markovian regime at leading order are unable to probe the temperature of the field even if both detectors do thermalize. This is because the Markovian regime where we need $\Omega/a\ll 1$ corresponds to the \textit{high-temperature limit}, hence the steady state solution only picks out the zeroth-order expansion in $\Omega/a$. Any dependence of the state on $\Omega/a$ can only appear at finite $\tau$.

Furthermore, the fact that the asymptotic final state is independent of $L$ means that the Markovian regime washes out the effect of field-mediated communication between the two detectors. Any $L$-dependent corrections to the steady-state solution can only appear at finite $\tau$ at this order in perturbation theory. It is worth emphasizing that the lack of $L$-dependence at late times on its own is not very surprising: already in other contexts such as entanglement harvesting, detectors are unable to get entangled by the quantum field vacuum when the energy gap is too small compared to other scales of the problem \cite{pozas2015harvesting,pozas2017degenerate,Simidzija2018no-go,tjoa2021entanglement}. Similarly, in perturbative short-time Dyson series expansions, accelerated detectors suffer entanglement degradation \cite{salton2015acceleration}. These older results already suggest (in hindsight) that one should not expect any entanglement in the late-time Markovian regime when both detectors are in their ground states.

That said, we should still be able to infer the Unruh temperature from the late-time dynamics indirectly. Indeed, since the solutions to the ODEs are given by Eqs.~\eqref{eq: general-sol-X-block} and \eqref{eq: general-sol-O-block} (we repeat here for convenience)
\begin{equation}
    \begin{aligned}
    \mathbf{x}(\tau) &= e^{(\mathbb{M}_0+g^2\mathbb{M}_2)\tau}(\mathbf{x}(0)-\mathbf{x}_\star) + \mathbf{x}_\star\,,\\
    \mathbf{y}(\tau) &= e^{(\mathbb{N}_0+g^2\mathbb{N}_2)\tau}\mathbf{y}(0)\,,
    \end{aligned}
\end{equation}
the eigenvalues of $\mathbb{M}_0+g^2\mathbb{M}_2$ and $\mathbb{N}_0+g^2\mathbb{N}_2$ set the scale for the thermalization process. In the limit of large separation ($aL\gg 1$), we have $C_\times/C_\s\ll 1$ and $K_\times/C_\s\ll 1$, so that the maximum of the real part of the eigenvalues is given by
\begin{align}
    \max_j \Re(\lambda^\text{X}_j) &\simeq  -2g^2C_\s = \frac{g^2 a}{2\pi^2} \equiv \frac{g^2 T_\mathsf{U}}{\pi}\,.
\end{align}
This is equal to the \textit{decay rate} of a single accelerating detector experiencing thermal bath at Unruh temperature $T_\mathsf{U}=a/(2\pi)$ found in the literature (see, e.g., \cite{Moustos2017nonmarkov}). Therefore, while the asymptotic final state cannot tell us about the Unruh temperature, we can still learn about the Unruh effect from its decay rates so long as the detectors are far enough apart. 

In contrast, for $aL \ll 1$ is not tractable in the Markovian regime since $\lambda^\text{X}_3$ approaches zero as $aL$ gets smaller. This can be attributed to field-mediated communication (from the commutator) between the two detectors at small separation which dominates. In effect, what is happening is that for $aL\ll 1$, the detectors can exchange information with one another via the field commutator, so one detector ``stores'' the memory of the other detector. Consequently, the decay process becomes much slower at small separation ($\Re[\lambda_3^\text{X}]\simeq 0^-$).

\section{Validity relations for Markovian limit}
\label{sec: validity}

In \S\ref{sec: markovian} we obtained the Markovian solution for the two-detector dynamics and showed that the evolution is CP without the need for RWA. This issue is largely ignored in the literature\footnote{One exception is \cite{Moustos2017nonmarkov} for a single detector, but the method is not very portable for two detectors.}, especially so when two detectors are considered. In the majority of past literature we are aware of, it is \textit{assumed} that the Markovian limit can be taken, without specifying \textit{when} it is valid in terms of the relevant scales for the problem at hand. We now find explicitly the requirements for the Markovian limit to be valid, and we show that similar two-detector calculations in the past can actually violate these requirements.

Recall that the Markovian approximation can be viewed in terms of the Taylor series (\ref{MarkovianTaylor}), repeated here,
\begin{eqnarray}\label{MarkovianTaylor_validity}
    && \mu_{j}^{\ssI}(\tau - s) \rho_{\textsc{ab}}^{\ssI}(\tau - s)  \\
    && \ \ \ \simeq  \mu^{\ssI}_{j}(\tau)  \rho_{\textsc{ab}}^{\ssI}(\tau)  - s \big( \Scale[0.93]{ \mu_{j}^{\ssI}(\tau) \dot{\rho}_{\textsc{ab}}^{\ssI}(\tau) + \dot{\mu}_j^{\ssI}(\tau) \rho_{\textsc{ab}}^{\ssI}(\tau) } \big) + \ldots\,. \notag 
\end{eqnarray}
This is physically motivated by the fact that the environment correlators $\mathcal{W}_{\s,\times}$ are strongly peaked about $s=0$ in the master equation. Inserting (\ref{MarkovianTaylor_validity}) into (\ref{eq: NZ-final}), we get
\begin{widetext}
\begin{eqnarray} \label{Markov_NLO}
    \frac{\dd \rho^{\ssI}_\textsc{ab}}{\dd \tau} &\simeq & g^2 \sum_{j,k \in \{\mathrm{A},\mathrm{B}\}} \; \int_0^\infty\dd s\; \Bigr( \; \mathcal{W}_{jk}(s) \bigr[\mu_{j}^\ssI(\tau)\rho^\ssI_\textsc{ab}(\tau), \mu^\ssI_{k}(\tau)\bigr] + \mathrm{H.c.} \; \Bigr)  \\
   & \  &  - g^2 \sum_{j,k \in \{\mathrm{A},\mathrm{B}\}} \; \int_0^\infty\dd s\; \Bigr( \; s \mathcal{W}_{jk}(s) \bigr[\mu_{j}^{\ssI}(\tau) \dot{\rho}_{\textsc{ab}}^{\ssI}(\tau) + \dot{\mu}_j^{\ssI}(\tau) \rho_{\textsc{ab}}^{\ssI}(\tau) , \mu^\ssI_{k}(\tau)\bigr] + \mathrm{H.c.} \; \Bigr)  \ + \ \ldots \notag \ .
\end{eqnarray}
Using $\dot{\mu}_j^{\ssI}(\tau) = - \Omega(-i \sigma_{+}^{j} e^{+ i \Omega \tau} + i  \sigma_{-}^{j} e^{- i \Omega \tau})$ we can put Eq.~(\ref{Markov_NLO}) into the Lindblad-like form, which in the Schr\"odinger picture reads
 {\begin{eqnarray} \label{Lindblad_NLO}
    \frac{\dd \rho_\textsc{ab}}{\dd \tau} &\simeq  -i[\mathfrak{h}_{\mathrm{eff}},\rho_\ab(\tau)] + \mathcal{D}_{\boldsymbol{\gamma}}[ \rho_{\textsc{ab}}(\tau) ]\  - i \big[ \mathfrak{z}_{\mathrm{eff}}, \rho_{\textsc{ab}}(\tau) \big] 
    + \mathcal{D}_{\boldsymbol{\eta}}\big[  \Scale[0.90]{ - i [ \mathfrak{h}, \rho_{\textsc{ab}}(\tau)] + \dot{\rho}_{\textsc{ab}}(\tau) } \; \big] 
    + \mathcal{D}_{\boldsymbol{\zeta}} {\big[ \rho_{\textsc{ab}}(\tau) \big]} \,.
\end{eqnarray}}
 {The first two terms of Eq.~\eqref{Lindblad_NLO} are the original terms in the GKSL master equation: the effective Hamiltonian $\mathfrak{h}_{\text{eff}}$ is given in Eq.\eqref{heff_ours}, while the dissipator $\mathcal{D}_{\boldsymbol{\gamma}}$ is given by Eq.~\eqref{dissipator}, with Kossakowski matrix $\boldsymbol{\gamma}$ computed in Eqs.~(\ref{eq: kossakowski-matrix-2}-\ref{eq: kossakowski-matrix-3}). The next three terms contain the subleading corrections to the GKSL equation}: first, we have $\mathfrak{z}_{\mathrm{eff}}$ defined by
\begin{equation}
\mathfrak{z}_{\mathrm{eff}} \coloneqq - \sfrac{g^2 \Omega D_{\s}'}{2} \big( \sigma_{3}^{\textsc{a}} + \sigma_{3}^{\textsc{b}} \big) - \sfrac{ g^2  \Omega \mathcal{S}'_{\times} }{2} \big( \sigma_{1}^{\textsc{a}} \sigma_{2}^{\textsc{b}} + \sigma_{2}^{\textsc{a}} \sigma_{1}^{\textsc{b}} \big)\,.
\end{equation}
The last two terms are extra ``dissipation terms'' $\mathcal{D}_{\boldsymbol\eta},\mathcal{D}_{\boldsymbol\zeta}$ with the corresponding ``Kossakowski matrices''
$\boldsymbol{\eta}$ and $\boldsymbol{\zeta}$ given by
\begin{equation}
  \begin{split}
    [ \eta_{\textsc{aa}}^{\alpha \beta} ] = [ \eta_{\textsc{bb}}^{\alpha \beta} ]  &= \left[ \begin{matrix} D'_{\s} & 0 & 0 \\ 0 & 0 & 0 \\ 0 & 0 & 0 \end{matrix} \right] \\
    [ \eta_{\textsc{ab}}^{\alpha \beta} ] = [ \eta_{\textsc{ba}}^{\alpha \beta} ]  &= \left[ \begin{matrix} D'_{\times} & 0 & 0 \\ 0 & 0 & 0 \\ 0 & 0 & 0 \end{matrix} \right] \\
  \end{split}
\qquad \qquad \mathrm{and} \qquad \qquad
  \begin{split}
    [ \zeta_{\textsc{aa}}^{\alpha \beta} ] = [ \zeta_{\textsc{bb}}^{\alpha \beta} ]  &= \left[ \begin{matrix} 0 & \frac{1}{2}( D_{\s}' - i S_{\s}' )   & 0 \\ \frac{1}{2}( D_{\s}' + i S_{\s}' ) & 0 & 0 \\ 0 & 0 & 0 \end{matrix} \right] \\
    [ \zeta_{\textsc{ab}}^{\alpha \beta} ] = [ \zeta_{\textsc{ba}}^{\alpha \beta} ]  &= \left[ \begin{matrix} 0 & \frac{1}{2}( D_{\times}' - i S_{\times}' )   & 0 \\ \frac{1}{2}( D_{\times}' + i S_{\times}' ) & 0 & 0 \\ 0 & 0 & 0 \end{matrix} \right] \ .
  \end{split}
\end{equation}
\end{widetext}
The constant coefficients $D_{\s,\times}'$ and $S_{\s,\times}'$ are given by
\begin{subequations}
\begin{align}
    D_{\s, \times}' &\coloneqq  2 \int_0^\infty \exd s \; \mathrm{Re}\left[ \mathcal{W}_{\s, \times}(s) \right] s \label{Dxp_def} \\
    S_{\s, \times}' & \coloneqq 2 \int_0^\infty \exd s \; \mathrm{Im}\left[ \mathcal{W}_{\s, \times}(s) \right] s \label{Sxp_def} 
\end{align}
\end{subequations}
which evaluate to (see Appendix \ref{subsec: Integrals}) 
\begin{subequations}
\begin{align}
&\hspace{-0.2cm} D_{\s}' = \frac{\log(a\epsilon)}{2\pi^2}\,,  \quad D_{\times}' =  \frac{ \mathrm{Re}\left[ \mathrm{Li}_{2}\left( \ell_-^2 \right) - \mathrm{Li}_{2}\left( \ell_+^2\right) \right] }{4 \pi^2 \;  aL \; \sqrt{1+  (aL/2)^2} }   \label{Dxp_answer} \,,\\
&\hspace{-0.13cm}S_{\s}' = - \frac{1}{4\pi}\,,\quad 
\quad S_{\times'}  = \frac{\sinh^{-1}\left( a L / 2 \right)}{ 2 \pi a L \sqrt{ 1 + (aL/2)^2 } }\,, \label{Sxp_answer}
\end{align}
\end{subequations}
where $\text{Li}_2(z)$ is the polylogarithm of order 2 \cite{olver2010nist} and we used the shorthand
\begin{equation} \label{aL_short}
\ell_\pm \coloneqq \tfrac{aL}{2} \pm \sqrt{1+(\tfrac{aL}{2})^2} \,.
\end{equation}
Note that $D_{\s}'$ is a UV-regulated function with the infinitesimal $\epsilon>0$ appearing in the $i \epsilon$-prescription of the environment correlators $\mathcal{W}_{\s, \times}$. It is worth stressing that the UV regulator is needed because sharp switching is not compatible with pointlike limits in (3+1) dimensions: the UV regulator is given the interpretation of a position-space cutoff on the size of the detector below which we cannot resolve\footnote{Note that in other contexts such as entanglement harvesting, pointlike detector models may still work because some initial states are insensitive to these UV issues. In the validity relations derived here, the UV regulators will appear explicitly as we consider arbitrary qubit initial states.}.

\subsection{Matrix ODE derivation of validity bounds}

The idea of finding where the Markovian approximation applies is to bound the last three terms in (\ref{Lindblad_NLO}) involving $\mathfrak{z}_{\mathrm{eff}}$, $\boldsymbol{\eta}$ and $\boldsymbol{\zeta}$ to be parametrically small compared to the first two terms involving $\mathfrak{h}_{\mathrm{eff}}$ and $\boldsymbol{\gamma}$. Wherever in parameter space this is true, the Lindbladian dynamics studied in this work are valid --- this results in a set of bounds which must be satisfied involving functions of $g$, $a$, $\Omega$ and $L$ (as well as a UV cutoff $\epsilon$) which we call {\it validity bounds}.

The simplest way of doing this is to again split \eqref{Lindblad_NLO} into the $X$- and $O$-blocks, keeping the subleading non-Markovian correction as in (\ref{Lindblad_NLO}) and find the analogous matrix ODE to Eq.~\eqref{eq: matrix-ODE-markov}: the result is an equation of the form
\begin{align} \label{eq: ODE_NLO}
    \frac{\dd \mathbf{u}}{\dd \tau} \simeq  \underbrace{\mathbb{A}\mathbf{u}+\mathbf{v}}_{\text{Markov.}}   \underbrace{-\mathbb{Y}\frac{\dd \mathbf{u}}{\dd \tau} - \mathbb{Z} \mathbf{u}}_{\text{lead. non-Markov.}}\,.
\end{align}
The matrix $\mathbb{Y}$ can be obtained from $\mathbb{A}$ by the replacement $C_{\s,\times} \to D'_{\s,\times}$ and $K_{\s,\times} \to S'_{\s,\times}$ (with the rest of the entries zero), while the matrix $\mathbb{Z}$ depends only on combinations of $\Omega D'_j$ and $\Omega S'_j$. Iteratively plugging (\ref{eq: matrix-ODE-markov}) into the RHS of (\ref{eq: ODE_NLO}) gives us
\begin{align} \label{eq: ODE_NLO2}
    \frac{\dd \mathbf{u}}{\dd \tau} &\simeq (\openone -\mathbb{Y})(\openone-\mathbb{Z}\mathbb{A}^{-1})\mathbb{A}\mathbf{u}+(\openone -\mathbb{Y})\mathbf{v}\,,
\end{align}
where we have used the fact that $\mathbb{A}$ is invertible (for both the $X$- and $O$-blocks, so long as $\Omega, L>0$). Eq.~\eqref{eq: ODE_NLO2} tells us that for the Markovian approximation to be valid, we need
\begin{align}
    \openone-\mathbb{Y} \simeq \openone\,,\quad \openone-\mathbb{Z}\mathbb{A}^{-1}\simeq \openone\,,
    \label{eq: matrix-bound-validity}
\end{align}
i.e., each matrix element satisfies
\begin{equation} \label{eq: matrix-bound-validity2}
    |\mathbb{Y}_{nm}|\ll 1 \qquad \mathrm{and} \qquad |(\mathbb{Z}\mathbb{A}^{-1})_{nm}|\ll 1\,.
\end{equation}
Eq.~\eqref{eq: matrix-bound-validity2} provides us with a very compact way of stating the constraints required for validity of Markovian approximation.

What remains to be done is to write down the information contained in \eqref{eq: matrix-bound-validity2} in terms of physical parameters $g$, $\Omega$, $a$, $L$ and $\epsilon$. Studying the $X$-block in the above manner results in entries of the matrix $\mathbb{Y}$ which from $|\mathbb{Y}_{nm}|\ll 1$ yield
\begin{subequations}
\begin{align}
    g^2|D_{\s}'|   &=  g^2 \sfrac{\log (a\epsilon)}{2\pi^2} \ll 1\,,\\
    g^2|D_\times'| &= g^2 \sfrac{\text{Re}\left[\mathrm{Li}_2(\ell_-^2)-\mathrm{Li}_2(\ell_+^2)\right]}{4\pi^2  aL \sqrt{1 + (aL/2)^2}}
    \ll 1\,,\\
    g^2 |S_\times'| & = g^2 \sfrac{\sinh^{-1}\left( a L / 2 \right)}{ 2 \pi a L \sqrt{ 1 + (aL/2)^2 } } \ll 1\,,
\end{align}
\end{subequations}
where $\ell_\pm$ is defined in (\ref{aL_short}). It is straightforward to see that $|S_\times'|\ll 1$ for all $aL\geq 0$, hence the third bound is automatically satisfied. The first bound a single-detector bound (also encountered in \cite{kaplanek2020hot}) and ensures the smallness of the coupling $g$ must compensate for the largeness of the UV cutoff $\epsilon$.  The non-trivial bound ends being the second one, since it depends on the detector separation: for $aL\gtrsim 1$ we always have $|D'_\times|\ll 1$, however
\begin{align}
    g^2|D'_\times|\simeq \frac{g^2(1-\log (aL))}{2 \pi ^2} \qquad \mathrm{for}\ aL \ll 1 \,.
\end{align}
What this means is that small $aL$ must be compensated by weaker coupling $g$, and so one cannot make $aL$ arbitrarily small. This bound is distinct from the UV cutoff requirement that demands $\epsilon \ll L$.

The condition $|(\mathbb{Z}\mathbb{A}^{-1})_{nm}|\ll 1$ for the $X$-block introduces more bounds, which are generally very complicated due to the matrix inverse. However, all the bounds involving $aL$ are only non-trivial when $aL\ll 1$ (i.e., they can be easily satisfied for $aL\gtrsim 1$), so below we restrict our attention only for $aL\ll 1$. The $g$-dependent bounds are
\begin{equation} \label{Z_valbounds1}
  \begin{split} 
    \frac{g^2 |\log(a\epsilon)|}{2\pi aL} & \ll 1\ , \\
    \frac{3 \pi  \Omega / a - g^2 aL}{(aL)^2} & \ll 1 \ ,
  \end{split}
\quad
  \begin{split}
    \frac{g^2}{aL} & \ll 1\ , \\
    \frac{g^2}{4aL}+g^2|\log (a\epsilon)| & \ll 1 \ ,
  \end{split}
\end{equation}
while the $g$-independent bounds are 
\begin{equation} \label{Z_valbounds2}
    \frac{\pi\Omega}{a} \ll 1 \ , \quad \frac{\pi\Omega}{a (aL)^2} \ll 1\ , \quad \frac{\Omega |\log(a^2 \epsilon L)|}{a(aL)^2}\ll 1\ .
\end{equation}
One of the main takeaways from this analysis is that  $aL$, which measures detector separation in units of $a$, cannot be arbitrarily small: it is bounded below by all other parameters involving $a\epsilon$, $\Omega/a$ and $g^2$. Very small $aL$ amounts to very closely-spaced detectors and the field-mediated communication makes memoryless approximation harder to satisfy.

We can perform the same kind of (tedious) analysis for the $O$-block and it turns out that up to irrelevant numerical factors and linear combinations of the above conditions, the $O$-block does not contain any new information about Markovian validity.

\subsection{Summary: when is Markovian approximation valid?}

The short story is that the Markovian validity favours weak coupling $g\ll 1$ (due to perturbation theory), as well as small detector gaps $\Omega\ \ll a$ and large separation $aL \gtrsim 1$. Overall, we can summarize the validity relations for Markovian limit as follows:
\begin{enumerate}[label=(\roman*),leftmargin=*]
    \item By default, the Markovian approximation requires that $\Omega/a\ll 1$. This requirement is often implicit or ignored in the literature.
    
    \item In general the pointlike limit is ill-defined for arbitrary qubit initial states when one considers sharp switching, which is the usual approach in open quantum systems (with rare exceptions such as \cite{Benito2019asymptotic}). Consequently, the UV divergences encountered are to be interpreted as ignorance about the detectors finite spatial extent. The validity bounds require
    \begin{align}
        \frac{g^2|\log (a\epsilon)|}{2\pi^2} \ll 1\, .
    \end{align}
    That is, $a\epsilon$ cannot be arbitrarily small: either we probe the ``high temperature'' regime\footnote{At the same time, no finite-sized realistic detector can maintain its rigid shape for arbitrarily large accelerations, e.g., due to ionization. So in practice high-temperature regime is highly non-trivial.} (large $a$) relative to effective size of the detector prescribed by the UV cutoff $\epsilon$), or that detector size cannot be arbitrarily small.
    
    \item In the presence of two detectors, we require that $g^2|D_\times'|\ll 1$, where $D_\times'$ depends on the dimensionless detector separation $aL$. For large $aL$ this is automatically satisfied if (i) and (ii) are properly satisfied; however, for small $aL\ll 1$ we require that $g$, $a$, $\Omega$, $L$ and $\epsilon$ work together to obey
    \begin{equation}
        \begin{aligned}
        \hspace{1cm}&\frac{\pi\Omega}{a(aL)^2} \ll 1 \,,\quad \frac{\Omega|\log(a^2 \epsilon L )|}{a(aL)^2}\ll 1\,,\quad g^2 \ll a L 
        \label{eq: bound-iii}
        \end{aligned}
    \end{equation}
    which are simplified versions of the $aL$-dependent validity bounds given in (\ref{Z_valbounds1}) and (\ref{Z_valbounds2}). The crucial point is that $aL$ cannot be arbitrarily small: it is bounded below by quantities involving $g,\Omega/a,a \epsilon$. Thus this condition favours $aL\gtrsim 1$; for $aL \ll 1$ one has to more carefully tune $\Omega/a,a\epsilon$ and $g$ in order to compensate for the non-Markovianity this introduces.
    
    \item Finally, the matrices governing the evolution for the $X$- and $O$-blocks are treated as perturbative in the coupling $g$ (for nonzero $\Omega >0$ where matrix determinants are nonzero), this means that the matrices $\mathbb{M}_0, \mathbb{N}_0$ are large compared to the perturbations $g^2 \mathbb{M}_2, g^2 \mathbb{N}_2$ in the matrix ODEs (\ref{eq: matrix-ODE-X-block}) and (\ref{eq: matrix-ODE-O-block}). What this amounts to is remaining in a regime where the eigenvalues $\lambda^{X,O}$ given in (\ref{lambdaX}) and (\ref{lambdaO}) are perturbative in the coupling --- this means that the oscillation scale $\Omega>0$ must be large compared to the $\mathcal{O}(g^2)$ corrections. At the end of this day this enforces
    \begin{equation} \label{nonDegen}
        g^2 \ll \frac{\Omega}{a} \qquad \mathrm{and} \qquad g^2 \ll \Omega L  
    \end{equation}
    which roughly speaking ensures that $g^2$ is the smallest parameter in the problem\footnote{Strictly speaking, enforcing the validity bounds (\ref{nonDegen}) is about remaining in the perturbative/non-degenerate regime as opposed to just being Markovian.}. Notice that the second relation in is yet another manifestation that $aL$ cannot be arbitrarily small.
    
    It is worth noting that \eqref{nonDegen} has two important implications: (1) the gapless limit $\Omega=0$ must be treated separately and not simply set $\Omega=0$ in the results we have gotten so far --- the reason has to do with the fact that the matrices $\mathbb{M}_0+g^2\mathbb{M}_2$ (also for $\mathbb{N}_0+g^2\mathbb{N}_2$) have vanishing determinant in this limit, so the density matrix does not generically decohere properly. This is not a real problem because for gapless regime we can fully solve the dynamics non-perturbatively (see \cite{Landulfo2016communication}); (2) since we also have $\Omega/a\ll1 $, it must mean that
    \begin{align}
        \Omega/a\sim g
    \end{align}
    so that $\Omega/a$ is small enough for Markovianity to hold, but large enough for the perturbative calculation to work.
    
\end{enumerate}

While the conditions (i)-(iii) are not prohibitively restrictive, they do imply that several calculations in the literature for the past two decades are strictly-speaking invalid or unreliable. These include (1) the stacked trajectory limit\footnote{ The fact that (iii) implies that $L\neq 0$ is disallowed is unsurprising firstly because the environment-induced interaction diverges in this limit. It is also well-known that divergences associated with pointlike detector models have nothing to do with open quantum systems: for finite-sized detectors, \textit{by construction} we do not allow centres of mass to overlap for this reason.} ($L=0$) considered in \cite{Benatti2004Unruh} and (2) calculations using RWA-based GKSL equation considered in \cite{Hu2015twodetectorent,Hu2022loss-anti} in the regime where $\Omega/a\gtrsim 1$, which already violate (i).

\section{Comparison with Rotating-Wave Approximation}
\label{sec: comparison}

The fact that the steady state at late times \eqref{eq: separable-final-state} is separable and maximally mixed is not in itself very surprising since a fast/hot environment should generically be expected to scramble any information contained in an arbitrary initial joint detector state. This section addresses the fact that there are several different results in the literature which seem to conflict with the results of this work (see, e.g.,
\cite{Benatti2004Unruh,BENATTI2005review,Hu2015twodetectorent,Hu2011blackhole,Hu2013desitter,huang2017dynamics,huang2019boundary,Zhou2021massiveentanglement,Zhou2021massiveentanglement2,Menezes2018entanglementKerr,Hu2022loss-anti}), most notably that sometimes the two detectors can end up entangled.

What makes comparison to these works difficult is that the microscopically derived Lindblad equations used all apply an additional approximation relative to our work: the RWA. Beginning with a Born-Markov approximation where only the reduced density matrix is slowly-varying (as described in (\ref{MarkovianTaylor_lit})), these works would find a GKSL equation whose Kossakowski matrix has in general some negative eigenvalues (hence spoiling complete positivity) as well as explicit dependence on divergences which cannot be renormalized into an effective Hamiltonian. The RWA is then applied to rectify this apparent CP-violation, dropping certain terms in the GKSL equation under the guise that they should not be important when the system oscillates quickly. After applying the RWA, the resulting Kossakowski matrix then ends up having non-negative eigenvalues for any sizes of parameters in the problem, including large detector gaps $\Omega \gtrsim a$, and the expression is free of any non-renormalizable divergences. 

As we have already argued in the preceding sections (motivated by an Effective Field Theory (EFT) way of thinking), the Markovian approximation has a domain of validity that restricts the parameter space that we can use. Once this is recognized and we strictly remain in this subset of the parameter space, the resulting Markovian dynamics \textit{is} CP without further approximations. Applying RWA at best will only restrict the domain of validity even further, making the resulting master equation even more restrictive in its use. Our results demonstrate that not only is RWA unnecessary, but in general one should always track the regime of validity of all approximations involved, otherwise one risks obtaining nonsensical master equations and output states that do not reflect the physical problem at hand.

In this last section, we investigate what happens if we apply the RWA anyway, {\it within the domain of validity of the (``correct'') Markovian approximation}, so as to make an easier point of comparison to the aforementioned literature. The point of this exercise is two-fold: first to emphasize that both the dynamics studied in this paper and the RWA yields late-time states which are {\it not} entangled (when constrained to be within the regime of validity of Markovianity). Furthermore, we show that the dynamics between the two approaches can differ notably, which means that carelessly applying the RWA alters physical predictions significantly.

\subsection{The RWA-based solution}

Let us now check what happens if we were to perform the RWA and see if the differences between this and the dynamics reported in this work are significant. Following for example \cite{breuer2002theory}, taking the RWA amounts to dropping all terms coming from 
\begin{align}
    \sigma_\pm^j\sigma_\pm^k\rho_\ab(\tau)\,,\hspace{0.5cm} \sigma_\pm^j\rho_\ab(\tau) \sigma_\pm^k\, .
\end{align}
In the interaction picture, these terms arise from products of the monopole operators with a phase $e^{\pm i(E+E')\tau}$ that correspond to fast ``counter-rotating'' terms (here $E, E' \in \{\Omega, 0 , 0, -\Omega\}$ denotes the spectrum of the system Hamiltonian $\mathfrak{h} = \mathfrak{h}_{\textsc{a}} +\mathfrak{h}_{\textsc{b}}$). The standard lore is that these terms oscillate much more quickly compared to the slow ``co-rotating'' terms that come with $e^{\pm i(E-E')\tau}$, and so should be neglected. This procedure, more rigorously described by Davies \cite{davies1974markovian,davies1976markovian}, yields a Lindblad equation of the form (\ref{eq: master-equation-dissipator}) with Kossakowski matrix $\boldsymbol{\gamma}^{(\textsc{rwa})}$, with components
\begin{subequations}
\begin{align}
    [\gamma_{\textsc{aa}}^{\alpha\beta}]^{(\textsc{rwa})} &= [ \gamma_{\textsc{bb}}^{\alpha\beta}]^{(\textsc{rwa})} = 
    \left[\begin{array}{ccc}
        \frac{g^2 C_\s}{2}  & 0 & 0\\
        0 & \frac{g^2 C_\s}{2} & 0 \\
        0 & 0 & 0
    \end{array}
    \right] \ , \label{eq: kossakowski-matrix-2RWA}\\
    [\gamma_{\textsc{ab}}^{\alpha\beta}]^{(\textsc{rwa})} &= [\gamma_{\textsc{ba}}^{\alpha\beta}]^{(\textsc{rwa})} =
    \left[\begin{array}{ccc}
        \frac{ g^2C_\times}{2} & 0 & 0\\
        0 & \frac{g^2C_\times }{2} & 0 \\
        0 & 0 & 0 
    \end{array}
    \right] \ , \label{eq: kossakowski-matrix-3RWA}
\end{align}
\end{subequations}
and an effective Hamiltonian
\begin{equation}
\mathfrak{h}_{\mathrm{eff}}^{(\textsc{rwa})} \ = \ \sfrac{g^2 K_{\times}}{2} \big( \sigma_{1}^{\textsc{a}}  \sigma_{1}^{\textsc{b}} + \sigma_{2}^{\textsc{a}}  \sigma_{2}^{\textsc{b}}  \big) 
\end{equation}
c.f. equations (\ref{eq: kossakowski-matrix-2}), (\ref{eq: kossakowski-matrix-3}) and (\ref{heff_ours}). There are now \textit{four} nonzero eigenvalues (with repetition) of the new Kossakowski matrix $\boldsymbol\gamma^{(\textsc{rwa})}$, given by
\begin{align}
    \lambda_1[\boldsymbol{\gamma}^{(\textsc{rwa})}] &= \lambda_2[\boldsymbol{\gamma}^{(\textsc{rwa})}] =  \frac{g^2}{2}\rr{C_\s+C_\times}\,,\quad\\ 
    \lambda_3[\boldsymbol{\gamma}^{(\textsc{rwa})}] &= 
    \lambda_4[ \boldsymbol{\gamma}^{(\textsc{rwa})}] = \frac{g^2}{2}\rr{C_\s-C_\times}\,.
\end{align}
The RWA yields almost identical equations of motion as the ones without RWA for both the $X$-block and $O$-block, with $\mathbb{M}_0^{\textsc{(rwa)}} = \mathbb{M}_0$, $\mathbb{N}_0^{\textsc{(rwa)}} = \mathbb{N}_0$, but with the perturbative correction  $\mathbb{M}_2^{\textsc{(rwa)}}$ and  $\mathbb{N}_2^{\textsc{(rwa)}}$ being sparser matrices:
\begin{widetext}
\begin{align}
    \mathbb{M}_2^{(\textsc{rwa})} = \Scale[0.80]{  \left[
    \begin{array}{ccccccc}
     -2 C_\s & C_\s & C_\s & 0 & 0 & 2 C_\times & 0 \\
     0 & -3 C_\s & -C_\s & 0 & 0 & -2 C_\times & -2 K_\times \\
     0 & -C_\s & -3 C_\s & 0 & 0 & -2 C_\times & 2 K_\times \\
     0 & 0 & 0 & -2 C_\s & 0 & 0 & 0 \\
     0 & 0 & 0 &0 & -2 C_\s & 0 & 0 \\
     0 & -2 C_\times & -2 C_\times & 2 C_\s & 0 & -2 C_\s & 0 \\
     0 & K_\times & -K_\times & 0 & 0 & 0 & -2 C_\s \\
    \end{array}
    \right] } \,, \; 
    \mathbb{N}^{(\textsc{rwa})}_2 &=  \Scale[0.80]{ \left[
    \begin{array}{cccccccc}
     -2 C_\s & -\alpha_- & C_\times & C_\s & 0 & 0 & 0 & 0 \\
     -\alpha_- & -2 C_\s & C_\s & C_\times & 0 & 0 & 0 & 0 \\
     C_\times & C_\s & -2 C_\s & x-\alpha_+ & 0 & 0 & 0 & 0 \\
     C_\s & C_\times & -\alpha_+ & -2 C_\s & 0 & 0 & 0 & 0 \\
     0 & 0 & 0 & 0 & -2 C_\s & -\alpha_+ & C_\times & C_\s \\
     0 & 0 & 0 & 0 & -\alpha_+ & -2 C_\s & C_\s & C_\times \\
     0 & 0 & 0 & 0 & C_\times & C_\s & -2 C_\s & -\alpha_- \\
     0 & 0 & 0 & 0 & C_\s & C_\times & -\alpha_- & -2 C_\s \\
    \end{array}
    \right] }
\end{align}
\end{widetext}
\textit{c.f.} Eqs.~(\ref{eq: matrix-ODE-X-block}) and (\ref{eq: matrix-ODE-O-block}). The matrix $\mathbb{M}_0+g^2\mathbb{M}_2^{(\textsc{rwa})}$ can be inverted and the general solution for the $X$-block is
\begin{align}
    \mathbf{x}^{(\textsc{rwa})}(\tau) = e^{(\mathbb{M}_0+g^2\mathbb{M}_2^{(\textsc{rwa})})\tau}(\mathbf{x}(0)-\mathbf{x}_\star)+\mathbf{x}_\star \ ,
    \label{eq: x-vector-RWA}
\end{align}
where $\mathbf{x}_\star$ ends up evaluating to be {\it exactly} the same steady-state solution as given earlier in Eq.~(\ref{eq: steady-x-vector}) without applying RWA.

For the $O$-block one can check that the block-diagonal matrix $\mathbb{N}_0+\mathbb{N}_2^{(\textsc{rwa})}$ is invertible with all eigenvalues having negative real parts. This proves that at late times the $O$-block decays to zero for arbitrary initial states of the field. Together, we have shown that for any initial state of the field, the steady state solution at late times is \textit{exactly} the same as without RWA: it is also a separable mixed state at leading order in perturbation theory:
\begin{align}
    \rho_{\textsc{ab}}^{(\textsc{rwa})}(\infty)=\frac{\openone}{2}\otimes\frac{\openone}{2} + \mathcal{O}(g^4) = \rho_\textsc{ab}(\infty)\,.
\end{align}
Being explicit about the dynamics in the RWA case, it turns out that at $\mathcal{O}(g^2)$ the matrices $\mathbb{M}_0 + g^2 \mathbb{M}_{2}^{(\textsc{rwa})}$ and $\mathbb{N}_0 + g^2 \mathbb{N}_{2}^{(\textsc{rwa})}$ have the same eigenvalues as the earlier (non-RWA) eigenvalues listed in Eqs.~(\ref{lambdaX}) and (\ref{lambdaO}). Since the real part of these eigenvalues are all negative this confirms that RWA also sinks towards a separable mixed state. 

The main difference between applying RWA and the earlier Markovian description arises when one tracks the finite-time dependence of the components of the density matrix. For example, let us consider $\rho_{14}^{(\textsc{rwa})}(\tau)$, which can be obtained from two of the components of $\mathbf{x}^{\textsc{rwa}}(\tau)$ in Eq.~\eqref{eq: x-vector-RWA}. We find that
\begin{equation} 
    \label{14_RWA_sol}
    \rho_{14}^{(\textsc{rwa})}(\tau) \ \simeq \ \rho_{14}(0) e^{ ( - 2 i \Omega - 2 g^2 C_{\mathrm{s}} ) \tau }\,.
\end{equation}
In contrast, for the non-RWA version the answer is somewhat more complicated: one way to proceed is to compute the (right) eigenvectors $\mathbf{r}_{j}^{X}$ of the matrix $\mathbb{M}_0+g^2\mathbb{M}_2$ (as a series in $g$), so that the solution (\ref{eq: general-sol-X-block}) is equivalent to the ansatz
\begin{equation}
    \mathbf{x}(\tau) = \sum_{j=1}^7 c_{j} e^{\lambda_j^{X}\tau} \mathbf{r}_{j}^{X} + \mathbf{x}_\star \ ,
\end{equation}
where the coefficients $c_{j}$ may also be computed as a series in $g$. This gives
\begin{widetext}
\begin{equation}
    \label{14_sol_noRWA}
    \begin{aligned}
        \rho_{14}(\tau) & \simeq \bigg( \rho_{14}(0) + \sfrac{i g^2 C_{\mathrm{s}} \mathrm{Re}[ \rho_{14}(0) ] }{\Omega} + \sfrac{i g^2 C_{\times}  [ 2 \rho_{22}(0) + 2 \rho_{33}(0) - 1 ] }{2\Omega} - \sfrac{g^2 K_{\times} [ 2 \rho_{11}(0) + \rho_{22}(0)+ \rho_{33}(0) - 1 ]}{2\Omega}  \bigg) e^{ ( - 2 i \Omega - 2 g^2 C_{\mathrm{s}} ) \tau}  \\
    &  \quad + \; \sfrac{g^2 K_{\times} [ 2 \rho_{11}(0) + \rho_{22}(0) + \rho_{33}(0) - 1  ] }{2\Omega}  e^{- 2 g^2 C_{\mathrm{s}} \tau }  \\
    & \quad - \; \tfrac{i g^2 C_{\times} \left( 4 - \frac{C_{\mathrm{s}}^2}{C_{\times}^2} + \frac{C_{\mathrm{s}}}{C_{\times}} \sqrt{8 + \frac{C_{\mathrm{s}}^2}{C_{\times}^2} } \right) \left[ 2 \rho_{22}(0) +2 \rho_{33}(0) - 1 + \left( - \frac{C_{\mathrm{s}}}{C_{\times}} + \sqrt{8 + \frac{C_{\mathrm{s}}^2}{C_{\times}^2}} \right) \mathrm{Re}[\rho_{14}(0)] \right) }{2 \Omega \left( 8 + \frac{C_{\mathrm{s}}^2}{C_{\times}^2}- \frac{C_{\mathrm{s}}}{C_{\times}} \sqrt{8 + \frac{C_{\mathrm{s}}^2}{C_{\times}^2} } \right] } e^{g^2\left( - 3 C_{\mathrm{s}} - \sqrt{ C_{\mathrm{s}}^2 + 8 C_{\times}^2 } \right) \tau } \\
    & \quad - \; \tfrac{i g^2 C_{\times} \left( - \frac{3 C_{\mathrm{s}}}{C_{\times}} + \sqrt{8 + \frac{C_{\mathrm{s}}^2 }{C_{\times}^2} } \right) \left[ \left( - \frac{C_{\mathrm{s}}}{C_{\times}} + \sqrt{8 + \frac{C_{\mathrm{s}}^2}{C_{\times}^2}  } \right) \left( 2 \rho_{22}(0) +2 \rho_{33}(0) - 1 \right) -8 \mathrm{Re}[\rho_{14}(0)] \right] }{4 \Omega \left( 8 + \frac{C_{\mathrm{s}}^2}{C_{\times}^2}- \frac{C_{\mathrm{s}}}{C_{\times}} \sqrt{8 + \frac{C_{\mathrm{s}}^2}{C_{\times}^2} } \right) } e^{g^2\left( - 3 C_{\mathrm{s}} + \sqrt{ C_{\mathrm{s}}^2 + 8 C_{\times}^2 } \right) \tau } 
    \end{aligned}
\end{equation}
\end{widetext}
where $\mathcal{O}(g^4)$ effects have been neglected. In both cases,  for any given initial data  $\rho_{14}(0)$ we have $\rho_{14}(\tau)\to 0$ and $\rho_{14}^{\textsc{(rwa)}}\to 0$ at very late times. In other words, both non-RWA and RWA solutions have the same late-time behaviour for $\rho_{14}$, which we already know since the stationary state at late time is diagonal in the uncoupled energy eigenbasis.

The main difference between Eqs.~(\ref{14_sol_noRWA}) and (\ref{14_RWA_sol}) can be understood in the context of the theorems outlined by Davies \cite{davies1974markovian,davies1976markovian}, and amounts to the simultaneous limit $g^2 \to 0$ and $\tau \to \infty$ while keeping $g^2 a \tau \sim \mathcal{O}(1)$. What is important to note is that Davies' limit in fact includes taking RWA: while the result of Davies is of course mathematically sound, the master equation obtained via Davies' approach cannot be used if we insist on not applying RWA. To put it another way, Davies' theorem does not account for the free parameters in the microscopic Hamiltonian which will vary from problem to problem. Since we started from a UDW interaction (which is the microscopic description of the setup), we are obliged to restrict our attention to a subset of parameter space where Born-Markov approximations apply. This in turn requires us to restrict to ``high-temperature'' regime $\Omega/a\ll 1$ (and all the complicated validity relations found earlier). We only get the same result as Davies' approach if we also apply RWA, and from an EFT perspective this means that we have to add more constraints to the parameter space \textit{in addition} to the validity relations we have found earlier. These restrictions are not given by Davies' theorem and depends on the system under consideration.

\subsection{Entanglement generation and degradation: with RWA vs without RWA}

As shown in the previous subsection, at {\it finite} times the two procedures yield different predictions since the density matrices $\rho_\textsc{ab}^{\textsc{RWA}}(\tau)\neq \rho_\textsc{ab}(\tau)$ (compare for example (\ref{14_sol_noRWA}) and (\ref{14_RWA_sol})). In particular, it can be shown that the RWA result approaches the full result when $\Omega L$ is large. This is consistent with the fact that RWA may lead to superluminal signalling in relativistic settings, yet large $\Omega L$ is precisely the regime where the detectors are so widely separated that the superluminal (but finite-time) propagation becomes negligible. This means that RWA and the full result particularly disagrees precisely when causal relations between the two detectors matter.

In this subsection we underline this point by studying the finite-time entanglement between the two detectors, showing RWA evolution can in general differ significantly from the non-RWA effective Markovian evolution studied in this work. We choose to study negativity \cite{Vidal2002negativity} as the entanglement monotone, defined by
\begin{equation} \label{negativity}
    \mathcal{N}[ \rho_{\textsc{ab}} ] = \frac{ ||\rho_{\textsc{ab}}^{\Gamma_{\mathrm{A}}}||_1-1}{2}\,,
\end{equation}
where $\Gamma_{\textsc{a}}$ denotes the partial transpose with respect to subsystem A and $||\cdot||_1$ is the trace norm. {Negativity turns out to vanish for separable states, and being an entanglement monotone takes its maximum value for maximally entangled bipartite states (such as Bell states for qubits, corresponding to a  negativity of $1/2$).}

We now show what happens to the entanglement generation and degradation by plotting $\mathcal{N}[ \rho_{\textsc{ab}}(\tau) ]$ for various initial states $\rho_{\textsc{ab}}(0)$ by comparing RWA-based vs non-RWA based solutions. We do not attempt to reproduce the plots in \cite{Hu2015twodetectorent,Hu2022loss-anti} since they are outside the domain of validity of the Born-Markov approximations (namely $\Omega/a \ll 1$ must always be enforced in what follows).  

\begin{figure*}[tp]
    \centering
    \includegraphics[scale=0.9]{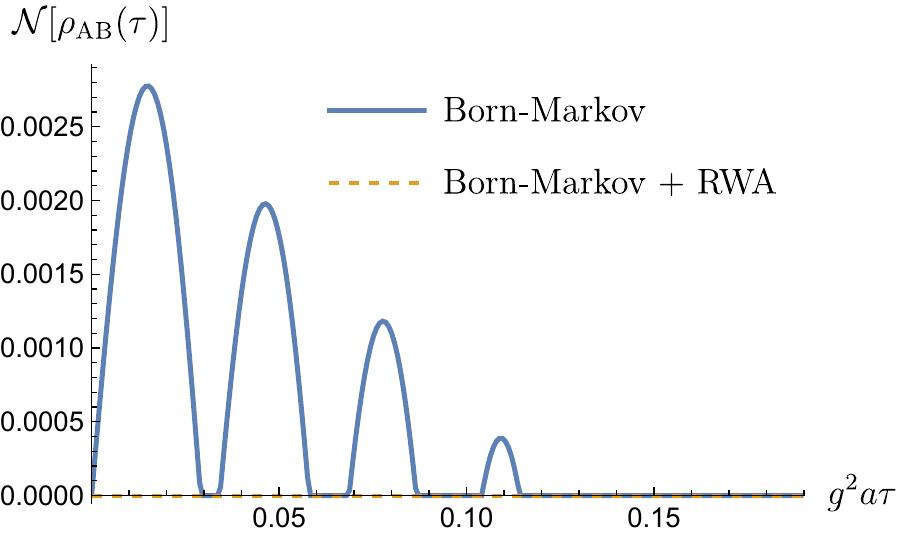}
    \includegraphics[scale=0.9]{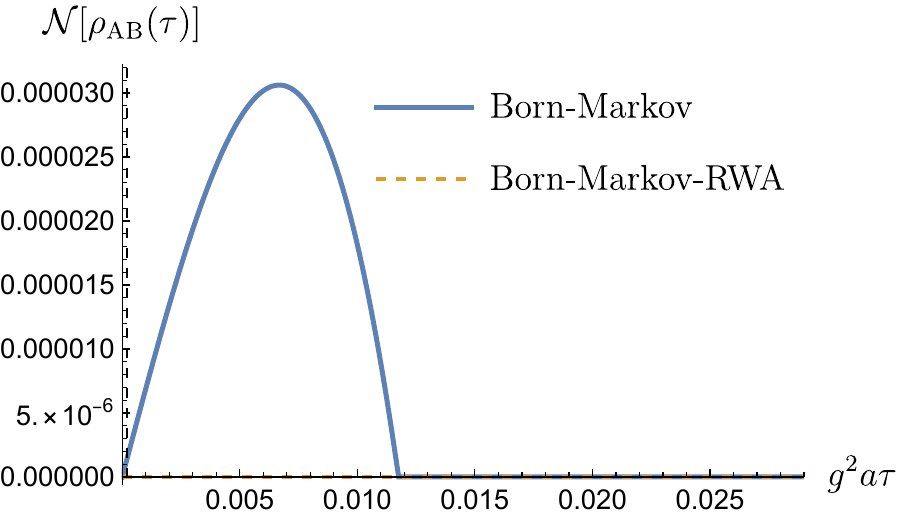}
    \caption{Initially prepared in ground state $\ket{\downarrow\downarrow}$ with or without RWA. The common parameter choices are $\Omega/a=0.01$ and $g = 0.01$.  \textit{Left}: $aL=0.25$. \textit{Right}: $aL=2$. Note that for RWA-based solution the detectors \textit{cannot} get entangled from the ground state even though the non-RWA solution could.}
    \label{fig: ground-state}
\end{figure*}

\begin{figure*}[tp]
    \centering
    \includegraphics[scale=0.9]{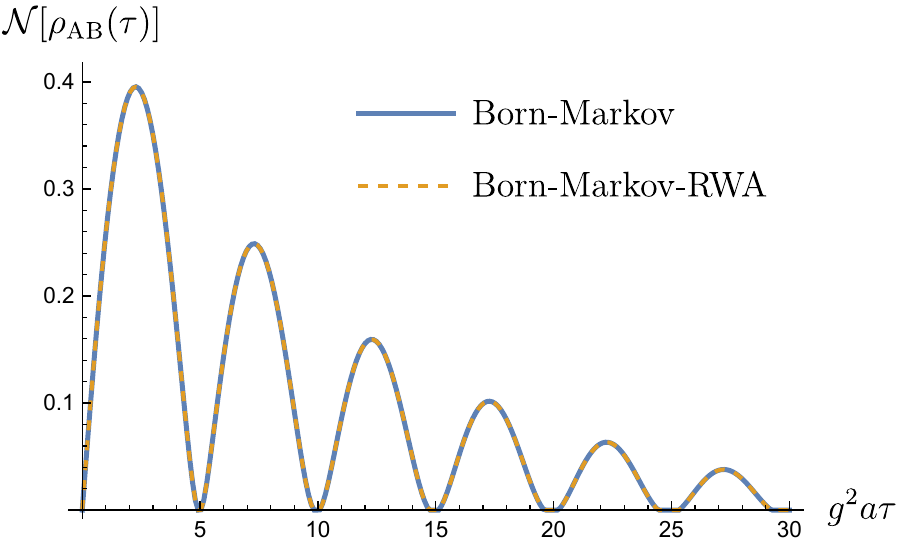}
    \includegraphics[scale=0.9]{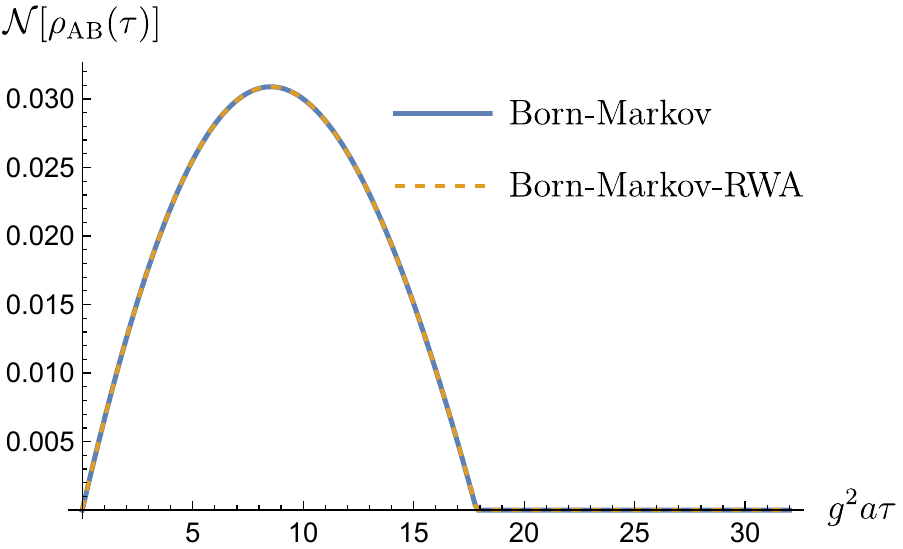}
    \caption{Initially prepared in states $\ket{\downarrow\uparrow}$ based on the choice of states in \cite{Hu2015twodetectorent}, with or without RWA. The common parameter choices are $\Omega/a=0.01$ and $g = 0.01$. \textit{Left}: $\ket{\downarrow\uparrow}$, $aL=0.25$. \textit{Right}: $\ket{\downarrow\uparrow}$, $aL=2$. The difference between the negativities with or without RWA is of order $10^{-9}$ to $10^{-7}$.}
    \label{fig: excited-state}
\end{figure*}

\begin{figure}[tp]
    \centering
    \includegraphics[scale=0.9]{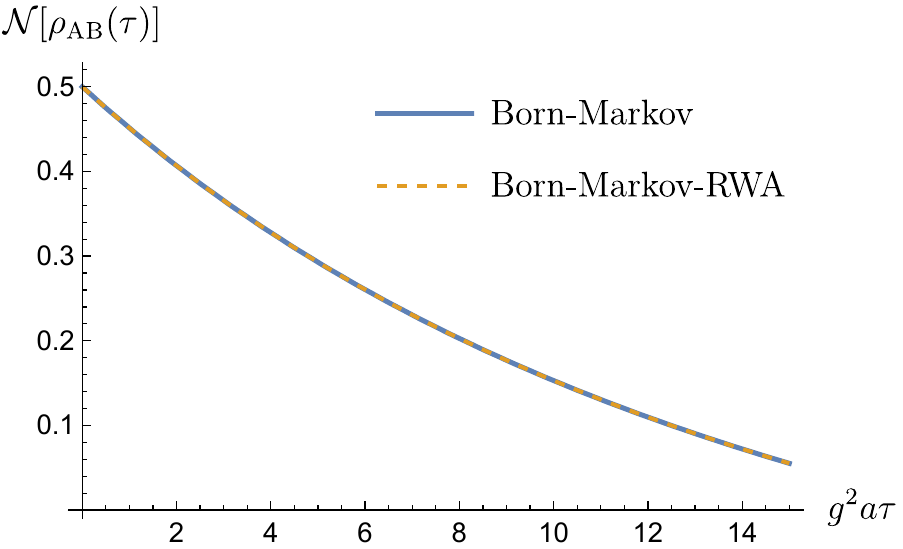}
    \includegraphics[scale=0.9]{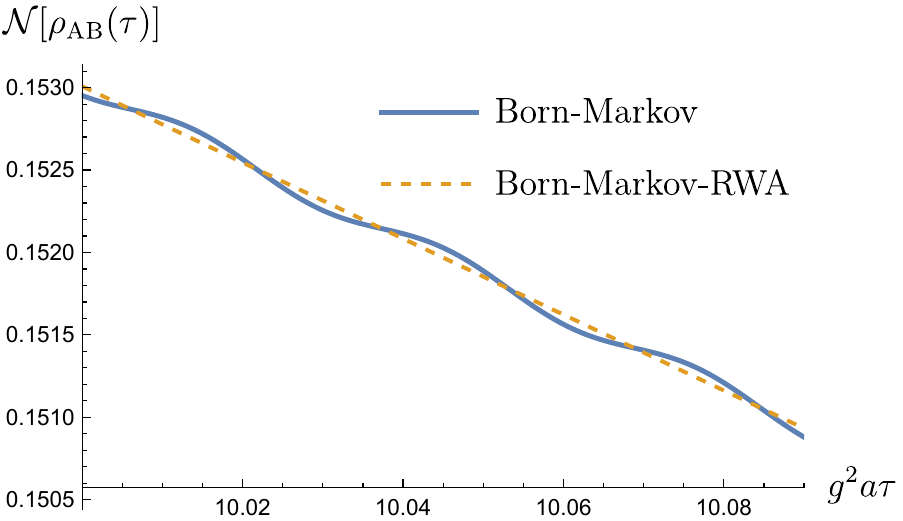}
    \includegraphics[scale=0.9]{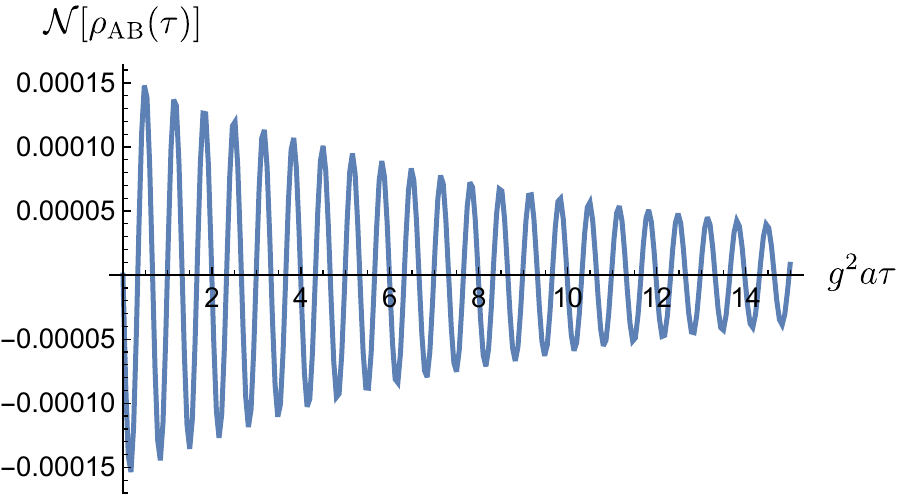}
    \caption{ Initially prepared in the Bell state $\ket{\Phi^+}=\frac{1}{\sqrt{2}}(\ket{\downarrow\downarrow}+\ket{\uparrow\uparrow})$, based on the choice of states in \cite{Hu2015twodetectorent} and \cite{Hu2022loss-anti} respectively, with or without RWA. The common parameter choices are $\Omega/a=0.01$, $g = 0.01$ and $aL=2$. \textit{Left:} negativity as a function of time. \textit{Middle:} Zoomed-in version around $g^2a\tau=10$, showing oscillatory behaviour for the solution without RWA. \textit{Right:} The difference in negativity $\Delta\mathcal{N} = \mathcal{N}-\mathcal{N}^{(\textsc{rwa})}$ is $\mathcal{O}(g^2)$. It shows oscillatory behaviour that decays at later times, with the non-RWA negativity approaching RWA version at very late times.}
    \label{fig: bell-state}
\end{figure}

In Figure~\ref{fig: ground-state} we show the case when the detectors begin the evolution in their ground states $\ket{\downarrow\downarrow}$. The common parameter choices are $\Omega/a=0.01$, $g = 0.01$ and we consider both smaller $(aL=0.25)$ and larger ($aL=2$) separations. We see a stark difference between the RWA-based solution and the one without RWA: at all times the RWA-based solution \textit{cannot} generate entanglement, while the full Born-Markov solution studied in this work (without RWA) can get temporarily entangled at later times. The smaller the detector separation, the longer the entanglement persists for non-RWA solutions. Above $aL \sim 2$ both solutions remain separable at all times. 

We emphasize that our results are completely within the domain of validity of the approximations (see Section~\ref{sec: validity} for how to check its validity). This can be compared with Figure~\ref{fig: excited-state}, where we show the case when the detectors are initially prepared in states $\ket{\downarrow\uparrow}$, based on the choice of states in \cite{Hu2015twodetectorent} with or without RWA for the same set of parameters. This time, the amount of entanglement is larger, however the differences between RWA and non-RWA are very slight: the negativity are very similar, and one can check that their differences $\mathcal{N}-\mathcal{N}^{(\textsc{rwa})}$ is of order $10^{-9}$ to $10^{-7}$ (i.e., way below $\sim g^2$). 

In Figure~\ref{fig: bell-state} we show the case when the detectors are in the Bell state $\ket{\Phi^\pm}=\frac{1}{\sqrt{2}}(\ket{\downarrow\downarrow}+ \ket{\uparrow\uparrow})$ with the same parameter choices. Since the Bell state is maximally entangled, there is no surprise in having entanglement degradation especially after interacting with the environment. This scenario is particularly interesting because it highlights what RWA actually does to entanglement dynamics: it ``smoothens out'' oscillatory behaviour of negativity obtained with only Born-Markov approximation, as can be seen from the zoomed-in version in the middle figure of Figure~\ref{fig: bell-state}. The oscillatory behaviour also is modulated by a decaying function at late times (the rightmost figure) and the differences in negativity is $\mathcal{O}(g^2$). Therefore, we see that the non-RWA solution approaches RWA result at very late times. Figure~\ref{fig: bell-state} gives an explicit demonstration of how RWA is a form of averaging/coarse-graining, essentially by removing the oscillatory components and get the overall large-timescale behaviour right. It is worth noting that Figure~\ref{fig: bell-state} has similar feature as the one shown in Figure 1 of \cite{benatti2022local}, even though there the parameters chosen are strictly speaking outside the validity Markovian approximation (due to large $\Omega/a\gg 1$).

To summarize, the results we obtained so far show that indeed analyses obtained with or without RWA can drastically differ depending on the physical setups and initial data at hand. In Figure~\ref{fig: ground-state}, the non-RWA solution can generate entanglement while the RWA solution cannot; in Figure~\ref{fig: bell-state}, we see the opposite where the entanglement dynamics is hardly distinguishable. On a closer look, which is best demonstrated using Bell states, we see that the RWA indeed applies averaging/coarse-graining procedure so that any oscillatory behaviour at short timescales are smoothened out. We have made sure that all our plots are within the regime of validity in which the dynamics is manifestly CP and Markovian even without RWA.

\section{Conclusion}

In this work we argue that the lack of CP property after performing the Born-Markov approximation in the standard open systems approach is problematic, as there is residual memory in the system operators that was not removed when performing the Markovian approximation. This leads to spurious divergences and lack of complete positivity, which typically ends up being resolved by \textit{ad hoc} procedures such as the RWA. We show this explicitly in the context of the oft-studied model of two accelerating detectors interacting with a quantized massless scalar field in flat spacetime{, using a different Markovian approximation first used in \cite{Martin2018cosmologydecohere}}. The fact that the bath (quantum field) is relativistic is important as it is well-known that rotating wave approximations can lead to important causality violations \cite{Funai2019rotatingwave}: this has to do with the fact that for fields in their vacuum states, both the co-rotating and counter-rotating terms are important and there is no single ``rotating frame'' that can counter all frequencies of the bath modes (i.e. the field is a continuum of infinitely many modes, as opposed to quantum optics settings where the bath is a laser tuned to a single particular frequency).

Our work has a wider implication to generic open system framework: more specifically, it suggests that the ``infamous'' property of being non-CP for Redfield-type equations seen in standard literature (see, e.g., \cite{breuer2002theory,lidar2020lecture}) is not quite correct. The problem is that when one performs the Born-Markov approximations, one has to restrict the parameter spaces for which the resulting equation is valid. Our example shows that the CP property is already guaranteed simply by faithfully staying in the regime where the Markovian approximation is valid. Applying RWA to fix the non-CP problem amounts in some sense to ``shifting the goalpost'': while the resulting GKSL-RWA equation \textit{is} manifestly CP even for large $\Omega/a \gtrsim 1$, we are \textit{not} allowed to do so because we needed $\Omega/a\ll 1$ to even arrive at this step. Therefore any calculation for $\Omega/a \gtrsim 1$ is automatically not reliable. 

This problem of ``misusing'' certain open systems approximations is actually worse than it appears because there is a general expectation that accelerating detectors should thermalize and entangle in some form or another. While both expectations are fine, showing that they do hold properly is rather tricky. For example, we here show that \textit{both} the RWA-based and non-RWA-based dynamics have the same stationary solutions given by the maximally mixed state: this is to be understood as a high-temperature thermal state since we are in the regime where $\Omega/a\ll 1$. Therefore, the asymptotic state cannot easily be used to test the validity of approximations taken. This is exacerbated by the fact that in the RWA-based approach, one can obtain an actual thermal Gibbs state with arbitrarily-sized Unruh temperature without properly removing the memory effect when taking the Markovian limit. This is a very tempting outcome, but we view this as an example of getting an expected result via incorrect reasoning.

It is interesting to see that there is merit in approaching problems in open quantum systems by treating the problem as an EFT (i.e. in the Open EFT framework \cite{Grishchuk:1989ss,Brandenberger:1990b,Calzetta:1995ys,Burgess:1996mz,Kiefer:1998qe,Agon:2014uxa,Burgess:2014eoa,Boyanovsky:2015tba,Boyanovsky:2015jen,Boyanovsky:2015xoa,Burgess:2015ajz,Braaten:2016sja,Hollowood:2017bil,Shandera:2017qkg,Agon:2017oia,Baidya:2017eho,Martin2018cosmologydecohere,Burrage:2018pyg,Martin:2018lin,burgess2020introduction,Burgess:2021luo,Brahma:2021mng,Brahma:2022yxu,Kading:2022jjl} typically used for studying quantum fields themselves as opposed to qubits). From an EFT perspective, when a hierarchy of scales can be utilized (in our case between the timescales of the environment and system), relative ``effective'' simplicity can arise. Furthermore, every EFT has its domain of validity and every approximation shrinks this domain; everything works so long as one remains strictly within the regime where the approximation is expected to work. Conversely, the breakdown of any approximate equation necessarily arises due to being outside the domain of validity of that approximation. Following this EFT line of thought, applying the RWA/secular approximation to fix CP-violations is an odd thing to do because it amounts to moving the target by changing the original problem to another (possibly unrelated) problem. It is also worth noting that a similar model to ours has been studied in \cite{Charis2020nonmarkov}, using harmonic oscillators as detectors in a static bath, and the results are consistent with what we found here.

We suspect that there are broader classes of problems within the open system framework that have these sort of spurious issues arising from apparent non-CP properties of approximate master equations. One line of investigation we are pursuing is to identify whether such problems arise in more general open systems settings simply due to forcing the resulting equation to to work outside the domain of applicability of approximations taken. If these issues can be solved by properly accounting for memory effects then this would make the open system framework more reliable, robust and more widely applicable for making physical predictions.

\section*{Acknowledgment}

We thank Thomas Colas and Jose de Ram\'{o}n for helpful discussions. G.K.~is supported by the Simons Foundation award ID 555326 under the Simons Foundation Origins of the Universe initiative, Cosmology Beyond Einstein's Theory as well as by the European Union Horizon 2020 Research Council grant 724659 MassiveCosmo ERC2016COG. E.T.~acknowledges generous support from Mike and Ophelia Lazaridis Fellowship during the period this work was undertaken. The authors would like to acknowledge Canadian Student and Postdoc Conference on Gravity (CSPCG) 2021 for creating opportunity for both authors to meet for the first time, hence making this work possible. Research at Perimeter Institute is supported in part by the Government of Canada through the Department of Innovation, Science and Economic Development Canada and by the Province of Ontario through the Ministry of Colleges and Universities. This work is partially conducted at University of Waterloo, Perimeter Institute for Theoretical Physics and Institute for Quantum Computing, which lies on the traditional territory of the Neutral, Anishnaabeg, and Haudenosaunee Peoples. The University of Waterloo and the Institute for Quantum Computing are situated on the Haldimand Tract, land that was promised to Six Nations, which includes six miles on each side of the Grand River. McMaster University is located on the traditional territories of the Mississauga and Haudenosaunee nations, and within the lands protected by the ``Dish with One Spoon'' wampum agreement.

\appendix

\begin{widetext}

\section{Nakajima-Zwanzig evolution of the $X$-block}
\label{sec: NZ-ME2_X}

In this Appendix we give the explicit evolution equations for the $X$-block $\rho^\ssI_\textsc{ab,X}(\tau)$ defined in (\ref{XO_split}) for the Nakajima-Zwanzig master equation (\ref{eq: NZ-final}) at $\mathcal{O}(g^2)$ given in \S\ref{sec: markovian}. The $X$-block contains five independent matrix elements. The coupled integro-differential equations for these matrix elements are given by:
{\allowdisplaybreaks
\begin{align}
    \frac{\partial \rho_{11}^{\ssI}}{\partial \tau} 
    & =  2 g^2 \int_0^\tau \exd s\, \mathrm{Re}[ \Ws(s) ] \cos(\Omega s) \bigg( - 2 \rho_{11}^{\ssI}(\tau- s) + \rho_{22}^{\ssI}(\tau - s) + \rho^{\ssI}_{33}(\tau - s) \bigg) \notag \\
    & \quad +2 g^2 \int_0^\tau \exd s\, \mathrm{Im}[ \Ws(s) ] \sin(\Omega s) \bigg( 2 \rho_{11}^{\ssI}(\tau- s) + \rho_{22}^{\ssI}(\tau - s) + \rho^{\ssI}_{33}(\tau - s) \bigg)  \notag \\
    & \quad - 2 g^2 e^{- 2 i \Omega \tau} \int_0^\tau \exd s \ e^{+ i \Omega s } \mathcal{W}^{\ast}_{\times}(s) \rho_{14}^{\ssI}(\tau - s) - 2 g^2 e^{+ 2 i \Omega \tau} \int_0^\tau \exd s \ e^{- i \Omega s } \Wc(s) \rho_{14}^{\ssI\ast}(\tau - s) \notag \\
    & \quad + 4 g^2 \int_0^\tau \exd s \ \bigg( \mathrm{Re}[\Wc(s)] \cos(\Omega s) + \mathrm{Im}[\Wc(s)] \sin(\Omega s) \bigg) \mathrm{Re}[ \rho_{23}^{\ssI}(\tau - s) ]  \\
\notag \\
    \frac{\partial \rho_{22}^{\ssI}}{\partial \tau} 
    & \quad =  2 g^2 \int_0^\tau \exd s \  \mathrm{Re}[ \Ws(s) ] \cos(\Omega s) \bigg( 1 - 3 \rho_{22}^{\ssI}(\tau - s) - \rho^{\ssI}_{33}(\tau - s) \bigg) \notag\\
    & \quad +  2 g^2 \int_0^\tau \exd s \ \mathrm{Im}[ \Ws(s) ] \sin(\Omega s) \bigg( 1 - 2 \rho_{11}^{\ssI}(\tau- s) - \rho_{22}^{\ssI}(\tau - s) - \rho^{\ssI}_{33}(\tau - s) \bigg)   \notag \\
    & \quad + 2 g^2 e^{- 2 i \Omega \tau} \int_0^\tau \exd s \ e^{+ i \Omega s } \mathrm{Re}[\Wc(s)] \rho_{14}^{\ssI}(\tau - s) + 2 g^2 e^{+ 2 i \Omega \tau} \int_0^\tau \exd s \ e^{- i \Omega s } \mathrm{Re}[\Wc(s)] \rho_{14}^{\ssI\ast}(\tau - s) \notag \\
    & \quad - 4 g^2 \int_0^\tau \exd s \ \bigg( \mathrm{Re}[\Wc(s)] \mathrm{Re}[ \rho_{23}^{\ssI}(\tau - s) ]  + \mathrm{Im}[\Wc(s)] \mathrm{Im}[ \rho_{23}^{\ssI}(\tau - s) ] \bigg) \cos(\Omega s) \\
\notag \\
    \frac{\partial \rho_{33}^{\ssI}}{\partial \tau} 
    & =  2 g^2 \int_0^\tau \exd s \ \mathrm{Re}[ \Ws(s) ] \cos(\Omega s) \bigg( 1 - \rho_{22}^{\ssI}(\tau - s) - 3\rho^{\ssI}_{33}(\tau - s) \bigg) \notag \\
    &+ 2 g^2 \int_0^\tau \exd s \ \mathrm{Im}[ \Ws(s) ] \sin(\Omega s) \bigg( 1 - 2 \rho_{11}^{\ssI}(\tau- s) - \rho_{22}^{\ssI}(\tau - s) - \rho^{\ssI}_{33}(\tau - s) \bigg)  \notag \\
    & + 2 g^2 e^{- 2 i \Omega \tau} \int_0^\tau \exd s \ e^{+ i \Omega s } \mathrm{Re}[\Wc(s)] \rho_{14}^{\ssI}(\tau - s) \ + \ 2 g^2 e^{+ 2 i \Omega \tau} \int_0^\tau \exd s \ e^{- i \Omega s } \mathrm{Re}[\Wc(s)] \rho_{14}^{\ssI\ast}(\tau - s) \notag \\
    & - 4 g^2 \int_0^\tau \exd s \ \bigg( \mathrm{Re}[\Wc(s)] \mathrm{Re}[ \rho_{23}^{\ssI}(\tau - s) ] - \mathrm{Im}[\Wc(s)] \mathrm{Im}[ \rho_{23}^{\ssI}(\tau - s) ] \bigg) \cos(\Omega s)  \\
    \notag \\
    \frac{\partial \rho_{14}^{\ssI}}{\partial \tau} 
    & =  - 4 g^2 \int_0^{\tau} \exd s\; e^{+ i \Omega s} \mathrm{Re}[\Ws(s)] \rho_{14}^{\ssI}(\tau - s) + 4 g^2 e^{+ 2 i \Omega \tau} \int_0^{\tau} \exd s\; e^{- i \Omega s} \mathrm{Re}[\Ws(s)] \mathrm{Re}\big[\rho_{23}^{\ssI}(\tau - s)\big] \notag \\
    &  - 2 g^2 e^{+ 2 i \Omega \tau} \int_0^\tau \exd s \  \mathrm{Re}[ \Wc(s) ] e^{- i \Omega s} \bigg( 1 - 2 \rho_{22}^{\ssI}(\tau - s) - 2\rho^{\ssI}_{33}(\tau - s) \bigg)  \notag \\
    & - 2 i g^2 e^{+ 2 i \Omega \tau} \int_0^\tau \exd s \ \mathrm{Im}[ \Wc(s) ] e^{- i \Omega s} \bigg( 1 - 2 \rho_{11}^{\ssI}(\tau - s) - \rho^{\ssI}_{22}(\tau - s) - \rho^{\ssI}_{33}(\tau - s) \bigg)   \\
    \notag\\
    \frac{\partial \rho_{23}^{\ssI}}{\partial \tau} 
    & =  2 g^2 e^{- 2 i \Omega \tau} \int_0^{\tau} \exd s\; e^{+ i \Omega s} \mathrm{Re}[\Ws(s)] \rho_{14}^{\ssI}(\tau - s) + 2 g^2 e^{+ 2 i \Omega \tau} \int_0^{\tau} \exd s\; e^{- i \Omega s} \mathrm{Re}[\Ws(s)] \rho_{14}^{\ssI\ast}(\tau - s) \notag \\
    &  - 4 g^2 \int_0^{\tau} \exd s\;  \mathrm{Re}[\Ws(s)] \cos(\Omega s) \rho_{23}^{\ssI}(\tau - s) + 2 g^2 \int_0^\tau \exd s\;  \mathrm{Re}[ \Wc(s) ] \cos(\Omega s) \bigg( 1 - 2 \rho_{22}^{\ssI}(\tau - s) - 2\rho^{\ssI}_{33}(\tau - s) \bigg) \notag \\
    &  + 2 g^2 \int_0^\tau \exd s\;  \mathrm{Im}[ \Wc(s) ] \sin(\Omega s)  \bigg( 1 - 2 \rho_{11}^{\ssI}(\tau - s) - 2{\rho^{\ssI}_{22}(\tau - s)} \bigg) 
\end{align}
}

\section{Nakajima-Zwanzig evolution of the $O$-block}
\label{sec: NZ-ME2_O}

In this Appendix we give the explicit evolution equations for the $O$-block $\rho^\ssI_\textsc{ab,O}(\tau)$ defined in (\ref{XO_split}) for the Nakajima-Zwanzig master equation (\ref{eq: NZ-final}) at $\mathcal{O}(g^2)$ given in \S\ref{sec: markovian}. The $O$-block contains four independent matrix elements. The coupled integro-differential equations for these matrix elements are given by:
{\allowdisplaybreaks
\begin{align} 
\label{square12NZ}
    \frac{\partial\rho^{\ssI}_{12}}{\partial\tau} 
    &= -2g^2\int_0^\tau \exd s\, \bigg[ \text{Re}[\mathcal{W}_{\s}(s)]\cos(\Omega s) \big( \rho^{\ssI}_{12}(\tau-s)-\rho^{\ssI}_{34}(\tau-s) \big) -\text{Im}[\mathcal{W}_{\s}(s)]\sin(\Omega s) \big( \rho^{\ssI}_{12}(\tau-s)+\rho^{\ssI}_{34}(\tau-s) \big) \bigg] \notag\\
    & \quad  -2g^2\int_0^\tau \exd s\,\text{Re}[\mathcal{W}_{\s}(s)]e^{i\Omega s}\rho^{\ssI}_{12}(\tau-s) +2g^2e^{2i\Omega\tau}\int_0^\tau \exd s\, \text{Re}[\mathcal{W}_{\s}(s)]e^{-i\Omega s}\rho_{12}^{\ssI*}(\tau-s)\notag\\
    & \quad -2g^2e^{2i\Omega \tau}\int_0^\tau \exd s\,e^{-i\Omega s}\mathcal{W}_\times(s)\rho_{24}^{\ssI\ast}(\tau-s)+ {   2 } g^2\int_0^\tau \exd s\,\rho^{\ssI}_{24}(\tau-s)\bigg( \Scale[0.90]{ \text{Re}[\mathcal{W}_\times(s)]\cos(\Omega s) + \text{Im}[\mathcal{W}_\times(s)]\sin(\Omega s) } \bigg) \notag\\
    &  \quad + 2g^2e^{2i\Omega\tau}\int_0^\tau \exd s\,e^{-i\Omega s}\text{Re}[\mathcal{W}_\times(s)]\rho_{13}^{\ssI*}(\tau-s) -2g^2\int_0^\tau \exd s\,\mathcal{W}^{{ \ast}}_\times(s) \cos(\Omega s) \rho^{\ssI}_{13}(\tau-s) \\
\notag \\
    \frac{\partial\rho^{\ssI}_{13}}{\partial\tau} 
    &= -4g^2\int_0^\tau \exd s\, \text{Re}[\mathcal{W}_{\s}(s)] e^{i\Omega s}\rho^{\ssI}_{13}(\tau-s) + {   2 i g^2 \int_0^{\tau} \exd s\; \mathcal{W}^{\ast}_{\mathrm{s}}(s) \sin(\Omega s) \rho^{\ssI}_{13}(\tau - s)  }\notag\\
    & \quad +2g^2\int_0^\tau \exd s\,\rho_{24}^{\ssI}(\tau-s)\bigg( \Scale[0.90]{ \text{Re}[\mathcal{W}_{\s}(s)]\cos(\Omega s)+\text{Im}[\mathcal{W}_{\s}(s)]\sin(\Omega s) } \bigg) +2g^2e^{2i\Omega \tau}\int_0^\tau \exd s\, \text{Re}[\mathcal{W}_{\s}(s)] e^{-i\Omega s} \rho_{13}^{\ssI \ast}(\tau - s) \notag\\
    & \quad  -2g^2e^{2i\Omega \tau}\int_0^\tau \exd s\, \mathcal{W}_\times(s) e^{-i\Omega s} \rho_{34}^{\ssI\ast}(\tau-s)
    +2g^2\int_0^\tau \exd s\,\rho^{\ssI}_{34}(\tau-s)\bigg(\Scale[0.90]{ \text{Re}[\mathcal{W}_\times(s)]\cos(\Omega s)+\text{Im}[\mathcal{W}_\times(s)]\sin(\Omega s) } \bigg) \notag \\
    &  \quad +2g^2e^{2i\Omega \tau}\int_0^\tau \exd s\, \text{Re}[\mathcal{W}_\times(s)] e^{-i\Omega s} \rho_{12}^{\ssI\ast}(\tau-s) -2g^2\int_0^\tau \exd s\,\mathcal{W}^{{ \ast}}_\times(s) \cos(\Omega s) \rho_{12}^{\ssI}(\tau-s)  \\
\notag \\
    \frac{\partial\rho^{\ssI}_{24}}{\partial\tau} 
    &= -2g^2\int_0^\tau \exd s\, \text{Re}[\mathcal{W}_{\s}(s)] e^{i\Omega s} \rho^{\ssI}_{24}(\tau-s)-2g^2\int_0^\tau \exd s\, \bigg( \text{Re}[\mathcal{W}_{\s}(s)]\cos(\Omega s)+\text{Im}[\mathcal{W}_{\s}(s)]\sin(\Omega s) \bigg) \rho^{\ssI}_{24}(\tau-s) \notag\\
    & \quad +2g^2\int_0^\tau \exd s\, \bigg( \Scale[0.90]{ \text{Re}[\mathcal{W}_{\s}(s)]\cos(\Omega s) -\text{Im}[\mathcal{W}_{\s}(s)]\sin(\Omega s) } \bigg) \rho^{\ssI}_{13}(\tau-s) +2g^2e^{2i\Omega\tau}\int_0^\tau \exd s\, \text{Re}[\mathcal{W}_{\s}(s)] e^{-i\Omega s} \rho^{\ssI\ast}_{24}(\tau-s) \notag\\
    & \quad -2g^2e^{2i\Omega \tau} \int_0^\tau \exd s\,{   \mathcal{W}_\times^{\ast}(s) } e^{-i\Omega s}  \rho_{12}^{\ssI\ast}(\tau-s) + 2 g^2 \int_0^\tau \exd s\,\rho^{\ssI}_{12}(\tau-s)\bigg( \Scale[0.90]{  \text{Re}[\mathcal{W}_\times(s)]\cos(\Omega s)-\text{Im}[\mathcal{W}_\times(s)]\sin(\Omega s ) } \bigg) \notag\\
    & \quad +2g^2e^{2i\Omega\tau}\int_0^\tau \exd s\,e^{-i\Omega s}\text{Re}[\mathcal{W}_\times(s)]\rho_{34}^{\ssI\ast}(\tau-s) -2g^2\int_0^\tau \exd s\, {  \mathcal{W}_\times(s) } \cos(\Omega s) \rho^{\ssI}_{34}(\tau-s) \\
\notag \\
    \frac{\partial\rho^{\ssI}_{34}}{\partial\tau} 
    &= -2 g^2 \int_0^\tau \exd s\, \bigg( \Scale[0.90]{ \text{Re}[\mathcal{W}_{\s}(s)]\cos(\Omega s)+\text{Im}[\mathcal{W}_{\s}(s)]\sin(\Omega s) }\bigg) \rho^{\ssI}_{34}(\tau-s) - 2g^2\int_0^\tau \exd s\,\text{Re}[\mathcal{W}_{\s}(s)] e^{i\Omega s}\rho^{\ssI}_{34}(\tau-s) \notag\\
    & \quad +2g^2\int_0^\tau \exd s\ \bigg( \Scale[0.90]{ \text{Re}[\mathcal{W}_{\s}(s)]\cos(\Omega s) - \text{Im}[\mathcal{W}_{\s}(s)]\sin(\Omega s) } \bigg) \rho^{\ssI}_{12}(\tau-s) +2g^2e^{2i\Omega\tau}\int_0^\tau \exd s\,\text{Re}[\mathcal{W}_{\s}(s)] e^{-i\Omega s}\rho_{34}^{\ssI\ast}(\tau-s) \notag\\
    & \quad + 2g^2\int_0^\tau \exd s\, \bigg( \Scale[0.90]{ \text{Re}[\mathcal{W}_\times(s)]\cos(\Omega s) - \text{Im}[\mathcal{W}_\times(s)]\sin(\Omega s) } \bigg) \rho^{\ssI}_{13}(\tau-s) -2g^2e^{2i\Omega \tau}\int_0^\tau \exd s\, \mathcal{W}_\times^*(s) e^{-i\Omega s}\rho_{13}^{\ssI\ast}(\tau-s) \notag \\
    & \quad - 2 g^2 \int_0^\tau \exd s\, \mathcal{W}_\times(s) { \cos(\Omega s) } \rho^{\ssI}_{24}(\tau-s)+2g^2e^{2i\Omega\tau}\int_0^\tau \exd s\, \text{Re}[\mathcal{W}_\times(s)]  e^{-i\Omega s}\rho_{24}^{\ssI\ast}(\tau-s)\label{square34NZ}
\end{align}
}

\section{When is Alice's detector Markovian?}
\label{App:Alice}

In this section we expand the argument leading to Eq.~(\ref{Alice_whentrue}), which shows that Alice's detector is only Markovian when $\Omega / a \ll 1$ even in the ``standard'' Markovian approach. Recall that the Nakajima-Zwanzig equation for the off-diagonal components of Alice's detector is (\ref{NZ_Alice_off}), repeated here for convenience:
\begin{subequations}
    \label{Markov_NZ_off_App}
\begin{align}
    \frac{\dd \rho^\ssI_{\textsc{a},12}}{\dd \tau} &\simeq \Scale[0.90]{ i g^2 \mathcal{D} \rho_{\textsc{a},12}^{\ssI}(\tau)  + F_{\mathrm{NZ}}(\tau) } \,,\\
    F_{\mathrm{NZ}}(\tau) &\coloneqq  { - 2g^2 \int\limits_0^\tau \dd s\ \mathrm{Re}\left[ \mathcal{W}_{\s}(s) \right] e^{+ i \Omega s} \rho^\ssI_{\textsc{a},12}(\tau - s) + 2g^2 e^{+ 2 i \Omega \tau} \int\limits_0^\tau \dd s\ \mathrm{Re}\left[ \mathcal{W}_{\s}(s) \right] e^{- i \Omega s} \rho^{\ssI\ast}_{\textsc{a},12}(\tau - s) }
\end{align}
\end{subequations}
with a counter-term $ g^2 \mathcal{D} \rho_{\textsc{a} 12}^{\ssI}(\tau)$ (see the definition (\ref{calD_def}) below) included to ensure that the Markovian solution oscillates at the physical detector gap \cite{kaplanek2020hot}. Applying the ``standard'' Markovian approximation  $\varrho^{\ssI}_{\textsc{a}}(\tau - s) \simeq \varrho^{\ssI}_{\textsc{a}}(\tau)$ commonly employed in the literature yields Eq.~(\ref{Markov_Alice_off}), we get 
\begin{subequations}    \label{Markov_Alice_off_App}
\begin{align}
    \frac{\dd \rho^\ssI_{\textsc{a},12}}{\dd \tau} &\simeq \Scale[0.90]{ i g^2 \mathcal{D} \rho_{\textsc{a},12}^{\ssI}(\tau)  + F_{\mathrm{M}}(\tau) } \quad \\
    F_{\mathrm{M}}(\tau) &\coloneqq { - 2 g^2 \int\limits_0^\infty \dd s\ \mathrm{Re}\left[ \mathcal{W}_{\s}(s) \right] e^{+ i \Omega s} \rho^\ssI_{\textsc{a},12}(\tau) + 2 g^2 e^{+ 2 i \Omega \tau} \int\limits_0^\infty \dd s\ \mathrm{Re}\left[ \mathcal{W}_{\s}(s) \right] e^{- i \Omega s} \rho^{\ssI\ast}_{\textsc{a},12}(\tau) }
\end{align}
\end{subequations}
with the counter-term also included. Here we have defined a constant $\mathcal{D}$ that is UV-regulated (i.e. UV-divergent in the limit $\epsilon\to 0^+$):
\begin{equation} \label{calD_def}
\mathcal{D} \coloneqq 2 \int_0^\infty \exd s \ \mathrm{Re}\left[ \mathcal{W}_{\s}(s) \right] \sin(\Omega s) \ = \ \frac{ \Omega }{ 2\pi^2 }  \log(e^{\gamma} a\epsilon)  + \frac{ \Omega }{ 2\pi^2 }  \mathrm{Re}\left[ \psi^{(0)}\left( - \sfrac{i \Omega}{a} \right) \right]\,,
\end{equation}
with $\gamma$ the Euler-Mascheroni constant, $\psi^{(0)}(z) = \Gamma'(z) / \Gamma(z)$ the digamma function \cite{olver2010nist}, and $\epsilon$ the finite UV cutoff associated with the $i \epsilon$-prescription of the correlator $\mathcal{W}_{\s}(s)$ defined in (\ref{eq: Wightman-functions1}). Using the definition of constant $\mathcal{C}$ from (\ref{mathcalC_def}), Eq.~(\ref{Markov_Alice_off_App}) simplifies to
\begin{equation} \label{Markov_Alice_off_App2}
\frac{\dd \rho^\ssI_{\textsc{a},12}}{\dd \tau} \simeq - g^2 \mathcal{C} \rho_{\textsc{a},12}^{\ssI}(\tau) + g^2 ( \mathcal{C} - i \mathcal{D} ) e^{+ 2i \Omega \tau} \rho_{\textsc{a},12}^{\ssI\ast}(\tau)\,.
\end{equation}
\noindent The Markovian solution of Eq.~\eqref{Markov_Alice_off_App} is given by
\begin{equation} \label{Alice_sol_App}
\rho^\ssI_{\textsc{a},12}(\tau) \simeq \mathcal{A} e^{ - g^2 \mathcal{C} \tau} + \mathcal{B} e^{(- g^2 \mathcal{C} + 2 i \Omega ) \tau}
\end{equation}
where the constant coefficients $\mathcal{A}$ and $\mathcal{B}$ are given by
\begin{equation}
\mathcal{A} =  \rho^{\ssI}_{\textsc{a},12}(0) +  \rho^{\ssI\ast}_{\textsc{a},12}(0)  \frac{g^2 ( \mathcal{D} + i \mathcal{C} )}{2 \Omega} \qquad \mathrm{and} \qquad \mathcal{B} = - \rho^{\ssI\ast}_{\textsc{a},12}(0) \frac{g^2 ( \mathcal{D} + i \mathcal{C} )}{2 \Omega} \ .
\end{equation}
We are interested in the conditions under which the RHS of (\ref{Markov_NZ_off_App}) is approximately equal the RHS of (\ref{Markov_Alice_off_App}), i.e., when $F_{\mathrm{NZ}}(\tau) \simeq F_{\mathrm{M}}(\tau)$. To self-consistently answer this question we here insert the Markovian solution (\ref{Alice_sol_App}) into both $F_{\mathrm{NZ}}(\tau)$ and $F_{\mathrm{M}}(\tau)$ and explore constraints on the parameters that yield $F_{\mathrm{NZ}}(\tau) \simeq F_{\mathrm{M}}(\tau)$. Substituting the Markovian solution (\ref{Alice_sol_App}) into $F_{\mathrm{NZ}}(\tau)$ and $F_{\mathrm{M}}(\tau)$ we get
\begin{align}
    \label{FNZ}
    F_{\mathrm{NZ}}(\tau) &= \Scale[0.90]{ 2 g^2 e^{- g^2 \mathcal{C} \tau} } \int_0^\tau \!\!\!\exd s \; \Scale[0.90]{ \mathrm{Re}\left[ \mathcal{W}(s) \right] e^{+ i \Omega s} \left( \mathcal{B}^{\ast}- \mathcal{A} \right) e^{+ g^2 \mathcal{C} s } } + \Scale[0.90]{ 2 g^2 e^{( - g^2 \mathcal{C} + 2 i \Omega ) \tau} } \int_0^\tau \!\!\!\exd s \; \Scale[0.90]{ \mathrm{Re}\left[ \mathcal{W}(s) \right] e^{- i \Omega s} \left( - \mathcal{B} + \mathcal{A}^{\ast} \right) e^{+ g^2 \mathcal{C} s } } \,,\\
    \label{FM}
    F_{\mathrm{M}}(\tau) &= \Scale[0.88]{ 2 g^2 e^{- g^2 \mathcal{C} \tau} } \int_0^\infty\!\!\!\exd s \; \Scale[0.90]{  \mathrm{Re}\left[ \mathcal{W}(s) \right] \left(  e^{- i \Omega s}  \mathcal{B}^{\ast} - e^{+ i \Omega s} \mathcal{A} \right) }  + \Scale[0.88]{ 2 g^2 e^{( - g^2 \mathcal{C} + 2 i \Omega ) \tau} } \int_0^\infty\!\!\! \exd s \; \Scale[0.88]{  \mathrm{Re}\left[ \mathcal{W}(s) \right] \left( - e^{+ i \Omega s}  \mathcal{B} + \mathcal{A}^{\ast} e^{- i \Omega s} \right) } \,. 
\end{align}
Now, if we assume that $g^2 \mathcal{C} \ll a$ and $a \tau \gg 1$ then (\ref{FNZ}) becomes approximately 
\begin{eqnarray}
F_{\mathrm{NZ}}(\tau) & \simeq & 2 g^2 e^{- g^2 \mathcal{C} \tau} \int_0^\infty \exd s \; \Scale[0.90]{  \mathrm{Re}\left[ \mathcal{W}(s) \right] e^{+ i \Omega s} \left( - \mathcal{A} + \mathcal{B}^{\ast} \right) } + 2 g^2 e^{( - g^2 \mathcal{C} + 2 i \Omega ) \tau} \int_0^\tau \exd s \;  \Scale[0.90]{  \mathrm{Re}\left[ \mathcal{W}(s) \right] e^{- i \Omega s} \left( - \mathcal{B} + \mathcal{A}^{\ast} \right) } \ . \ \
\end{eqnarray}
However, for arbitrary $\Omega>0$ this is in general {\it not} approximately equal to $F_{\mathrm{M}}(\tau)$, since
\begin{equation}
F_{\mathrm{NZ}}(\tau) - F_{\mathrm{M}}(\tau) \simeq \Scale[0.88]{ 2 g^2 e^{- g^2 \mathcal{C} \tau} \mathcal{B}^{\ast} } \int_0^\infty \exd s \; \Scale[0.88]{ \mathrm{Re}\left[ \mathcal{W}(s) \right] ( e^{+ i \Omega s} - e^{- i \Omega s} ) } + \Scale[0.88]{ 2 g^2 e^{( - g^2 \mathcal{C} + 2 i \Omega ) \tau} \mathcal{B} } \int_0^\tau \exd s \; \Scale[0.88]{ \mathrm{Re}\left[ \mathcal{W}(s) \right] ( e^{- i \Omega s} - e^{+ i \Omega s} ) } \ .
\end{equation}
The only way we can have $F_{\mathrm{NZ}}(\tau) - F_{\mathrm{M}}(\tau) \simeq 0$ in the above is if we additionally assume that $\Omega / a \ll 1$ so that $e^{+ i \Omega s} \simeq e^{- i \Omega s} \simeq 1$ under the integral sign, as claimed in the main text surrounding Eq.~\eqref{Alice_whentrue}.

\end{widetext}

\section{Some useful integrals}
\label{subsec: Integrals}

In this appendix we compute the functions $C_{\times}$ and $K_{\times}$ defined in (\ref{Cx_def}) and (\ref{Kx_def}), as well as $D_{\times}'$ and $S_{\times}'$ defined in (\ref{Sxp_def}) and (\ref{Dxp_def}) (where we note that $C_{\s}$, $D_{\s}'$ and $S_{\s}'$ have all been computed in \cite{kaplanek2020hot,kaplanek2021qubits}). 

First we compute $C_{\times}$ and $S_{\times}$, and to this end we define the related Fourier cosine and sine transforms
\begin{align}
    \mathcal{C}_{\times}(\Omega) &\coloneqq 2\int_0^\infty\dd s \,\Re[\mathcal{W}_{\s,\times}(s)] \cos(\Omega s)\ , \\
    \mathcal{S}_{\times} (\Omega) &\coloneqq 2\int_0^\infty\dd s \,\Im[\mathcal{W}_{\s,\times}(s)] \sin(\Omega s)\  .
\end{align}
The values of these integrals depend on the proper acceleration $a$, detector separation $L$ as well as the energy gap $\Omega$. These integrals have the properties
\begin{equation} \label{CxSx_limits}
C_{\times} = \lim_{\Omega \to 0^{+}} \mathcal{C}_{\times}(\Omega) \ , \quad S_{\times}' = \lim_{\Omega \to 0^{+}} \frac{\dd \mathcal{S}_{\times}(\Omega)}{\dd \Omega}
\end{equation}
and are useful definitions because the thermality encoded in the correlator $\mathcal{W}_{\times}$ implies that $ \mathcal{C}_{\times}(\Omega)$ and $\mathcal{S}_{\times}(\Omega)$ are related for arbitrary $\Omega >0$. In particular, $\mathcal{W}_{\times}$ satisfies the KMS condition,
\begin{align}
    \mathcal{W}_{\times}(\tau - i \beta) = \mathcal{W}^{\ast}_{\times}(\tau)\,,
\end{align}
with $\beta = 2 \pi / a = T_{\mathrm{U}}^{-1}$ the inverse Unruh temperature, which in turn implies the detailed balance relationship
\begin{equation} \label{det_bal}
\mathcal{C}_{\times}(\Omega) = - \coth \left( \sfrac{\pi\Omega}{a} \right) \mathcal{S}_{\times}(\Omega) \ .
\end{equation}
Since the imaginary part of the Wightman function is precisely half the expectation value of the field commutator, we have (see \cite{tjoa2021entanglement})
\begin{align} 
    &\Im[\langle 0 |\phi(t,\bx)\phi(t',\bx)| 0 \rangle] \notag\\
    &= \frac{i}{4\pi^2|\Delta\bx|}\Bigr[\delta(\Delta t+|\Delta\bx|)-\delta(\Delta t-|\Delta\bx|)\Bigr]\,,
\end{align}
where $\Delta\bx=\bx-\bx'$ and $\Delta t = t-t'$. Substituting the trajectories \eqref{eq: trajectories} and setting $s=\tau-\tau'$ gives
\begin{align} \label{ImWx}
    \Im[\mathcal{W}_\times(s)] = -\frac{a\delta\rr{s-\frac{2}{a}\sinh^{-1}(aL/2)}}{8\pi L \sqrt{1+(aL/2)^2}} \ .
\end{align}
With this one can compute $\mathcal{S}_{\times}(\Omega)$ to give
\begin{equation}
    \mathcal{S}_\times(\Omega) = -\frac{\sin\left(\frac{2\Omega}{a} \sinh^{-1}\left(aL/2\right)\right)}{4\pi L\sqrt{1+(aL/2)^2}} \ ,
\end{equation}
which using (\ref{det_bal}) in turn implies 
\begin{equation}
    \mathcal{C}_{\times}(\Omega) = \frac{\coth\left( \frac{\pi \Omega}{a} \right) \sin\left(\frac{2\Omega}{a} \sinh^{-1}\left(aL/2\right)\right)}{4\pi L \sqrt{ 1 +  (aL/2)^2 } }   \,.
\end{equation}
Using the above formulae along with (\ref{CxSx_limits}) then straightforwardly gives the answers for $C_{\times}$ and $S_{\times}'$ given in (\ref{Cx_def}) and (\ref{Sxp_answer}). Using the expression (\ref{ImWx}) for $\Im[\mathcal{W}_\times(s)]$ in the integral definition (\ref{Kx_def}) for $K_{\times}$ also easily gives the result quoted in the main text. 

Finally we compute $D_{\times}'$ defined in (\ref{Dxp_def}). Using the form \eqref{eq: Wightman-functions2} of $\mathcal{W}_{\times}(s)$ and then switching the integration variable $z\coloneqq as/2$ turns (\ref{Dxp_def}) into
\begin{equation}
D_{\times}' = - \frac{1}{2\pi^2} \mathrm{Re} \bigg[  \lim_{\delta \to 0^{+}} \int_0^\infty\!\!\! \exd z\; \sfrac{z}{\sinh^2(z)- (\frac{aL}{2})^2 - i \delta}  \; \bigg] \ ,
\end{equation}
where we have defined the dimensionless infinitesimal $\delta = a\epsilon/2$ (and taken it out of the argument of the $\sinh$ function noting that $z>0$). Performing a partial fraction expansion on the integrand yields
\begin{equation}
D_{\times}'=  \mathrm{Re} \bigg[  \lim_{\delta \to 0^{+}} \sfrac{\mathcal{I}(\sqrt{(\frac{aL}{2})^2 + i \delta}\; ) -  \mathcal{I}(- \sqrt{(\frac{aL}{2})^2 + i \delta} \; )  }{4\pi^2 \sqrt{(\frac{aL}{2})^2 + i \delta}}  \; \bigg]\ ,
\end{equation}
with $\mathcal{I}(b)$ defined by
\begin{equation}
    \mathcal{I}(b) : = \int_0^\infty \exd z\; \frac{z}{\sinh(z) + b }
\end{equation}
for $b \in \mathbb{C} \setminus (- \infty, 0 ]$. This evaluates to
\begin{equation}
    \mathcal{I}(b) = \frac{\mathrm{Li}_{2}(-b+\sqrt{b^2 + 1}) - \mathrm{Li}_{2}(-b-\sqrt{b^2 + 1})}{\sqrt{b^2 + 1}} 
\end{equation}
where $\mathrm{Li}_{2}$ is the polylogarithm of order 2 \cite{olver2010nist},
\begin{equation}
    \mathrm{Li}_{2}(z) \ = \ - \int_0^z \exd t \; \frac{\log(1-t)}{t} 
\end{equation}
defined for $z \in \mathbb{C} \setminus [ 1, \infty) $. With the above formula we safely can take the limit $\delta \to 0^{+}$ giving the result (\ref{Dxp_answer}) quoted in the main text.

\bibliography{ref-NZ.bib}

\begin{thebibliography}{107}%
\makeatletter
\providecommand \@ifxundefined [1]{%
 \@ifx{#1\undefined}
}%
\providecommand \@ifnum [1]{%
 \ifnum #1\expandafter \@firstoftwo
 \else \expandafter \@secondoftwo
 \fi
}%
\providecommand \@ifx [1]{%
 \ifx #1\expandafter \@firstoftwo
 \else \expandafter \@secondoftwo
 \fi
}%
\providecommand \natexlab [1]{#1}%
\providecommand \enquote  [1]{``#1''}%
\providecommand \bibnamefont  [1]{#1}%
\providecommand \bibfnamefont [1]{#1}%
\providecommand \citenamefont [1]{#1}%
\providecommand \href@noop [0]{\@secondoftwo}%
\providecommand \href [0]{\begingroup \@sanitize@url \@href}%
\providecommand \@href[1]{\@@startlink{#1}\@@href}%
\providecommand \@@href[1]{\endgroup#1\@@endlink}%
\providecommand \@sanitize@url [0]{\catcode `\\12\catcode `\$12\catcode
  `\&12\catcode `\#12\catcode `\^12\catcode `\_12\catcode `\%12\relax}%
\providecommand \@@startlink[1]{}%
\providecommand \@@endlink[0]{}%
\providecommand \url  [0]{\begingroup\@sanitize@url \@url }%
\providecommand \@url [1]{\endgroup\@href {#1}{\urlprefix }}%
\providecommand \urlprefix  [0]{URL }%
\providecommand \Eprint [0]{\href }%
\providecommand \doibase [0]{https://doi.org/}%
\providecommand \selectlanguage [0]{\@gobble}%
\providecommand \bibinfo  [0]{\@secondoftwo}%
\providecommand \bibfield  [0]{\@secondoftwo}%
\providecommand \translation [1]{[#1]}%
\providecommand \BibitemOpen [0]{}%
\providecommand \bibitemStop [0]{}%
\providecommand \bibitemNoStop [0]{.\EOS\space}%
\providecommand \EOS [0]{\spacefactor3000\relax}%
\providecommand \BibitemShut  [1]{\csname bibitem#1\endcsname}%
\let\auto@bib@innerbib\@empty
\bibitem [{\citenamefont {Summers}\ and\ \citenamefont
  {Werner}(1985)}]{summers1985bell}%
  \BibitemOpen
  \bibfield  {author} {\bibinfo {author} {\bibfnamefont {S.~J.}\ \bibnamefont
  {Summers}}\ and\ \bibinfo {author} {\bibfnamefont {R.}~\bibnamefont
  {Werner}},\ }\bibfield  {title} {\bibinfo {title} {{The vacuum violates
  Bell's inequalities}},\ }\href
  {https://doi.org/https://doi.org/10.1016/0375-9601(85)90093-3} {\bibfield
  {journal} {\bibinfo  {journal} {Phys. Lett. A}\ }\textbf {\bibinfo {volume}
  {110}},\ \bibinfo {pages} {257 } (\bibinfo {year} {1985})}\BibitemShut
  {NoStop}%
\bibitem [{\citenamefont {Summers}\ and\ \citenamefont
  {Werner}(1987)}]{summers1987bell}%
  \BibitemOpen
  \bibfield  {author} {\bibinfo {author} {\bibfnamefont {S.~J.}\ \bibnamefont
  {Summers}}\ and\ \bibinfo {author} {\bibfnamefont {R.}~\bibnamefont
  {Werner}},\ }\bibfield  {title} {\bibinfo {title} {Bell’s inequalities and
  quantum field theory. i. general setting},\ }\href
  {https://doi.org/10.1063/1.527733} {\bibfield  {journal} {\bibinfo  {journal}
  {J. Math. Phys.}\ }\textbf {\bibinfo {volume} {28}},\ \bibinfo {pages} {2440}
  (\bibinfo {year} {1987})}\BibitemShut {NoStop}%
\bibitem [{\citenamefont {Higuchi}\ \emph {et~al.}(2017)\citenamefont
  {Higuchi}, \citenamefont {Iso}, \citenamefont {Ueda},\ and\ \citenamefont
  {Yamamoto}}]{Higuchi2017wedges}%
  \BibitemOpen
  \bibfield  {author} {\bibinfo {author} {\bibfnamefont {A.}~\bibnamefont
  {Higuchi}}, \bibinfo {author} {\bibfnamefont {S.}~\bibnamefont {Iso}},
  \bibinfo {author} {\bibfnamefont {K.}~\bibnamefont {Ueda}},\ and\ \bibinfo
  {author} {\bibfnamefont {K.}~\bibnamefont {Yamamoto}},\ }\bibfield  {title}
  {\bibinfo {title} {Entanglement of the vacuum between left, right, future,
  and past: The origin of entanglement-induced quantum radiation},\ }\href
  {https://doi.org/10.1103/PhysRevD.96.083531} {\bibfield  {journal} {\bibinfo
  {journal} {Phys. Rev. D}\ }\textbf {\bibinfo {volume} {96}},\ \bibinfo
  {pages} {083531} (\bibinfo {year} {2017})}\BibitemShut {NoStop}%
\bibitem [{\citenamefont {Unruh}(1976)}]{Unruh1979evaporation}%
  \BibitemOpen
  \bibfield  {author} {\bibinfo {author} {\bibfnamefont {W.~G.}\ \bibnamefont
  {Unruh}},\ }\bibfield  {title} {\bibinfo {title} {Notes on black-hole
  evaporation},\ }\href {https://doi.org/10.1103/PhysRevD.14.870} {\bibfield
  {journal} {\bibinfo  {journal} {Phys. Rev. D}\ }\textbf {\bibinfo {volume}
  {14}},\ \bibinfo {pages} {870} (\bibinfo {year} {1976})}\BibitemShut
  {NoStop}%
\bibitem [{\citenamefont {{DeWitt}}(1979)}]{DeWitt1979}%
  \BibitemOpen
  \bibfield  {author} {\bibinfo {author} {\bibfnamefont {B.~S.}\ \bibnamefont
  {{DeWitt}}},\ }\bibfield  {title} {\bibinfo {title} {{Quantum gravity: the
  new synthesis}},\ }in\ \href@noop {} {\emph {\bibinfo {booktitle} {General
  Relativity: An Einstein centenary survey}}},\ \bibinfo {editor} {edited by\
  \bibinfo {editor} {\bibfnamefont {S.~W.}\ \bibnamefont {{Hawking}}}\ and\
  \bibinfo {editor} {\bibfnamefont {W.}~\bibnamefont {{Israel}}}}\ (\bibinfo
  {year} {1979})\ pp.\ \bibinfo {pages} {680--745}\BibitemShut {NoStop}%
\bibitem [{\citenamefont {Reznik}(2003)}]{reznik2003entanglement}%
  \BibitemOpen
  \bibfield  {author} {\bibinfo {author} {\bibfnamefont {B.}~\bibnamefont
  {Reznik}},\ }\bibfield  {title} {\bibinfo {title} {Entanglement from the
  vacuum},\ }\href {https://doi.org/https://doi.org/10.1023/A:1022875910744}
  {\bibfield  {journal} {\bibinfo  {journal} {Found. Phys.}\ }\textbf {\bibinfo
  {volume} {33}},\ \bibinfo {pages} {167} (\bibinfo {year} {2003})}\BibitemShut
  {NoStop}%
\bibitem [{\citenamefont {Valentini}(1991)}]{Valentini1991nonlocalcorr}%
  \BibitemOpen
  \bibfield  {author} {\bibinfo {author} {\bibfnamefont {A.}~\bibnamefont
  {Valentini}},\ }\bibfield  {title} {\bibinfo {title} {Non-local correlations
  in quantum electrodynamics},\ }\href
  {https://doi.org/https://doi.org/10.1016/0375-9601(91)90952-5} {\bibfield
  {journal} {\bibinfo  {journal} {Phys. Lett. A}\ }\textbf {\bibinfo {volume}
  {153}},\ \bibinfo {pages} {321 } (\bibinfo {year} {1991})}\BibitemShut
  {NoStop}%
\bibitem [{\citenamefont {Steeg}\ and\ \citenamefont
  {Menicucci}(2009)}]{VerSteeg2009entangling}%
  \BibitemOpen
  \bibfield  {author} {\bibinfo {author} {\bibfnamefont {G.~V.}\ \bibnamefont
  {Steeg}}\ and\ \bibinfo {author} {\bibfnamefont {N.~C.}\ \bibnamefont
  {Menicucci}},\ }\bibfield  {title} {\bibinfo {title} {Entangling power of an
  expanding universe},\ }\href {https://doi.org/10.1103/PhysRevD.79.044027}
  {\bibfield  {journal} {\bibinfo  {journal} {Phys. Rev. D}\ }\textbf {\bibinfo
  {volume} {79}},\ \bibinfo {pages} {044027} (\bibinfo {year}
  {2009})}\BibitemShut {NoStop}%
\bibitem [{\citenamefont {Pozas-Kerstjens}\ and\ \citenamefont
  {Mart\'{\i}n-Mart\'{\i}nez}(2015)}]{pozas2015harvesting}%
  \BibitemOpen
  \bibfield  {author} {\bibinfo {author} {\bibfnamefont {A.}~\bibnamefont
  {Pozas-Kerstjens}}\ and\ \bibinfo {author} {\bibfnamefont {E.}~\bibnamefont
  {Mart\'{\i}n-Mart\'{\i}nez}},\ }\bibfield  {title} {\bibinfo {title}
  {Harvesting correlations from the quantum vacuum},\ }\href
  {https://doi.org/10.1103/PhysRevD.92.064042} {\bibfield  {journal} {\bibinfo
  {journal} {Phys. Rev. D}\ }\textbf {\bibinfo {volume} {92}},\ \bibinfo
  {pages} {064042} (\bibinfo {year} {2015})}\BibitemShut {NoStop}%
\bibitem [{\citenamefont {Pozas-Kerstjens}\ and\ \citenamefont
  {Mart\'{\i}n-Mart\'{\i}nez}(2016)}]{pozas2016entanglement}%
  \BibitemOpen
  \bibfield  {author} {\bibinfo {author} {\bibfnamefont {A.}~\bibnamefont
  {Pozas-Kerstjens}}\ and\ \bibinfo {author} {\bibfnamefont {E.}~\bibnamefont
  {Mart\'{\i}n-Mart\'{\i}nez}},\ }\bibfield  {title} {\bibinfo {title}
  {Entanglement harvesting from the electromagnetic vacuum with hydrogenlike
  atoms},\ }\href {https://doi.org/10.1103/PhysRevD.94.064074} {\bibfield
  {journal} {\bibinfo  {journal} {Phys. Rev. D}\ }\textbf {\bibinfo {volume}
  {94}},\ \bibinfo {pages} {064074} (\bibinfo {year} {2016})}\BibitemShut
  {NoStop}%
\bibitem [{\citenamefont {Grimmer}\ \emph {et~al.}(2021)\citenamefont
  {Grimmer}, \citenamefont {Torres},\ and\ \citenamefont
  {Mart\'{\i}n-Mart\'{\i}nez}}]{Grimmer2021algebraic}%
  \BibitemOpen
  \bibfield  {author} {\bibinfo {author} {\bibfnamefont {D.}~\bibnamefont
  {Grimmer}}, \bibinfo {author} {\bibfnamefont {B.~d. S.~L.}\ \bibnamefont
  {Torres}},\ and\ \bibinfo {author} {\bibfnamefont {E.}~\bibnamefont
  {Mart\'{\i}n-Mart\'{\i}nez}},\ }\bibfield  {title} {\bibinfo {title}
  {Measurements in qft: Weakly coupled local particle detectors and
  entanglement harvesting},\ }\href
  {https://doi.org/10.1103/PhysRevD.104.085014} {\bibfield  {journal} {\bibinfo
   {journal} {Phys. Rev. D}\ }\textbf {\bibinfo {volume} {104}},\ \bibinfo
  {pages} {085014} (\bibinfo {year} {2021})}\BibitemShut {NoStop}%
\bibitem [{\citenamefont {Tjoa}\ and\ \citenamefont
  {Mart\'{\i}n-Mart\'{\i}nez}(2021)}]{tjoa2021entanglement}%
  \BibitemOpen
  \bibfield  {author} {\bibinfo {author} {\bibfnamefont {E.}~\bibnamefont
  {Tjoa}}\ and\ \bibinfo {author} {\bibfnamefont {E.}~\bibnamefont
  {Mart\'{\i}n-Mart\'{\i}nez}},\ }\bibfield  {title} {\bibinfo {title} {When
  entanglement harvesting is not really harvesting},\ }\href
  {https://doi.org/10.1103/PhysRevD.104.125005} {\bibfield  {journal} {\bibinfo
   {journal} {Phys. Rev. D}\ }\textbf {\bibinfo {volume} {104}},\ \bibinfo
  {pages} {125005} (\bibinfo {year} {2021})}\BibitemShut {NoStop}%
\bibitem [{\citenamefont {Henderson}\ \emph {et~al.}(2018)\citenamefont
  {Henderson}, \citenamefont {Hennigar}, \citenamefont {Mann}, \citenamefont
  {Smith},\ and\ \citenamefont {Zhang}}]{henderson2018harvestingBH}%
  \BibitemOpen
  \bibfield  {author} {\bibinfo {author} {\bibfnamefont {L.~J.}\ \bibnamefont
  {Henderson}}, \bibinfo {author} {\bibfnamefont {R.~A.}\ \bibnamefont
  {Hennigar}}, \bibinfo {author} {\bibfnamefont {R.~B.}\ \bibnamefont {Mann}},
  \bibinfo {author} {\bibfnamefont {A.~R.~H.}\ \bibnamefont {Smith}},\ and\
  \bibinfo {author} {\bibfnamefont {J.}~\bibnamefont {Zhang}},\ }\bibfield
  {title} {\bibinfo {title} {Harvesting entanglement from the black hole
  vacuum},\ }\bibfield  {journal} {\bibinfo  {journal} {Class. Quantum
  Gravity}\ }\textbf {\bibinfo {volume} {35}},\ \href
  {https://doi.org/10.1088/1361-6382/aae27e} {10.1088/1361-6382/aae27e}
  (\bibinfo {year} {2018})\BibitemShut {NoStop}%
\bibitem [{\citenamefont {Mart\'{\i}n-Mart\'{\i}nez}\ \emph
  {et~al.}(2016)\citenamefont {Mart\'{\i}n-Mart\'{\i}nez}, \citenamefont
  {Smith},\ and\ \citenamefont {Terno}}]{smith2016topology}%
  \BibitemOpen
  \bibfield  {author} {\bibinfo {author} {\bibfnamefont {E.}~\bibnamefont
  {Mart\'{\i}n-Mart\'{\i}nez}}, \bibinfo {author} {\bibfnamefont {A.~R.~H.}\
  \bibnamefont {Smith}},\ and\ \bibinfo {author} {\bibfnamefont {D.~R.}\
  \bibnamefont {Terno}},\ }\bibfield  {title} {\bibinfo {title} {Spacetime
  structure and vacuum entanglement},\ }\href
  {https://doi.org/10.1103/PhysRevD.93.044001} {\bibfield  {journal} {\bibinfo
  {journal} {Phys. Rev. D}\ }\textbf {\bibinfo {volume} {93}},\ \bibinfo
  {pages} {044001} (\bibinfo {year} {2016})}\BibitemShut {NoStop}%
\bibitem [{\citenamefont {Perche}\ \emph {et~al.}(2022)\citenamefont {Perche},
  \citenamefont {Lima},\ and\ \citenamefont
  {Mart\'{\i}n-Mart\'{\i}nez}}]{Caroline2022harvesting}%
  \BibitemOpen
  \bibfield  {author} {\bibinfo {author} {\bibfnamefont {T.~R.}\ \bibnamefont
  {Perche}}, \bibinfo {author} {\bibfnamefont {C.}~\bibnamefont {Lima}},\ and\
  \bibinfo {author} {\bibfnamefont {E.}~\bibnamefont
  {Mart\'{\i}n-Mart\'{\i}nez}},\ }\bibfield  {title} {\bibinfo {title}
  {Harvesting entanglement from complex scalar and fermionic fields with
  linearly coupled particle detectors},\ }\href
  {https://doi.org/10.1103/PhysRevD.105.065016} {\bibfield  {journal} {\bibinfo
   {journal} {Phys. Rev. D}\ }\textbf {\bibinfo {volume} {105}},\ \bibinfo
  {pages} {065016} (\bibinfo {year} {2022})}\BibitemShut {NoStop}%
\bibitem [{\citenamefont {Tjoa}\ and\ \citenamefont
  {Mann}(2020)}]{Tjoa2020vaidya}%
  \BibitemOpen
  \bibfield  {author} {\bibinfo {author} {\bibfnamefont {E.}~\bibnamefont
  {Tjoa}}\ and\ \bibinfo {author} {\bibfnamefont {R.~B.}\ \bibnamefont
  {Mann}},\ }\bibfield  {title} {\bibinfo {title} {{Harvesting correlations in
  Schwarzschild and collapsing shell spacetimes}},\ }\href
  {https://doi.org/https://doi.org/10.1007/JHEP08(2020)155} {\bibfield
  {journal} {\bibinfo  {journal} {J. High Energy Phys.}\ }\textbf {\bibinfo
  {volume} {2020}}\bibinfo  {number} { (8)},\ \bibinfo {pages} {1}}\BibitemShut
  {NoStop}%
\bibitem [{\citenamefont {Gallock-Yoshimura}\ \emph {et~al.}(2021)\citenamefont
  {Gallock-Yoshimura}, \citenamefont {Tjoa},\ and\ \citenamefont
  {Mann}}]{Gallock-Yoshimura2021freefall}%
  \BibitemOpen
\bibfield  {number} {  }\bibfield  {author} {\bibinfo {author} {\bibfnamefont
  {K.}~\bibnamefont {Gallock-Yoshimura}}, \bibinfo {author} {\bibfnamefont
  {E.}~\bibnamefont {Tjoa}},\ and\ \bibinfo {author} {\bibfnamefont {R.~B.}\
  \bibnamefont {Mann}},\ }\bibfield  {title} {\bibinfo {title} {Harvesting
  entanglement with detectors freely falling into a black hole},\ }\href
  {https://doi.org/10.1103/PhysRevD.104.025001} {\bibfield  {journal} {\bibinfo
   {journal} {Phys. Rev. D}\ }\textbf {\bibinfo {volume} {104}},\ \bibinfo
  {pages} {025001} (\bibinfo {year} {2021})}\BibitemShut {NoStop}%
\bibitem [{\citenamefont {Maeso-Garc{\'\i}a}\ \emph {et~al.}(2022)\citenamefont
  {Maeso-Garc{\'\i}a}, \citenamefont {Perche},\ and\ \citenamefont
  {Mart{\'\i}n-Mart{\'\i}nez}}]{maeso2022entanglement}%
  \BibitemOpen
  \bibfield  {author} {\bibinfo {author} {\bibfnamefont {H.}~\bibnamefont
  {Maeso-Garc{\'\i}a}}, \bibinfo {author} {\bibfnamefont {T.~R.}\ \bibnamefont
  {Perche}},\ and\ \bibinfo {author} {\bibfnamefont {E.}~\bibnamefont
  {Mart{\'\i}n-Mart{\'\i}nez}},\ }\bibfield  {title} {\bibinfo {title}
  {Entanglement harvesting: detector gap and field mass optimization},\
  }\href@noop {} {\bibfield  {journal} {\bibinfo  {journal} {arXiv preprint
  arXiv:2206.06381}\ } (\bibinfo {year} {2022})}\BibitemShut {NoStop}%
\bibitem [{\citenamefont {Salton}\ \emph {et~al.}(2015)\citenamefont {Salton},
  \citenamefont {Mann},\ and\ \citenamefont
  {Menicucci}}]{salton2015acceleration}%
  \BibitemOpen
  \bibfield  {author} {\bibinfo {author} {\bibfnamefont {G.}~\bibnamefont
  {Salton}}, \bibinfo {author} {\bibfnamefont {R.~B.}\ \bibnamefont {Mann}},\
  and\ \bibinfo {author} {\bibfnamefont {N.~C.}\ \bibnamefont {Menicucci}},\
  }\bibfield  {title} {\bibinfo {title} {Acceleration-assisted entanglement
  harvesting and rangefinding},\ }\href
  {https://doi.org/10.1088/1367-2630/17/3/035001} {\bibfield  {journal}
  {\bibinfo  {journal} {New J. Phys.}\ }\textbf {\bibinfo {volume} {17}},\
  \bibinfo {pages} {035001} (\bibinfo {year} {2015})}\BibitemShut {NoStop}%
\bibitem [{\citenamefont {Cong}\ \emph {et~al.}(2019)\citenamefont {Cong},
  \citenamefont {Tjoa},\ and\ \citenamefont {Mann}}]{cong2019entanglement}%
  \BibitemOpen
  \bibfield  {author} {\bibinfo {author} {\bibfnamefont {W.}~\bibnamefont
  {Cong}}, \bibinfo {author} {\bibfnamefont {E.}~\bibnamefont {Tjoa}},\ and\
  \bibinfo {author} {\bibfnamefont {R.~B.}\ \bibnamefont {Mann}},\ }\bibfield
  {title} {\bibinfo {title} {Entanglement harvesting with moving mirrors},\
  }\href {https://doi.org/https://doi.org/10.1007/JHEP06(2019)021} {\bibfield
  {journal} {\bibinfo  {journal} {J. High Energ. Phys}\ }\textbf {\bibinfo
  {volume} {2019}},\ \bibinfo {pages} {21} (\bibinfo {year}
  {2019})}\BibitemShut {NoStop}%
\bibitem [{\citenamefont {Ng}\ \emph {et~al.}(2018)\citenamefont {Ng},
  \citenamefont {Mann},\ and\ \citenamefont
  {Mart\'{\i}n-Mart\'{\i}nez}}]{Ng2018newtechniques}%
  \BibitemOpen
  \bibfield  {author} {\bibinfo {author} {\bibfnamefont {K.~K.}\ \bibnamefont
  {Ng}}, \bibinfo {author} {\bibfnamefont {R.~B.}\ \bibnamefont {Mann}},\ and\
  \bibinfo {author} {\bibfnamefont {E.}~\bibnamefont
  {Mart\'{\i}n-Mart\'{\i}nez}},\ }\bibfield  {title} {\bibinfo {title} {New
  techniques for entanglement harvesting in flat and curved spacetimes},\
  }\href {https://doi.org/10.1103/PhysRevD.97.125011} {\bibfield  {journal}
  {\bibinfo  {journal} {Phys. Rev. D}\ }\textbf {\bibinfo {volume} {97}},\
  \bibinfo {pages} {125011} (\bibinfo {year} {2018})}\BibitemShut {NoStop}%
\bibitem [{\citenamefont {Cong}\ \emph {et~al.}(2020)\citenamefont {Cong},
  \citenamefont {Qian}, \citenamefont {Good},\ and\ \citenamefont
  {Mann}}]{cong2020horizon}%
  \BibitemOpen
  \bibfield  {author} {\bibinfo {author} {\bibfnamefont {W.}~\bibnamefont
  {Cong}}, \bibinfo {author} {\bibfnamefont {C.}~\bibnamefont {Qian}}, \bibinfo
  {author} {\bibfnamefont {M.~R.}\ \bibnamefont {Good}},\ and\ \bibinfo
  {author} {\bibfnamefont {R.~B.}\ \bibnamefont {Mann}},\ }\bibfield  {title}
  {\bibinfo {title} {Effects of horizons on entanglement harvesting},\ }\href
  {https://doi.org/10.1007/JHEP10(2020)067} {\bibfield  {journal} {\bibinfo
  {journal} {J. High Energy Phys.}\ }\textbf {\bibinfo {volume} {2020}}\bibinfo
   {number} { (10)},\ \bibinfo {pages} {67}}\BibitemShut {NoStop}%
\bibitem [{\citenamefont {Henderson}\ \emph {et~al.}(2019)\citenamefont
  {Henderson}, \citenamefont {Hennigar}, \citenamefont {Mann}, \citenamefont
  {Smith},\ and\ \citenamefont {Zhang}}]{henderson2019AdSharvesting}%
  \BibitemOpen
\bibfield  {number} {  }\bibfield  {author} {\bibinfo {author} {\bibfnamefont
  {L.~J.}\ \bibnamefont {Henderson}}, \bibinfo {author} {\bibfnamefont {R.~A.}\
  \bibnamefont {Hennigar}}, \bibinfo {author} {\bibfnamefont {R.~B.}\
  \bibnamefont {Mann}}, \bibinfo {author} {\bibfnamefont {A.~R.}\ \bibnamefont
  {Smith}},\ and\ \bibinfo {author} {\bibfnamefont {J.}~\bibnamefont {Zhang}},\
  }\bibfield  {title} {\bibinfo {title} {{Entangling detectors in anti-de
  Sitter space}},\ }\href
  {https://doi.org/https://doi.org/10.1007/JHEP05(2019)178} {\bibfield
  {journal} {\bibinfo  {journal} {J. High Energ. Phys.}\ }\textbf {\bibinfo
  {volume} {2019}}\bibinfo  {number} { (5)},\ \bibinfo {pages}
  {178}}\BibitemShut {NoStop}%
\bibitem [{\citenamefont {Stritzelberger}\ \emph {et~al.}(2021)\citenamefont
  {Stritzelberger}, \citenamefont {Henderson}, \citenamefont {Baccetti},
  \citenamefont {Menicucci},\ and\ \citenamefont
  {Kempf}}]{Nadine2021delocharvesting}%
  \BibitemOpen
\bibfield  {number} {  }\bibfield  {author} {\bibinfo {author} {\bibfnamefont
  {N.}~\bibnamefont {Stritzelberger}}, \bibinfo {author} {\bibfnamefont
  {L.~J.}\ \bibnamefont {Henderson}}, \bibinfo {author} {\bibfnamefont
  {V.}~\bibnamefont {Baccetti}}, \bibinfo {author} {\bibfnamefont {N.~C.}\
  \bibnamefont {Menicucci}},\ and\ \bibinfo {author} {\bibfnamefont
  {A.}~\bibnamefont {Kempf}},\ }\bibfield  {title} {\bibinfo {title}
  {Entanglement harvesting with coherently delocalized matter},\ }\href
  {https://doi.org/10.1103/PhysRevD.103.016007} {\bibfield  {journal} {\bibinfo
   {journal} {Phys. Rev. D}\ }\textbf {\bibinfo {volume} {103}},\ \bibinfo
  {pages} {016007} (\bibinfo {year} {2021})}\BibitemShut {NoStop}%
\bibitem [{\citenamefont {Simidzija}\ and\ \citenamefont
  {Mart\'{\i}n-Mart\'{\i}nez}(2018)}]{simidzija2018harvesting}%
  \BibitemOpen
  \bibfield  {author} {\bibinfo {author} {\bibfnamefont {P.}~\bibnamefont
  {Simidzija}}\ and\ \bibinfo {author} {\bibfnamefont {E.}~\bibnamefont
  {Mart\'{\i}n-Mart\'{\i}nez}},\ }\bibfield  {title} {\bibinfo {title}
  {Harvesting correlations from thermal and squeezed coherent states},\ }\href
  {https://doi.org/10.1103/PhysRevD.98.085007} {\bibfield  {journal} {\bibinfo
  {journal} {Phys. Rev. D}\ }\textbf {\bibinfo {volume} {98}},\ \bibinfo
  {pages} {085007} (\bibinfo {year} {2018})}\BibitemShut {NoStop}%
\bibitem [{\citenamefont {Xu}\ \emph {et~al.}(2020)\citenamefont {Xu},
  \citenamefont {Ali~Ahmad},\ and\ \citenamefont
  {Smith}}]{Smith2020harvestingGW}%
  \BibitemOpen
  \bibfield  {author} {\bibinfo {author} {\bibfnamefont {Q.}~\bibnamefont
  {Xu}}, \bibinfo {author} {\bibfnamefont {S.}~\bibnamefont {Ali~Ahmad}},\ and\
  \bibinfo {author} {\bibfnamefont {A.~R.~H.}\ \bibnamefont {Smith}},\
  }\bibfield  {title} {\bibinfo {title} {Gravitational waves affect vacuum
  entanglement},\ }\href {https://doi.org/10.1103/PhysRevD.102.065019}
  {\bibfield  {journal} {\bibinfo  {journal} {Phys. Rev. D}\ }\textbf {\bibinfo
  {volume} {102}},\ \bibinfo {pages} {065019} (\bibinfo {year}
  {2020})}\BibitemShut {NoStop}%
\bibitem [{\citenamefont {Benatti}\ and\ \citenamefont
  {Floreanini}(2004)}]{Benatti2004Unruh}%
  \BibitemOpen
  \bibfield  {author} {\bibinfo {author} {\bibfnamefont {F.}~\bibnamefont
  {Benatti}}\ and\ \bibinfo {author} {\bibfnamefont {R.}~\bibnamefont
  {Floreanini}},\ }\bibfield  {title} {\bibinfo {title} {Entanglement
  generation in uniformly accelerating atoms: Reexamination of the unruh
  effect},\ }\href {https://doi.org/10.1103/PhysRevA.70.012112} {\bibfield
  {journal} {\bibinfo  {journal} {Phys. Rev. A}\ }\textbf {\bibinfo {volume}
  {70}},\ \bibinfo {pages} {012112} (\bibinfo {year} {2004})}\BibitemShut
  {NoStop}%
\bibitem [{\citenamefont {Benatti}\ and\ \citenamefont
  {Floreanini}(2005)}]{BENATTI2005review}%
  \BibitemOpen
  \bibfield  {author} {\bibinfo {author} {\bibfnamefont {F.}~\bibnamefont
  {Benatti}}\ and\ \bibinfo {author} {\bibfnamefont {R.}~\bibnamefont
  {Floreanini}},\ }\bibfield  {title} {\bibinfo {title} {Open quantum dynamics:
  Complete positivity and entanglement},\ }\href
  {https://doi.org/10.1142/s0217979205032097} {\bibfield  {journal} {\bibinfo
  {journal} {International Journal of Modern Physics B}\ }\textbf {\bibinfo
  {volume} {19}},\ \bibinfo {pages} {3063–3139} (\bibinfo {year}
  {2005})}\BibitemShut {NoStop}%
\bibitem [{\citenamefont {Hu}\ and\ \citenamefont
  {Yu}(2015)}]{Hu2015twodetectorent}%
  \BibitemOpen
  \bibfield  {author} {\bibinfo {author} {\bibfnamefont {J.}~\bibnamefont
  {Hu}}\ and\ \bibinfo {author} {\bibfnamefont {H.}~\bibnamefont {Yu}},\
  }\bibfield  {title} {\bibinfo {title} {Entanglement dynamics for uniformly
  accelerated two-level atoms},\ }\href
  {https://doi.org/10.1103/PhysRevA.91.012327} {\bibfield  {journal} {\bibinfo
  {journal} {Phys. Rev. A}\ }\textbf {\bibinfo {volume} {91}},\ \bibinfo
  {pages} {012327} (\bibinfo {year} {2015})}\BibitemShut {NoStop}%
\bibitem [{\citenamefont {Zhou}\ \emph
  {et~al.}(2021{\natexlab{a}})\citenamefont {Zhou}, \citenamefont {Hu},\ and\
  \citenamefont {Yu}}]{Zhou2021massiveentanglement2}%
  \BibitemOpen
  \bibfield  {author} {\bibinfo {author} {\bibfnamefont {Y.}~\bibnamefont
  {Zhou}}, \bibinfo {author} {\bibfnamefont {J.}~\bibnamefont {Hu}},\ and\
  \bibinfo {author} {\bibfnamefont {H.}~\bibnamefont {Yu}},\ }\bibfield
  {title} {\bibinfo {title} {Entanglement dynamics for unruh-dewitt detectors
  interacting with massive scalar fields: the unruh and anti-unruh effects},\
  }\href@noop {} {\bibfield  {journal} {\bibinfo  {journal} {Journal of High
  Energy Physics}\ }\textbf {\bibinfo {volume} {2021}},\ \bibinfo {pages} {1}
  (\bibinfo {year} {2021}{\natexlab{a}})}\BibitemShut {NoStop}%
\bibitem [{\citenamefont {Menezes}(2018)}]{Menezes2018entanglementKerr}%
  \BibitemOpen
  \bibfield  {author} {\bibinfo {author} {\bibfnamefont {G.}~\bibnamefont
  {Menezes}},\ }\bibfield  {title} {\bibinfo {title} {Entanglement dynamics in
  a kerr spacetime},\ }\href {https://doi.org/10.1103/PhysRevD.97.085021}
  {\bibfield  {journal} {\bibinfo  {journal} {Phys. Rev. D}\ }\textbf {\bibinfo
  {volume} {97}},\ \bibinfo {pages} {085021} (\bibinfo {year}
  {2018})}\BibitemShut {NoStop}%
\bibitem [{\citenamefont {Huang}\ and\ \citenamefont
  {Tian}(2017)}]{huang2017dynamics}%
  \BibitemOpen
  \bibfield  {author} {\bibinfo {author} {\bibfnamefont {Z.}~\bibnamefont
  {Huang}}\ and\ \bibinfo {author} {\bibfnamefont {Z.}~\bibnamefont {Tian}},\
  }\bibfield  {title} {\bibinfo {title} {Dynamics of quantum entanglement in de
  sitter spacetime and thermal minkowski spacetime},\ }\href@noop {} {\bibfield
   {journal} {\bibinfo  {journal} {Nuclear Physics B}\ }\textbf {\bibinfo
  {volume} {923}},\ \bibinfo {pages} {458} (\bibinfo {year}
  {2017})}\BibitemShut {NoStop}%
\bibitem [{\citenamefont {Soares}\ \emph {et~al.}(2022)\citenamefont {Soares},
  \citenamefont {Svaiter},\ and\ \citenamefont {Menezes}}]{Soares2022open}%
  \BibitemOpen
  \bibfield  {author} {\bibinfo {author} {\bibfnamefont {M.~S.}\ \bibnamefont
  {Soares}}, \bibinfo {author} {\bibfnamefont {N.~F.}\ \bibnamefont
  {Svaiter}},\ and\ \bibinfo {author} {\bibfnamefont {G.}~\bibnamefont
  {Menezes}},\ }\bibfield  {title} {\bibinfo {title} {Entanglement dynamics:
  Generalized master equation for uniformly accelerated two-level systems},\
  }\href {https://doi.org/10.1103/PhysRevA.106.062440} {\bibfield  {journal}
  {\bibinfo  {journal} {Phys. Rev. A}\ }\textbf {\bibinfo {volume} {106}},\
  \bibinfo {pages} {062440} (\bibinfo {year} {2022})}\BibitemShut {NoStop}%
\bibitem [{\citenamefont {Benatti}\ \emph {et~al.}(2022)\citenamefont
  {Benatti}, \citenamefont {Chruści\'{n}ski},\ and\ \citenamefont
  {Floreanini}}]{benatti2022local}%
  \BibitemOpen
  \bibfield  {author} {\bibinfo {author} {\bibfnamefont {F.}~\bibnamefont
  {Benatti}}, \bibinfo {author} {\bibfnamefont {D.}~\bibnamefont
  {Chruści\'{n}ski}},\ and\ \bibinfo {author} {\bibfnamefont {R.}~\bibnamefont
  {Floreanini}},\ }\bibfield  {title} {\bibinfo {title} {Local generation of
  entanglement with redfield dynamics},\ }\href
  {https://doi.org/10.1142/S1230161222500019} {\bibfield  {journal} {\bibinfo
  {journal} {Open Systems \& Information Dynamics}\ }\textbf {\bibinfo {volume}
  {29}},\ \bibinfo {pages} {2250001} (\bibinfo {year} {2022})}\BibitemShut
  {NoStop}%
\bibitem [{\citenamefont {Zhang}\ and\ \citenamefont
  {Yu}(2007{\natexlab{a}})}]{Zhang2007reflect}%
  \BibitemOpen
  \bibfield  {author} {\bibinfo {author} {\bibfnamefont {J.}~\bibnamefont
  {Zhang}}\ and\ \bibinfo {author} {\bibfnamefont {H.}~\bibnamefont {Yu}},\
  }\bibfield  {title} {\bibinfo {title} {Unruh effect and entanglement
  generation for accelerated atoms near a reflecting boundary},\ }\href
  {https://doi.org/10.1103/PhysRevD.75.104014} {\bibfield  {journal} {\bibinfo
  {journal} {Phys. Rev. D}\ }\textbf {\bibinfo {volume} {75}},\ \bibinfo
  {pages} {104014} (\bibinfo {year} {2007}{\natexlab{a}})}\BibitemShut
  {NoStop}%
\bibitem [{\citenamefont {Zhang}\ and\ \citenamefont
  {Yu}(2007{\natexlab{b}})}]{Zhang2007reflect2}%
  \BibitemOpen
  \bibfield  {author} {\bibinfo {author} {\bibfnamefont {J.}~\bibnamefont
  {Zhang}}\ and\ \bibinfo {author} {\bibfnamefont {H.}~\bibnamefont {Yu}},\
  }\bibfield  {title} {\bibinfo {title} {Entanglement generation in atoms
  immersed in a thermal bath of external quantum scalar fields with a
  boundary},\ }\href {https://doi.org/10.1103/PhysRevA.75.012101} {\bibfield
  {journal} {\bibinfo  {journal} {Phys. Rev. A}\ }\textbf {\bibinfo {volume}
  {75}},\ \bibinfo {pages} {012101} (\bibinfo {year}
  {2007}{\natexlab{b}})}\BibitemShut {NoStop}%
\bibitem [{\citenamefont {Hu}\ and\ \citenamefont
  {Yu}(2011)}]{Hu2011blackhole}%
  \BibitemOpen
  \bibfield  {author} {\bibinfo {author} {\bibfnamefont {J.}~\bibnamefont
  {Hu}}\ and\ \bibinfo {author} {\bibfnamefont {H.}~\bibnamefont {Yu}},\
  }\bibfield  {title} {\bibinfo {title} {Entanglement generation outside a
  schwarzschild black hole and the hawking effect},\ }\bibfield  {journal}
  {\bibinfo  {journal} {Journal of High Energy Physics}\ }\textbf {\bibinfo
  {volume} {2011}},\ \href {https://doi.org/10.1007/jhep08(2011)137}
  {10.1007/jhep08(2011)137} (\bibinfo {year} {2011})\BibitemShut {NoStop}%
\bibitem [{\citenamefont {Hu}\ and\ \citenamefont {Yu}(2013)}]{Hu2013desitter}%
  \BibitemOpen
  \bibfield  {author} {\bibinfo {author} {\bibfnamefont {J.}~\bibnamefont
  {Hu}}\ and\ \bibinfo {author} {\bibfnamefont {H.}~\bibnamefont {Yu}},\
  }\bibfield  {title} {\bibinfo {title} {Quantum entanglement generation in de
  sitter spacetime},\ }\href {https://doi.org/10.1103/PhysRevD.88.104003}
  {\bibfield  {journal} {\bibinfo  {journal} {Phys. Rev. D}\ }\textbf {\bibinfo
  {volume} {88}},\ \bibinfo {pages} {104003} (\bibinfo {year}
  {2013})}\BibitemShut {NoStop}%
\bibitem [{\citenamefont {Huang}(2019)}]{huang2019boundary}%
  \BibitemOpen
  \bibfield  {author} {\bibinfo {author} {\bibfnamefont {Z.}~\bibnamefont
  {Huang}},\ }\bibfield  {title} {\bibinfo {title} {Behaviors of quantum
  correlation for accelerated atoms coupled with a fluctuating massless scalar
  field with a perfectly reflecting boundary},\ }\href@noop {} {\bibfield
  {journal} {\bibinfo  {journal} {Quantum Information Processing}\ }\textbf
  {\bibinfo {volume} {18}},\ \bibinfo {pages} {1} (\bibinfo {year}
  {2019})}\BibitemShut {NoStop}%
\bibitem [{\citenamefont {Zhou}\ \emph
  {et~al.}(2021{\natexlab{b}})\citenamefont {Zhou}, \citenamefont {Hu},\ and\
  \citenamefont {Yu}}]{Zhou2021massiveentanglement}%
  \BibitemOpen
  \bibfield  {author} {\bibinfo {author} {\bibfnamefont {Y.}~\bibnamefont
  {Zhou}}, \bibinfo {author} {\bibfnamefont {J.}~\bibnamefont {Hu}},\ and\
  \bibinfo {author} {\bibfnamefont {H.}~\bibnamefont {Yu}},\ }\bibfield
  {title} {\bibinfo {title} {Entanglement dynamics for two-level quantum
  systems coupled with massive scalar fields},\ }\href
  {https://doi.org/https://doi.org/10.1016/j.physleta.2021.127460} {\bibfield
  {journal} {\bibinfo  {journal} {Physics Letters A}\ }\textbf {\bibinfo
  {volume} {406}},\ \bibinfo {pages} {127460} (\bibinfo {year}
  {2021}{\natexlab{b}})}\BibitemShut {NoStop}%
\bibitem [{\citenamefont {Chen}\ \emph {et~al.}(2022)\citenamefont {Chen},
  \citenamefont {Hu},\ and\ \citenamefont {Yu}}]{Hu2022loss-anti}%
  \BibitemOpen
  \bibfield  {author} {\bibinfo {author} {\bibfnamefont {Y.}~\bibnamefont
  {Chen}}, \bibinfo {author} {\bibfnamefont {J.}~\bibnamefont {Hu}},\ and\
  \bibinfo {author} {\bibfnamefont {H.}~\bibnamefont {Yu}},\ }\bibfield
  {title} {\bibinfo {title} {Entanglement generation for uniformly accelerated
  atoms assisted by environment-induced interatomic interaction and the loss of
  the anti-unruh effect},\ }\href {https://doi.org/10.1103/PhysRevD.105.045013}
  {\bibfield  {journal} {\bibinfo  {journal} {Phys. Rev. D}\ }\textbf {\bibinfo
  {volume} {105}},\ \bibinfo {pages} {045013} (\bibinfo {year}
  {2022})}\BibitemShut {NoStop}%
\bibitem [{\citenamefont {Ju\'arez-Aubry}\ and\ \citenamefont
  {Moustos}(2019)}]{Benito2019asymptotic}%
  \BibitemOpen
  \bibfield  {author} {\bibinfo {author} {\bibfnamefont {B.~A.}\ \bibnamefont
  {Ju\'arez-Aubry}}\ and\ \bibinfo {author} {\bibfnamefont {D.}~\bibnamefont
  {Moustos}},\ }\bibfield  {title} {\bibinfo {title} {Asymptotic states for
  stationary unruh-dewitt detectors},\ }\href
  {https://doi.org/10.1103/PhysRevD.100.025018} {\bibfield  {journal} {\bibinfo
   {journal} {Phys. Rev. D}\ }\textbf {\bibinfo {volume} {100}},\ \bibinfo
  {pages} {025018} (\bibinfo {year} {2019})}\BibitemShut {NoStop}%
\bibitem [{\citenamefont {Moustos}\ and\ \citenamefont
  {Anastopoulos}(2017)}]{Moustos2017nonmarkov}%
  \BibitemOpen
  \bibfield  {author} {\bibinfo {author} {\bibfnamefont {D.}~\bibnamefont
  {Moustos}}\ and\ \bibinfo {author} {\bibfnamefont {C.}~\bibnamefont
  {Anastopoulos}},\ }\bibfield  {title} {\bibinfo {title} {Non-markovian time
  evolution of an accelerated qubit},\ }\href
  {https://doi.org/10.1103/PhysRevD.95.025020} {\bibfield  {journal} {\bibinfo
  {journal} {Phys. Rev. D}\ }\textbf {\bibinfo {volume} {95}},\ \bibinfo
  {pages} {025020} (\bibinfo {year} {2017})}\BibitemShut {NoStop}%
\bibitem [{\citenamefont {Ju{\'{a}}rez-Aubry}\ and\ \citenamefont
  {Louko}(2014)}]{Aubry2014derivative}%
  \BibitemOpen
  \bibfield  {author} {\bibinfo {author} {\bibfnamefont {B.~A.}\ \bibnamefont
  {Ju{\'{a}}rez-Aubry}}\ and\ \bibinfo {author} {\bibfnamefont
  {J.}~\bibnamefont {Louko}},\ }\bibfield  {title} {\bibinfo {title} {{Onset
  and decay of the 1 + 1 Hawking-Unruh effect: what the derivative-coupling
  detector saw}},\ }\href {https://doi.org/10.1088/0264-9381/31/24/245007}
  {\bibfield  {journal} {\bibinfo  {journal} {Class. and Quantum Gravity}\
  }\textbf {\bibinfo {volume} {31}},\ \bibinfo {pages} {245007} (\bibinfo
  {year} {2014})}\BibitemShut {NoStop}%
\bibitem [{\citenamefont {Louko}\ and\ \citenamefont
  {Satz}(2007)}]{Satz2007transitionrate}%
  \BibitemOpen
  \bibfield  {author} {\bibinfo {author} {\bibfnamefont {J.}~\bibnamefont
  {Louko}}\ and\ \bibinfo {author} {\bibfnamefont {A.}~\bibnamefont {Satz}},\
  }\bibfield  {title} {\bibinfo {title} {{ Transition rate of the Unruh-DeWitt
  detector in curved spacetime }},\ }\href {http://arxiv.org/abs/0710.5671}
  {\bibfield  {journal} {\bibinfo  {journal} {Class. Quantum Gravity}\ }\textbf
  {\bibinfo {volume} {25}},\ \bibinfo {pages} {055012} (\bibinfo {year}
  {2007})},\ \Eprint {https://arxiv.org/abs/Arxiv:0710.5671v3}
  {Arxiv:0710.5671v3} \BibitemShut {NoStop}%
\bibitem [{\citenamefont {Burgess}\ \emph {et~al.}(2018)\citenamefont
  {Burgess}, \citenamefont {Hainge}, \citenamefont {Kaplanek},\ and\
  \citenamefont {Rummel}}]{Kaplanek2018rindler}%
  \BibitemOpen
  \bibfield  {author} {\bibinfo {author} {\bibfnamefont {C.~P.}\ \bibnamefont
  {Burgess}}, \bibinfo {author} {\bibfnamefont {J.}~\bibnamefont {Hainge}},
  \bibinfo {author} {\bibfnamefont {G.}~\bibnamefont {Kaplanek}},\ and\
  \bibinfo {author} {\bibfnamefont {M.}~\bibnamefont {Rummel}},\ }\bibfield
  {title} {\bibinfo {title} {Failure of perturbation theory near horizons: the
  rindler example},\ }\bibfield  {journal} {\bibinfo  {journal} {Journal of
  High Energy Physics}\ }\textbf {\bibinfo {volume} {2018}},\ \href
  {https://doi.org/10.1007/jhep10(2018)122} {10.1007/jhep10(2018)122} (\bibinfo
  {year} {2018})\BibitemShut {NoStop}%
\bibitem [{\citenamefont {Kaplanek}\ and\ \citenamefont
  {Burgess}(2020{\natexlab{a}})}]{kaplanek2020hot}%
  \BibitemOpen
  \bibfield  {author} {\bibinfo {author} {\bibfnamefont {G.}~\bibnamefont
  {Kaplanek}}\ and\ \bibinfo {author} {\bibfnamefont {C.}~\bibnamefont
  {Burgess}},\ }\bibfield  {title} {\bibinfo {title} {Hot accelerated qubits:
  decoherence, thermalization, secular growth and reliable late-time
  predictions},\ }\href@noop {} {\bibfield  {journal} {\bibinfo  {journal}
  {Journal of High Energy Physics}\ }\textbf {\bibinfo {volume} {2020}},\
  \bibinfo {pages} {1} (\bibinfo {year} {2020}{\natexlab{a}})}\BibitemShut
  {NoStop}%
\bibitem [{\citenamefont {Kaplanek}\ and\ \citenamefont
  {Burgess}(2020{\natexlab{b}})}]{kaplanek2020hot2}%
  \BibitemOpen
  \bibfield  {author} {\bibinfo {author} {\bibfnamefont {G.}~\bibnamefont
  {Kaplanek}}\ and\ \bibinfo {author} {\bibfnamefont {C.}~\bibnamefont
  {Burgess}},\ }\bibfield  {title} {\bibinfo {title} {Hot cosmic qubits:
  late-time de sitter evolution and critical slowing down},\ }\href@noop {}
  {\bibfield  {journal} {\bibinfo  {journal} {Journal of High Energy Physics}\
  }\textbf {\bibinfo {volume} {2020}},\ \bibinfo {pages} {1} (\bibinfo {year}
  {2020}{\natexlab{b}})}\BibitemShut {NoStop}%
\bibitem [{\citenamefont {Kaplanek}\ and\ \citenamefont
  {Burgess}(2021)}]{kaplanek2021qubits}%
  \BibitemOpen
  \bibfield  {author} {\bibinfo {author} {\bibfnamefont {G.}~\bibnamefont
  {Kaplanek}}\ and\ \bibinfo {author} {\bibfnamefont {C.}~\bibnamefont
  {Burgess}},\ }\bibfield  {title} {\bibinfo {title} {Qubits on the horizon:
  decoherence and thermalization near black holes},\ }\href@noop {} {\bibfield
  {journal} {\bibinfo  {journal} {Journal of High Energy Physics}\ }\textbf
  {\bibinfo {volume} {2021}},\ \bibinfo {pages} {1} (\bibinfo {year}
  {2021})}\BibitemShut {NoStop}%
\bibitem [{\citenamefont {Louko}\ and\ \citenamefont
  {Satz}(2006)}]{Satz2006howoften}%
  \BibitemOpen
  \bibfield  {author} {\bibinfo {author} {\bibfnamefont {J.}~\bibnamefont
  {Louko}}\ and\ \bibinfo {author} {\bibfnamefont {A.}~\bibnamefont {Satz}},\
  }\bibfield  {title} {\bibinfo {title} {{ How often does the Unruh-DeWitt
  detector click? Regularisation by a spatial profile }},\ }\href
  {http://arxiv.org/abs/gr-qc/0606067} {\bibfield  {journal} {\bibinfo
  {journal} {Class. Quantum Gravity}\ }\textbf {\bibinfo {volume} {23}},\
  \bibinfo {pages} {6321} (\bibinfo {year} {2006})},\ \Eprint
  {https://arxiv.org/abs/Arxiv:gr-qc/0606067v3} {Arxiv:gr-qc/0606067v3}
  \BibitemShut {NoStop}%
\bibitem [{\citenamefont {Fewster}\ \emph {et~al.}(2016)\citenamefont
  {Fewster}, \citenamefont {Juárez-Aubry},\ and\ \citenamefont
  {Louko}}]{Fewster2016wait}%
  \BibitemOpen
  \bibfield  {author} {\bibinfo {author} {\bibfnamefont {C.~J.}\ \bibnamefont
  {Fewster}}, \bibinfo {author} {\bibfnamefont {B.~A.}\ \bibnamefont
  {Juárez-Aubry}},\ and\ \bibinfo {author} {\bibfnamefont {J.}~\bibnamefont
  {Louko}},\ }\bibfield  {title} {\bibinfo {title} {Waiting for unruh},\ }\href
  {https://doi.org/10.1088/0264-9381/33/16/165003} {\bibfield  {journal}
  {\bibinfo  {journal} {Classical and Quantum Gravity}\ }\textbf {\bibinfo
  {volume} {33}},\ \bibinfo {pages} {165003} (\bibinfo {year}
  {2016})}\BibitemShut {NoStop}%
\bibitem [{\citenamefont {Gaspard}\ and\ \citenamefont
  {Nagaoka}(1999)}]{Gaspard1999slippage}%
  \BibitemOpen
  \bibfield  {author} {\bibinfo {author} {\bibfnamefont {P.}~\bibnamefont
  {Gaspard}}\ and\ \bibinfo {author} {\bibfnamefont {M.}~\bibnamefont
  {Nagaoka}},\ }\bibfield  {title} {\bibinfo {title} {Slippage of initial
  conditions for the redfield master equation},\ }\href
  {https://doi.org/10.1063/1.479867} {\bibfield  {journal} {\bibinfo  {journal}
  {The Journal of Chemical Physics}\ }\textbf {\bibinfo {volume} {111}},\
  \bibinfo {pages} {5668} (\bibinfo {year} {1999})},\ \Eprint
  {https://arxiv.org/abs/https://doi.org/10.1063/1.479867}
  {https://doi.org/10.1063/1.479867} \BibitemShut {NoStop}%
\bibitem [{\citenamefont {Su\'arez}\ \emph {et~al.}(1992)\citenamefont
  {Su\'arez}, \citenamefont {Silbey},\ and\ \citenamefont
  {Oppenheim}}]{Suarez1992memory}%
  \BibitemOpen
  \bibfield  {author} {\bibinfo {author} {\bibfnamefont {A.}~\bibnamefont
  {Su\'arez}}, \bibinfo {author} {\bibfnamefont {R.}~\bibnamefont {Silbey}},\
  and\ \bibinfo {author} {\bibfnamefont {I.}~\bibnamefont {Oppenheim}},\
  }\bibfield  {title} {\bibinfo {title} {Memory effects in the relaxation of
  quantum open systems},\ }\href {https://doi.org/10.1063/1.463831} {\bibfield
  {journal} {\bibinfo  {journal} {The Journal of Chemical Physics}\ }\textbf
  {\bibinfo {volume} {97}},\ \bibinfo {pages} {5101} (\bibinfo {year}
  {1992})},\ \Eprint {https://arxiv.org/abs/https://doi.org/10.1063/1.463831}
  {https://doi.org/10.1063/1.463831} \BibitemShut {NoStop}%
\bibitem [{\citenamefont {Benatti}\ \emph {et~al.}(2010)\citenamefont
  {Benatti}, \citenamefont {Floreanini},\ and\ \citenamefont
  {Marzolino}}]{Benatti2010entanglement2}%
  \BibitemOpen
  \bibfield  {author} {\bibinfo {author} {\bibfnamefont {F.}~\bibnamefont
  {Benatti}}, \bibinfo {author} {\bibfnamefont {R.}~\bibnamefont
  {Floreanini}},\ and\ \bibinfo {author} {\bibfnamefont {U.}~\bibnamefont
  {Marzolino}},\ }\bibfield  {title} {\bibinfo {title} {Entangling two unequal
  atoms through a common bath},\ }\href
  {https://doi.org/10.1103/PhysRevA.81.012105} {\bibfield  {journal} {\bibinfo
  {journal} {Phys. Rev. A}\ }\textbf {\bibinfo {volume} {81}},\ \bibinfo
  {pages} {012105} (\bibinfo {year} {2010})}\BibitemShut {NoStop}%
\bibitem [{\citenamefont {Anderloni}\ \emph {et~al.}(2007)\citenamefont
  {Anderloni}, \citenamefont {Benatti},\ and\ \citenamefont
  {Floreanini}}]{Anderloni2007redfield}%
  \BibitemOpen
  \bibfield  {author} {\bibinfo {author} {\bibfnamefont {S.}~\bibnamefont
  {Anderloni}}, \bibinfo {author} {\bibfnamefont {F.}~\bibnamefont {Benatti}},\
  and\ \bibinfo {author} {\bibfnamefont {R.}~\bibnamefont {Floreanini}},\
  }\bibfield  {title} {\bibinfo {title} {Redfield reduced dynamics and
  entanglement},\ }\href {https://doi.org/10.1088/1751-8113/40/7/013}
  {\bibfield  {journal} {\bibinfo  {journal} {Journal of Physics A:
  Mathematical and Theoretical}\ }\textbf {\bibinfo {volume} {40}},\ \bibinfo
  {pages} {1625} (\bibinfo {year} {2007})}\BibitemShut {NoStop}%
\bibitem [{\citenamefont {Breuer}\ \emph {et~al.}(2002)\citenamefont {Breuer},
  \citenamefont {Petruccione},\ and\ \citenamefont
  {Petruccione}}]{breuer2002theory}%
  \BibitemOpen
  \bibfield  {author} {\bibinfo {author} {\bibfnamefont {H.}~\bibnamefont
  {Breuer}}, \bibinfo {author} {\bibfnamefont {F.}~\bibnamefont
  {Petruccione}},\ and\ \bibinfo {author} {\bibfnamefont {S.}~\bibnamefont
  {Petruccione}},\ }\href {https://books.google.ca/books?id=0Yx5VzaMYm8C}
  {\emph {\bibinfo {title} {The Theory of Open Quantum Systems}}}\ (\bibinfo
  {publisher} {Oxford University Press},\ \bibinfo {year} {2002})\BibitemShut
  {NoStop}%
\bibitem [{\citenamefont {Lidar}(2020)}]{lidar2020lecture}%
  \BibitemOpen
  \bibfield  {author} {\bibinfo {author} {\bibfnamefont {D.~A.}\ \bibnamefont
  {Lidar}},\ }\href@noop {} {\bibinfo {title} {Lecture notes on the theory of
  open quantum systems}} (\bibinfo {year} {2020}),\ \Eprint
  {https://arxiv.org/abs/1902.00967} {arXiv:1902.00967 [quant-ph]} \BibitemShut
  {NoStop}%
\bibitem [{\citenamefont {Davies}(1974)}]{davies1974markovian}%
  \BibitemOpen
  \bibfield  {author} {\bibinfo {author} {\bibfnamefont {E.~B.}\ \bibnamefont
  {Davies}},\ }\bibfield  {title} {\bibinfo {title} {Markovian master
  equations},\ }\href@noop {} {\bibfield  {journal} {\bibinfo  {journal}
  {Communications in mathematical Physics}\ }\textbf {\bibinfo {volume} {39}},\
  \bibinfo {pages} {91} (\bibinfo {year} {1974})}\BibitemShut {NoStop}%
\bibitem [{\citenamefont {Davies}(1976)}]{davies1976markovian}%
  \BibitemOpen
  \bibfield  {author} {\bibinfo {author} {\bibfnamefont {E.~B.}\ \bibnamefont
  {Davies}},\ }\bibfield  {title} {\bibinfo {title} {Markovian master
  equations. ii},\ }\href@noop {} {\bibfield  {journal} {\bibinfo  {journal}
  {Mathematische Annalen}\ }\textbf {\bibinfo {volume} {219}},\ \bibinfo
  {pages} {147} (\bibinfo {year} {1976})}\BibitemShut {NoStop}%
\bibitem [{\citenamefont {Martin}\ and\ \citenamefont
  {Vennin}(2018{\natexlab{a}})}]{Martin2018cosmologydecohere}%
  \BibitemOpen
  \bibfield  {author} {\bibinfo {author} {\bibfnamefont {J.}~\bibnamefont
  {Martin}}\ and\ \bibinfo {author} {\bibfnamefont {V.}~\bibnamefont
  {Vennin}},\ }\bibfield  {title} {\bibinfo {title} {Observational constraints
  on quantum decoherence during inflation},\ }\href
  {https://doi.org/10.1088/1475-7516/2018/05/063} {\bibfield  {journal}
  {\bibinfo  {journal} {Journal of Cosmology and Astroparticle Physics}\
  }\textbf {\bibinfo {volume} {2018}}\bibinfo  {number} { (05)},\ \bibinfo
  {pages} {063–063}}\BibitemShut {NoStop}%
\bibitem [{\citenamefont {Lindblad}(1976)}]{lindblad1976generators}%
  \BibitemOpen
\bibfield  {number} {  }\bibfield  {author} {\bibinfo {author} {\bibfnamefont
  {G.}~\bibnamefont {Lindblad}},\ }\bibfield  {title} {\bibinfo {title} {On the
  generators of quantum dynamical semigroups},\ }\href
  {https://doi.org/10.1007/BF01608499} {\bibfield  {journal} {\bibinfo
  {journal} {Communications in Mathematical Physics}\ }\textbf {\bibinfo
  {volume} {48}},\ \bibinfo {pages} {119} (\bibinfo {year} {1976})}\BibitemShut
  {NoStop}%
\bibitem [{\citenamefont {Gorini}\ \emph {et~al.}(1976)\citenamefont {Gorini},
  \citenamefont {Kossakowski},\ and\ \citenamefont
  {Sudarshan}}]{gorini1976completely}%
  \BibitemOpen
  \bibfield  {author} {\bibinfo {author} {\bibfnamefont {V.}~\bibnamefont
  {Gorini}}, \bibinfo {author} {\bibfnamefont {A.}~\bibnamefont
  {Kossakowski}},\ and\ \bibinfo {author} {\bibfnamefont {E.~C.~G.}\
  \bibnamefont {Sudarshan}},\ }\bibfield  {title} {\bibinfo {title} {Completely
  positive dynamical semigroups of n-level systems},\ }\href
  {https://doi.org/10.1063/1.522979} {\bibfield  {journal} {\bibinfo  {journal}
  {Journal of Mathematical Physics}\ }\textbf {\bibinfo {volume} {17}},\
  \bibinfo {pages} {821} (\bibinfo {year} {1976})}\BibitemShut {NoStop}%
\bibitem [{\citenamefont {Whitney}(2008)}]{whitney2008staying}%
  \BibitemOpen
  \bibfield  {author} {\bibinfo {author} {\bibfnamefont {R.~S.}\ \bibnamefont
  {Whitney}},\ }\bibfield  {title} {\bibinfo {title} {Staying positive: going
  beyond lindblad with perturbative master equations},\ }\href
  {https://doi.org/10.1088/1751-8113/41/17/175304} {\bibfield  {journal}
  {\bibinfo  {journal} {Journal of Physics A: Mathematical and Theoretical}\
  }\textbf {\bibinfo {volume} {41}},\ \bibinfo {pages} {175304} (\bibinfo
  {year} {2008})}\BibitemShut {NoStop}%
\bibitem [{\citenamefont {Funai}\ and\ \citenamefont
  {Mart\'{\i}n-Mart\'{\i}nez}(2019)}]{Funai2019rotatingwave}%
  \BibitemOpen
  \bibfield  {author} {\bibinfo {author} {\bibfnamefont {N.}~\bibnamefont
  {Funai}}\ and\ \bibinfo {author} {\bibfnamefont {E.}~\bibnamefont
  {Mart\'{\i}n-Mart\'{\i}nez}},\ }\bibfield  {title} {\bibinfo {title}
  {Faster-than-light signaling in the rotating-wave approximation},\ }\href
  {https://doi.org/10.1103/PhysRevD.100.065021} {\bibfield  {journal} {\bibinfo
   {journal} {Phys. Rev. D}\ }\textbf {\bibinfo {volume} {100}},\ \bibinfo
  {pages} {065021} (\bibinfo {year} {2019})}\BibitemShut {NoStop}%
\bibitem [{\citenamefont {Fleming}\ \emph {et~al.}(2010)\citenamefont
  {Fleming}, \citenamefont {Cummings}, \citenamefont {Anastopoulos},\ and\
  \citenamefont {Hu}}]{Fleming2010RWA}%
  \BibitemOpen
  \bibfield  {author} {\bibinfo {author} {\bibfnamefont {C.}~\bibnamefont
  {Fleming}}, \bibinfo {author} {\bibfnamefont {N.~I.}\ \bibnamefont
  {Cummings}}, \bibinfo {author} {\bibfnamefont {C.}~\bibnamefont
  {Anastopoulos}},\ and\ \bibinfo {author} {\bibfnamefont {B.~L.}\ \bibnamefont
  {Hu}},\ }\bibfield  {title} {\bibinfo {title} {The rotating-wave
  approximation: consistency and applicability from an open quantum system
  analysis},\ }\href {https://doi.org/10.1088/1751-8113/43/40/405304}
  {\bibfield  {journal} {\bibinfo  {journal} {Journal of Physics A:
  Mathematical and Theoretical}\ }\textbf {\bibinfo {volume} {43}},\ \bibinfo
  {pages} {405304} (\bibinfo {year} {2010})}\BibitemShut {NoStop}%
\bibitem [{\citenamefont {Dicke}(1954)}]{DickeModel}%
  \BibitemOpen
  \bibfield  {author} {\bibinfo {author} {\bibfnamefont {R.~H.}\ \bibnamefont
  {Dicke}},\ }\bibfield  {title} {\bibinfo {title} {Coherence in spontaneous
  radiation processes},\ }\href@noop {} {\bibfield  {journal} {\bibinfo
  {journal} {Phys. Rev.}\ }\textbf {\bibinfo {volume} {93}} (\bibinfo {year}
  {1954})}\BibitemShut {NoStop}%
\bibitem [{\citenamefont {Du}\ and\ \citenamefont {Mann}(2021)}]{du2021fisher}%
  \BibitemOpen
  \bibfield  {author} {\bibinfo {author} {\bibfnamefont {H.}~\bibnamefont
  {Du}}\ and\ \bibinfo {author} {\bibfnamefont {R.~B.}\ \bibnamefont {Mann}},\
  }\bibfield  {title} {\bibinfo {title} {Fisher information as a probe of
  spacetime structure: relativistic quantum metrology in (a) ds},\ }\href@noop
  {} {\bibfield  {journal} {\bibinfo  {journal} {Journal of High Energy
  Physics}\ }\textbf {\bibinfo {volume} {2021}},\ \bibinfo {pages} {1}
  (\bibinfo {year} {2021})}\BibitemShut {NoStop}%
\bibitem [{\citenamefont {Feng}\ and\ \citenamefont
  {Zhang}(2022)}]{feng2022quantum-fisher}%
  \BibitemOpen
  \bibfield  {author} {\bibinfo {author} {\bibfnamefont {J.}~\bibnamefont
  {Feng}}\ and\ \bibinfo {author} {\bibfnamefont {J.-J.}\ \bibnamefont
  {Zhang}},\ }\bibfield  {title} {\bibinfo {title} {Quantum fisher information
  as a probe for unruh thermality},\ }\href
  {https://doi.org/10.1016/j.physletb.2022.136992} {\bibfield  {journal}
  {\bibinfo  {journal} {Physics Letters B}\ }\textbf {\bibinfo {volume}
  {827}},\ \bibinfo {pages} {136992} (\bibinfo {year} {2022})}\BibitemShut
  {NoStop}%
\bibitem [{\citenamefont {Burgess}(2020)}]{burgess2020introduction}%
  \BibitemOpen
  \bibfield  {author} {\bibinfo {author} {\bibfnamefont {C.}~\bibnamefont
  {Burgess}},\ }\href@noop {} {\emph {\bibinfo {title} {Introduction to
  Effective Field Theory: Thinking effectively about hierarchies of scale}}}\
  (\bibinfo  {publisher} {Cambridge University Press},\ \bibinfo {year}
  {2020})\BibitemShut {NoStop}%
\bibitem [{\citenamefont {{Kaplanek, Gregory}}(2022)}]{kaplanek2022some}%
  \BibitemOpen
  \bibfield  {author} {\bibinfo {author} {\bibnamefont {{Kaplanek, Gregory}}},\
  }\emph {\bibinfo {title} {{Some Applications of Open Effective Field Theories
  to Gravitating Quantum Systems}}},\ \href@noop {} {Ph.D. thesis},\ \bibinfo
  {school} {{McMaster University}} (\bibinfo {year} {{2022}})\BibitemShut
  {NoStop}%
\bibitem [{\citenamefont {Nakajima}(1958)}]{Nakajima1958}%
  \BibitemOpen
  \bibfield  {author} {\bibinfo {author} {\bibfnamefont {S.}~\bibnamefont
  {Nakajima}},\ }\bibfield  {title} {\bibinfo {title} {{On Quantum Theory of
  Transport Phenomena: Steady Diffusion}},\ }\href
  {https://doi.org/10.1143/PTP.20.948} {\bibfield  {journal} {\bibinfo
  {journal} {Progress of Theoretical Physics}\ }\textbf {\bibinfo {volume}
  {20}},\ \bibinfo {pages} {948} (\bibinfo {year} {1958})},\ \Eprint
  {https://arxiv.org/abs/https://academic.oup.com/ptp/article-pdf/20/6/948/5440766/20-6-948.pdf}
  {https://academic.oup.com/ptp/article-pdf/20/6/948/5440766/20-6-948.pdf}
  \BibitemShut {NoStop}%
\bibitem [{\citenamefont {Zwanzig}(1960)}]{Zwanzig1960}%
  \BibitemOpen
  \bibfield  {author} {\bibinfo {author} {\bibfnamefont {R.}~\bibnamefont
  {Zwanzig}},\ }\bibfield  {title} {\bibinfo {title} {Ensemble method in the
  theory of irreversibility},\ }\href {https://doi.org/10.1063/1.1731409}
  {\bibfield  {journal} {\bibinfo  {journal} {The Journal of Chemical Physics}\
  }\textbf {\bibinfo {volume} {33}},\ \bibinfo {pages} {1338} (\bibinfo {year}
  {1960})},\ \Eprint {https://arxiv.org/abs/https://doi.org/10.1063/1.1731409}
  {https://doi.org/10.1063/1.1731409} \BibitemShut {NoStop}%
\bibitem [{\citenamefont {Sciama}\ \emph {et~al.}(1981)\citenamefont {Sciama},
  \citenamefont {Candelas},\ and\ \citenamefont {Deutsch}}]{sciama1981quantum}%
  \BibitemOpen
  \bibfield  {author} {\bibinfo {author} {\bibfnamefont {D.}~\bibnamefont
  {Sciama}}, \bibinfo {author} {\bibfnamefont {P.}~\bibnamefont {Candelas}},\
  and\ \bibinfo {author} {\bibfnamefont {D.}~\bibnamefont {Deutsch}},\
  }\bibfield  {title} {\bibinfo {title} {Quantum field theory, horizons and
  thermodynamics},\ }\href@noop {} {\bibfield  {journal} {\bibinfo  {journal}
  {Advances in Physics}\ }\textbf {\bibinfo {volume} {30}},\ \bibinfo {pages}
  {327} (\bibinfo {year} {1981})}\BibitemShut {NoStop}%
\bibitem [{\citenamefont {Ali}\ \emph {et~al.}(2010)\citenamefont {Ali},
  \citenamefont {Rau},\ and\ \citenamefont {Alber}}]{Ali2010discord}%
  \BibitemOpen
  \bibfield  {author} {\bibinfo {author} {\bibfnamefont {M.}~\bibnamefont
  {Ali}}, \bibinfo {author} {\bibfnamefont {A.~R.~P.}\ \bibnamefont {Rau}},\
  and\ \bibinfo {author} {\bibfnamefont {G.}~\bibnamefont {Alber}},\ }\bibfield
   {title} {\bibinfo {title} {Quantum discord for two-qubit $x$ states},\
  }\href {https://doi.org/10.1103/PhysRevA.81.042105} {\bibfield  {journal}
  {\bibinfo  {journal} {Phys. Rev. A}\ }\textbf {\bibinfo {volume} {81}},\
  \bibinfo {pages} {042105} (\bibinfo {year} {2010})}\BibitemShut {NoStop}%
\bibitem [{\citenamefont {Tana{\'{s}}}(2013)}]{Tanas2013}%
  \BibitemOpen
  \bibfield  {author} {\bibinfo {author} {\bibfnamefont {R.}~\bibnamefont
  {Tana{\'{s}}}},\ }\bibfield  {title} {\bibinfo {title} {Evolution of quantum
  correlations in a two-atom system},\ }\href
  {https://doi.org/10.1088/0031-8949/2013/t153/014059} {\bibfield  {journal}
  {\bibinfo  {journal} {Physica Scripta}\ }\textbf {\bibinfo {volume} {T153}},\
  \bibinfo {pages} {014059} (\bibinfo {year} {2013})}\BibitemShut {NoStop}%
\bibitem [{\citenamefont {Kossakowski}(1972)}]{kossakowski1972quantum}%
  \BibitemOpen
  \bibfield  {author} {\bibinfo {author} {\bibfnamefont {A.}~\bibnamefont
  {Kossakowski}},\ }\bibfield  {title} {\bibinfo {title} {On quantum
  statistical mechanics of non-hamiltonian systems},\ }\href
  {https://doi.org/10.1016/0034-4877(72)90010-9} {\bibfield  {journal}
  {\bibinfo  {journal} {Reports on Mathematical Physics}\ }\textbf {\bibinfo
  {volume} {3}},\ \bibinfo {pages} {247} (\bibinfo {year} {1972})}\BibitemShut
  {NoStop}%
\bibitem [{\citenamefont {Kaplanek}\ \emph {et~al.}(2021)\citenamefont
  {Kaplanek}, \citenamefont {Burgess},\ and\ \citenamefont
  {Holman}}]{kaplanek2021qubit}%
  \BibitemOpen
  \bibfield  {author} {\bibinfo {author} {\bibfnamefont {G.}~\bibnamefont
  {Kaplanek}}, \bibinfo {author} {\bibfnamefont {C.}~\bibnamefont {Burgess}},\
  and\ \bibinfo {author} {\bibfnamefont {R.}~\bibnamefont {Holman}},\
  }\bibfield  {title} {\bibinfo {title} {Qubit heating near a hotspot},\
  }\href@noop {} {\bibfield  {journal} {\bibinfo  {journal} {Journal of High
  Energy Physics}\ }\textbf {\bibinfo {volume} {2021}},\ \bibinfo {pages} {1}
  (\bibinfo {year} {2021})}\BibitemShut {NoStop}%
\bibitem [{\citenamefont {Gorini}\ \emph {et~al.}(1978)\citenamefont {Gorini},
  \citenamefont {Frigerio}, \citenamefont {Verri}, \citenamefont
  {Kossakowski},\ and\ \citenamefont {Sudarshan}}]{gorini1978properties}%
  \BibitemOpen
  \bibfield  {author} {\bibinfo {author} {\bibfnamefont {V.}~\bibnamefont
  {Gorini}}, \bibinfo {author} {\bibfnamefont {A.}~\bibnamefont {Frigerio}},
  \bibinfo {author} {\bibfnamefont {M.}~\bibnamefont {Verri}}, \bibinfo
  {author} {\bibfnamefont {A.}~\bibnamefont {Kossakowski}},\ and\ \bibinfo
  {author} {\bibfnamefont {E.}~\bibnamefont {Sudarshan}},\ }\bibfield  {title}
  {\bibinfo {title} {Properties of quantum markovian master equations},\ }\href
  {https://doi.org/10.1016/0034-4877(78)90050-2} {\bibfield  {journal}
  {\bibinfo  {journal} {Reports on Mathematical Physics}\ }\textbf {\bibinfo
  {volume} {13}},\ \bibinfo {pages} {149} (\bibinfo {year} {1978})}\BibitemShut
  {NoStop}%
\bibitem [{\citenamefont {Ingarden}\ and\ \citenamefont
  {Kossakowski}(1975)}]{ingarden1975connection}%
  \BibitemOpen
  \bibfield  {author} {\bibinfo {author} {\bibfnamefont {R.}~\bibnamefont
  {Ingarden}}\ and\ \bibinfo {author} {\bibfnamefont {A.}~\bibnamefont
  {Kossakowski}},\ }\bibfield  {title} {\bibinfo {title} {On the connection of
  nonequilibrium information thermodynamics with non-hamiltonian quantum
  mechanics of open systems},\ }\href@noop {} {\bibfield  {journal} {\bibinfo
  {journal} {Annals of Physics}\ }\textbf {\bibinfo {volume} {89}},\ \bibinfo
  {pages} {451} (\bibinfo {year} {1975})}\BibitemShut {NoStop}%
\bibitem [{\citenamefont {Pozas-Kerstjens}\ \emph {et~al.}(2017)\citenamefont
  {Pozas-Kerstjens}, \citenamefont {Louko},\ and\ \citenamefont
  {Mart\'{\i}n-Mart\'{\i}nez}}]{pozas2017degenerate}%
  \BibitemOpen
  \bibfield  {author} {\bibinfo {author} {\bibfnamefont {A.}~\bibnamefont
  {Pozas-Kerstjens}}, \bibinfo {author} {\bibfnamefont {J.}~\bibnamefont
  {Louko}},\ and\ \bibinfo {author} {\bibfnamefont {E.}~\bibnamefont
  {Mart\'{\i}n-Mart\'{\i}nez}},\ }\bibfield  {title} {\bibinfo {title}
  {Degenerate detectors are unable to harvest spacelike entanglement},\ }\href
  {https://doi.org/10.1103/PhysRevD.95.105009} {\bibfield  {journal} {\bibinfo
  {journal} {Phys. Rev. D}\ }\textbf {\bibinfo {volume} {95}},\ \bibinfo
  {pages} {105009} (\bibinfo {year} {2017})}\BibitemShut {NoStop}%
\bibitem [{\citenamefont {Simidzija}\ \emph {et~al.}(2018)\citenamefont
  {Simidzija}, \citenamefont {Jonsson},\ and\ \citenamefont
  {Mart\'{\i}n-Mart\'{\i}nez}}]{Simidzija2018no-go}%
  \BibitemOpen
  \bibfield  {author} {\bibinfo {author} {\bibfnamefont {P.}~\bibnamefont
  {Simidzija}}, \bibinfo {author} {\bibfnamefont {R.~H.}\ \bibnamefont
  {Jonsson}},\ and\ \bibinfo {author} {\bibfnamefont {E.}~\bibnamefont
  {Mart\'{\i}n-Mart\'{\i}nez}},\ }\bibfield  {title} {\bibinfo {title} {General
  no-go theorem for entanglement extraction},\ }\href
  {https://doi.org/10.1103/PhysRevD.97.125002} {\bibfield  {journal} {\bibinfo
  {journal} {Phys. Rev. D}\ }\textbf {\bibinfo {volume} {97}},\ \bibinfo
  {pages} {125002} (\bibinfo {year} {2018})}\BibitemShut {NoStop}%
\bibitem [{\citenamefont {Olver}\ \emph {et~al.}(2010)\citenamefont {Olver},
  \citenamefont {Lozier}, \citenamefont {Boisvert},\ and\ \citenamefont
  {Clark}}]{olver2010nist}%
  \BibitemOpen
  \bibfield  {author} {\bibinfo {author} {\bibfnamefont {F.~W.}\ \bibnamefont
  {Olver}}, \bibinfo {author} {\bibfnamefont {D.~W.}\ \bibnamefont {Lozier}},
  \bibinfo {author} {\bibfnamefont {R.~F.}\ \bibnamefont {Boisvert}},\ and\
  \bibinfo {author} {\bibfnamefont {C.~W.}\ \bibnamefont {Clark}},\ }\href@noop
  {} {\emph {\bibinfo {title} {NIST handbook of mathematical functions}}}\
  (\bibinfo  {publisher} {Cambridge university press},\ \bibinfo {year}
  {2010})\BibitemShut {NoStop}%
\bibitem [{\citenamefont {Landulfo}(2016)}]{Landulfo2016communication}%
  \BibitemOpen
  \bibfield  {author} {\bibinfo {author} {\bibfnamefont {A.~G.~S.}\
  \bibnamefont {Landulfo}},\ }\bibfield  {title} {\bibinfo {title}
  {Nonperturbative approach to relativistic quantum communication channels},\
  }\href {https://doi.org/10.1103/PhysRevD.93.104019} {\bibfield  {journal}
  {\bibinfo  {journal} {Phys. Rev. D}\ }\textbf {\bibinfo {volume} {93}},\
  \bibinfo {pages} {104019} (\bibinfo {year} {2016})}\BibitemShut {NoStop}%
\bibitem [{\citenamefont {Vidal}\ and\ \citenamefont
  {Werner}(2002)}]{Vidal2002negativity}%
  \BibitemOpen
  \bibfield  {author} {\bibinfo {author} {\bibfnamefont {G.}~\bibnamefont
  {Vidal}}\ and\ \bibinfo {author} {\bibfnamefont {R.~F.}\ \bibnamefont
  {Werner}},\ }\bibfield  {title} {\bibinfo {title} {Computable measure of
  entanglement},\ }\href {https://doi.org/10.1103/PhysRevA.65.032314}
  {\bibfield  {journal} {\bibinfo  {journal} {Phys. Rev. A}\ }\textbf {\bibinfo
  {volume} {65}},\ \bibinfo {pages} {032314} (\bibinfo {year}
  {2002})}\BibitemShut {NoStop}%
\bibitem [{\citenamefont {Grishchuk}\ and\ \citenamefont
  {Sidorov}(1989)}]{Grishchuk:1989ss}%
  \BibitemOpen
  \bibfield  {author} {\bibinfo {author} {\bibfnamefont {L.~P.}\ \bibnamefont
  {Grishchuk}}\ and\ \bibinfo {author} {\bibfnamefont {Y.~V.}\ \bibnamefont
  {Sidorov}},\ }\bibfield  {title} {\bibinfo {title} {{On the Quantum State of
  Relic Gravitons}},\ }\href {https://doi.org/10.1088/0264-9381/6/9/002}
  {\bibfield  {journal} {\bibinfo  {journal} {Class. Quant. Grav.}\ }\textbf
  {\bibinfo {volume} {6}},\ \bibinfo {pages} {L161} (\bibinfo {year}
  {1989})}\BibitemShut {NoStop}%
\bibitem [{\citenamefont {Brandenberger}\ \emph {et~al.}(1990)\citenamefont
  {Brandenberger}, \citenamefont {Laflamme},\ and\ \citenamefont
  {Mijic}}]{Brandenberger:1990b}%
  \BibitemOpen
  \bibfield  {author} {\bibinfo {author} {\bibfnamefont {R.~H.}\ \bibnamefont
  {Brandenberger}}, \bibinfo {author} {\bibfnamefont {R.}~\bibnamefont
  {Laflamme}},\ and\ \bibinfo {author} {\bibfnamefont {M.}~\bibnamefont
  {Mijic}},\ }\bibfield  {title} {\bibinfo {title} {{Classical Perturbations
  From Decoherence of Quantum Fluctuations in the Inflationary Universe}},\
  }\href {https://doi.org/10.1142/S0217732390002651} {\bibfield  {journal}
  {\bibinfo  {journal} {Mod. Phys. Lett. A}\ }\textbf {\bibinfo {volume} {5}},\
  \bibinfo {pages} {2311} (\bibinfo {year} {1990})}\BibitemShut {NoStop}%
\bibitem [{\citenamefont {Calzetta}\ and\ \citenamefont
  {Hu}(1995)}]{Calzetta:1995ys}%
  \BibitemOpen
  \bibfield  {author} {\bibinfo {author} {\bibfnamefont {E.}~\bibnamefont
  {Calzetta}}\ and\ \bibinfo {author} {\bibfnamefont {B.~L.}\ \bibnamefont
  {Hu}},\ }\bibfield  {title} {\bibinfo {title} {{Quantum fluctuations,
  decoherence of the mean field, and structure formation in the early
  universe}},\ }\href {https://doi.org/10.1103/PhysRevD.52.6770} {\bibfield
  {journal} {\bibinfo  {journal} {Phys. Rev. D}\ }\textbf {\bibinfo {volume}
  {52}},\ \bibinfo {pages} {6770} (\bibinfo {year} {1995})},\ \Eprint
  {https://arxiv.org/abs/gr-qc/9505046} {arXiv:gr-qc/9505046} \BibitemShut
  {NoStop}%
\bibitem [{\citenamefont {Burgess}\ and\ \citenamefont
  {Michaud}(1997)}]{Burgess:1996mz}%
  \BibitemOpen
  \bibfield  {author} {\bibinfo {author} {\bibfnamefont {C.~P.}\ \bibnamefont
  {Burgess}}\ and\ \bibinfo {author} {\bibfnamefont {D.}~\bibnamefont
  {Michaud}},\ }\bibfield  {title} {\bibinfo {title} {{Neutrino propagation in
  a fluctuating sun}},\ }\href {https://doi.org/10.1006/aphy.1996.5660}
  {\bibfield  {journal} {\bibinfo  {journal} {Annals Phys.}\ }\textbf {\bibinfo
  {volume} {256}},\ \bibinfo {pages} {1} (\bibinfo {year} {1997})},\ \Eprint
  {https://arxiv.org/abs/hep-ph/9606295} {arXiv:hep-ph/9606295} \BibitemShut
  {NoStop}%
\bibitem [{\citenamefont {Kiefer}\ \emph {et~al.}(1998)\citenamefont {Kiefer},
  \citenamefont {Polarski},\ and\ \citenamefont {Starobinsky}}]{Kiefer:1998qe}%
  \BibitemOpen
  \bibfield  {author} {\bibinfo {author} {\bibfnamefont {C.}~\bibnamefont
  {Kiefer}}, \bibinfo {author} {\bibfnamefont {D.}~\bibnamefont {Polarski}},\
  and\ \bibinfo {author} {\bibfnamefont {A.~A.}\ \bibnamefont {Starobinsky}},\
  }\bibfield  {title} {\bibinfo {title} {{Quantum to classical transition for
  fluctuations in the early universe}},\ }\href
  {https://doi.org/10.1142/S0218271898000292} {\bibfield  {journal} {\bibinfo
  {journal} {Int. J. Mod. Phys. D}\ }\textbf {\bibinfo {volume} {7}},\ \bibinfo
  {pages} {455} (\bibinfo {year} {1998})},\ \Eprint
  {https://arxiv.org/abs/gr-qc/9802003} {arXiv:gr-qc/9802003} \BibitemShut
  {NoStop}%
\bibitem [{\citenamefont {Agon}\ \emph {et~al.}(2018)\citenamefont {Agon},
  \citenamefont {Balasubramanian}, \citenamefont {Kasko},\ and\ \citenamefont
  {Lawrence}}]{Agon:2014uxa}%
  \BibitemOpen
  \bibfield  {author} {\bibinfo {author} {\bibfnamefont {C.}~\bibnamefont
  {Agon}}, \bibinfo {author} {\bibfnamefont {V.}~\bibnamefont
  {Balasubramanian}}, \bibinfo {author} {\bibfnamefont {S.}~\bibnamefont
  {Kasko}},\ and\ \bibinfo {author} {\bibfnamefont {A.}~\bibnamefont
  {Lawrence}},\ }\bibfield  {title} {\bibinfo {title} {{Coarse Grained Quantum
  Dynamics}},\ }\href {https://doi.org/10.1103/PhysRevD.98.025019} {\bibfield
  {journal} {\bibinfo  {journal} {Phys. Rev. D}\ }\textbf {\bibinfo {volume}
  {98}},\ \bibinfo {pages} {025019} (\bibinfo {year} {2018})},\ \Eprint
  {https://arxiv.org/abs/1412.3148} {arXiv:1412.3148 [hep-th]} \BibitemShut
  {NoStop}%
\bibitem [{\citenamefont {Burgess}\ \emph {et~al.}(2015)\citenamefont
  {Burgess}, \citenamefont {Holman}, \citenamefont {Tasinato},\ and\
  \citenamefont {Williams}}]{Burgess:2014eoa}%
  \BibitemOpen
  \bibfield  {author} {\bibinfo {author} {\bibfnamefont {C.~P.}\ \bibnamefont
  {Burgess}}, \bibinfo {author} {\bibfnamefont {R.}~\bibnamefont {Holman}},
  \bibinfo {author} {\bibfnamefont {G.}~\bibnamefont {Tasinato}},\ and\
  \bibinfo {author} {\bibfnamefont {M.}~\bibnamefont {Williams}},\ }\bibfield
  {title} {\bibinfo {title} {{EFT Beyond the Horizon: Stochastic Inflation and
  How Primordial Quantum Fluctuations Go Classical}},\ }\href
  {https://doi.org/10.1007/JHEP03(2015)090} {\bibfield  {journal} {\bibinfo
  {journal} {JHEP}\ }\textbf {\bibinfo {volume} {03}},\ \bibinfo {pages}
  {090}},\ \Eprint {https://arxiv.org/abs/1408.5002} {arXiv:1408.5002 [hep-th]}
  \BibitemShut {NoStop}%
\bibitem [{\citenamefont
  {Boyanovsky}(2015{\natexlab{a}})}]{Boyanovsky:2015tba}%
  \BibitemOpen
  \bibfield  {author} {\bibinfo {author} {\bibfnamefont {D.}~\bibnamefont
  {Boyanovsky}},\ }\bibfield  {title} {\bibinfo {title} {{Effective field
  theory during inflation: Reduced density matrix and its quantum master
  equation}},\ }\href {https://doi.org/10.1103/PhysRevD.92.023527} {\bibfield
  {journal} {\bibinfo  {journal} {Phys. Rev. D}\ }\textbf {\bibinfo {volume}
  {92}},\ \bibinfo {pages} {023527} (\bibinfo {year} {2015}{\natexlab{a}})},\
  \Eprint {https://arxiv.org/abs/1506.07395} {arXiv:1506.07395 [astro-ph.CO]}
  \BibitemShut {NoStop}%
\bibitem [{\citenamefont {Boyanovsky}(2016)}]{Boyanovsky:2015jen}%
  \BibitemOpen
  \bibfield  {author} {\bibinfo {author} {\bibfnamefont {D.}~\bibnamefont
  {Boyanovsky}},\ }\bibfield  {title} {\bibinfo {title} {{Effective field
  theory during inflation. II. Stochastic dynamics and power spectrum
  suppression}},\ }\href {https://doi.org/10.1103/PhysRevD.93.043501}
  {\bibfield  {journal} {\bibinfo  {journal} {Phys. Rev. D}\ }\textbf {\bibinfo
  {volume} {93}},\ \bibinfo {pages} {043501} (\bibinfo {year} {2016})},\
  \Eprint {https://arxiv.org/abs/1511.06649} {arXiv:1511.06649 [astro-ph.CO]}
  \BibitemShut {NoStop}%
\bibitem [{\citenamefont
  {Boyanovsky}(2015{\natexlab{b}})}]{Boyanovsky:2015xoa}%
  \BibitemOpen
  \bibfield  {author} {\bibinfo {author} {\bibfnamefont {D.}~\bibnamefont
  {Boyanovsky}},\ }\bibfield  {title} {\bibinfo {title} {{Effective Field
  Theory out of Equilibrium: Brownian quantum fields}},\ }\href
  {https://doi.org/10.1088/1367-2630/17/6/063017} {\bibfield  {journal}
  {\bibinfo  {journal} {New J. Phys.}\ }\textbf {\bibinfo {volume} {17}},\
  \bibinfo {pages} {063017} (\bibinfo {year} {2015}{\natexlab{b}})},\ \Eprint
  {https://arxiv.org/abs/1503.00156} {arXiv:1503.00156 [hep-ph]} \BibitemShut
  {NoStop}%
\bibitem [{\citenamefont {Burgess}\ \emph {et~al.}(2016)\citenamefont
  {Burgess}, \citenamefont {Holman},\ and\ \citenamefont
  {Tasinato}}]{Burgess:2015ajz}%
  \BibitemOpen
  \bibfield  {author} {\bibinfo {author} {\bibfnamefont {C.~P.}\ \bibnamefont
  {Burgess}}, \bibinfo {author} {\bibfnamefont {R.}~\bibnamefont {Holman}},\
  and\ \bibinfo {author} {\bibfnamefont {G.}~\bibnamefont {Tasinato}},\
  }\bibfield  {title} {\bibinfo {title} {{Open EFTs, IR effects
  \textbackslash{}\& late-time resummations: systematic corrections in
  stochastic inflation}},\ }\href {https://doi.org/10.1007/JHEP01(2016)153}
  {\bibfield  {journal} {\bibinfo  {journal} {JHEP}\ }\textbf {\bibinfo
  {volume} {01}},\ \bibinfo {pages} {153}},\ \Eprint
  {https://arxiv.org/abs/1512.00169} {arXiv:1512.00169 [gr-qc]} \BibitemShut
  {NoStop}%
\bibitem [{\citenamefont {Braaten}\ \emph {et~al.}(2016)\citenamefont
  {Braaten}, \citenamefont {Hammer},\ and\ \citenamefont
  {Lepage}}]{Braaten:2016sja}%
  \BibitemOpen
  \bibfield  {author} {\bibinfo {author} {\bibfnamefont {E.}~\bibnamefont
  {Braaten}}, \bibinfo {author} {\bibfnamefont {H.~W.}\ \bibnamefont
  {Hammer}},\ and\ \bibinfo {author} {\bibfnamefont {G.~P.}\ \bibnamefont
  {Lepage}},\ }\bibfield  {title} {\bibinfo {title} {{Open Effective Field
  Theories from Deeply Inelastic Reactions}},\ }\href
  {https://doi.org/10.1103/PhysRevD.94.056006} {\bibfield  {journal} {\bibinfo
  {journal} {Phys. Rev. D}\ }\textbf {\bibinfo {volume} {94}},\ \bibinfo
  {pages} {056006} (\bibinfo {year} {2016})},\ \Eprint
  {https://arxiv.org/abs/1607.02939} {arXiv:1607.02939 [hep-ph]} \BibitemShut
  {NoStop}%
\bibitem [{\citenamefont {Hollowood}\ and\ \citenamefont
  {McDonald}(2017)}]{Hollowood:2017bil}%
  \BibitemOpen
  \bibfield  {author} {\bibinfo {author} {\bibfnamefont {T.~J.}\ \bibnamefont
  {Hollowood}}\ and\ \bibinfo {author} {\bibfnamefont {J.~I.}\ \bibnamefont
  {McDonald}},\ }\bibfield  {title} {\bibinfo {title} {{Decoherence, discord
  and the quantum master equation for cosmological perturbations}},\ }\href
  {https://doi.org/10.1103/PhysRevD.95.103521} {\bibfield  {journal} {\bibinfo
  {journal} {Phys. Rev. D}\ }\textbf {\bibinfo {volume} {95}},\ \bibinfo
  {pages} {103521} (\bibinfo {year} {2017})},\ \Eprint
  {https://arxiv.org/abs/1701.02235} {arXiv:1701.02235 [gr-qc]} \BibitemShut
  {NoStop}%
\bibitem [{\citenamefont {Shandera}\ \emph {et~al.}(2018)\citenamefont
  {Shandera}, \citenamefont {Agarwal},\ and\ \citenamefont
  {Kamal}}]{Shandera:2017qkg}%
  \BibitemOpen
  \bibfield  {author} {\bibinfo {author} {\bibfnamefont {S.}~\bibnamefont
  {Shandera}}, \bibinfo {author} {\bibfnamefont {N.}~\bibnamefont {Agarwal}},\
  and\ \bibinfo {author} {\bibfnamefont {A.}~\bibnamefont {Kamal}},\ }\bibfield
   {title} {\bibinfo {title} {{Open quantum cosmological system}},\ }\href
  {https://doi.org/10.1103/PhysRevD.98.083535} {\bibfield  {journal} {\bibinfo
  {journal} {Phys. Rev. D}\ }\textbf {\bibinfo {volume} {98}},\ \bibinfo
  {pages} {083535} (\bibinfo {year} {2018})},\ \Eprint
  {https://arxiv.org/abs/1708.00493} {arXiv:1708.00493 [hep-th]} \BibitemShut
  {NoStop}%
\bibitem [{\citenamefont {Ag\'on}\ and\ \citenamefont
  {Lawrence}(2018)}]{Agon:2017oia}%
  \BibitemOpen
  \bibfield  {author} {\bibinfo {author} {\bibfnamefont {C.}~\bibnamefont
  {Ag\'on}}\ and\ \bibinfo {author} {\bibfnamefont {A.}~\bibnamefont
  {Lawrence}},\ }\bibfield  {title} {\bibinfo {title} {{Divergences in open
  quantum systems}},\ }\href {https://doi.org/10.1007/JHEP04(2018)008}
  {\bibfield  {journal} {\bibinfo  {journal} {JHEP}\ }\textbf {\bibinfo
  {volume} {04}},\ \bibinfo {pages} {008}},\ \Eprint
  {https://arxiv.org/abs/1709.10095} {arXiv:1709.10095 [hep-th]} \BibitemShut
  {NoStop}%
\bibitem [{\citenamefont {Baidya}\ \emph {et~al.}(2017)\citenamefont {Baidya},
  \citenamefont {Jana}, \citenamefont {Loganayagam},\ and\ \citenamefont
  {Rudra}}]{Baidya:2017eho}%
  \BibitemOpen
  \bibfield  {author} {\bibinfo {author} {\bibfnamefont {A.}~\bibnamefont
  {Baidya}}, \bibinfo {author} {\bibfnamefont {C.}~\bibnamefont {Jana}},
  \bibinfo {author} {\bibfnamefont {R.}~\bibnamefont {Loganayagam}},\ and\
  \bibinfo {author} {\bibfnamefont {A.}~\bibnamefont {Rudra}},\ }\bibfield
  {title} {\bibinfo {title} {{Renormalization in open quantum field theory.
  Part I. Scalar field theory}},\ }\href
  {https://doi.org/10.1007/JHEP11(2017)204} {\bibfield  {journal} {\bibinfo
  {journal} {JHEP}\ }\textbf {\bibinfo {volume} {11}},\ \bibinfo {pages}
  {204}},\ \Eprint {https://arxiv.org/abs/1704.08335} {arXiv:1704.08335
  [hep-th]} \BibitemShut {NoStop}%
\bibitem [{\citenamefont {Burrage}\ \emph {et~al.}(2019)\citenamefont
  {Burrage}, \citenamefont {K\"ading}, \citenamefont {Millington},\ and\
  \citenamefont {Min\'a\v{r}}}]{Burrage:2018pyg}%
  \BibitemOpen
  \bibfield  {author} {\bibinfo {author} {\bibfnamefont {C.}~\bibnamefont
  {Burrage}}, \bibinfo {author} {\bibfnamefont {C.}~\bibnamefont {K\"ading}},
  \bibinfo {author} {\bibfnamefont {P.}~\bibnamefont {Millington}},\ and\
  \bibinfo {author} {\bibfnamefont {J.}~\bibnamefont {Min\'a\v{r}}},\
  }\bibfield  {title} {\bibinfo {title} {{Open quantum dynamics induced by
  light scalar fields}},\ }\href {https://doi.org/10.1103/PhysRevD.100.076003}
  {\bibfield  {journal} {\bibinfo  {journal} {Phys. Rev. D}\ }\textbf {\bibinfo
  {volume} {100}},\ \bibinfo {pages} {076003} (\bibinfo {year} {2019})},\
  \Eprint {https://arxiv.org/abs/1812.08760} {arXiv:1812.08760 [hep-th]}
  \BibitemShut {NoStop}%
\bibitem [{\citenamefont {Martin}\ and\ \citenamefont
  {Vennin}(2018{\natexlab{b}})}]{Martin:2018lin}%
  \BibitemOpen
  \bibfield  {author} {\bibinfo {author} {\bibfnamefont {J.}~\bibnamefont
  {Martin}}\ and\ \bibinfo {author} {\bibfnamefont {V.}~\bibnamefont
  {Vennin}},\ }\bibfield  {title} {\bibinfo {title} {{Non Gaussianities from
  Quantum Decoherence during Inflation}},\ }\href
  {https://doi.org/10.1088/1475-7516/2018/06/037} {\bibfield  {journal}
  {\bibinfo  {journal} {JCAP}\ }\textbf {\bibinfo {volume} {06}},\ \bibinfo
  {pages} {037}},\ \Eprint {https://arxiv.org/abs/1805.05609} {arXiv:1805.05609
  [astro-ph.CO]} \BibitemShut {NoStop}%
\bibitem [{\citenamefont {Burgess}\ \emph {et~al.}(2022)\citenamefont
  {Burgess}, \citenamefont {Holman},\ and\ \citenamefont
  {Kaplanek}}]{Burgess:2021luo}%
  \BibitemOpen
  \bibfield  {author} {\bibinfo {author} {\bibfnamefont {C.~P.}\ \bibnamefont
  {Burgess}}, \bibinfo {author} {\bibfnamefont {R.}~\bibnamefont {Holman}},\
  and\ \bibinfo {author} {\bibfnamefont {G.}~\bibnamefont {Kaplanek}},\
  }\bibfield  {title} {\bibinfo {title} {{Quantum Hotspots: Mean Fields, Open
  EFTs, Nonlocality and Decoherence Near Black Holes}},\ }\href
  {https://doi.org/10.1002/prop.202200019} {\bibfield  {journal} {\bibinfo
  {journal} {Fortsch. Phys.}\ }\textbf {\bibinfo {volume} {70}},\ \bibinfo
  {pages} {2200019} (\bibinfo {year} {2022})},\ \Eprint
  {https://arxiv.org/abs/2106.10804} {arXiv:2106.10804 [hep-th]} \BibitemShut
  {NoStop}%
\bibitem [{\citenamefont {Brahma}\ \emph {et~al.}(2021)\citenamefont {Brahma},
  \citenamefont {Berera},\ and\ \citenamefont
  {Calder{\'o}n-Figueroa}}]{Brahma:2021mng}%
  \BibitemOpen
  \bibfield  {author} {\bibinfo {author} {\bibfnamefont {S.}~\bibnamefont
  {Brahma}}, \bibinfo {author} {\bibfnamefont {A.}~\bibnamefont {Berera}},\
  and\ \bibinfo {author} {\bibfnamefont {J.}~\bibnamefont
  {Calder{\'o}n-Figueroa}},\ }\bibfield  {title} {\bibinfo {title} {Universal
  signature of quantum entanglement across cosmological distances},\
  }\href@noop {} {\bibfield  {journal} {\bibinfo  {journal} {arXiv preprint
  arXiv:2107.06910}\ } (\bibinfo {year} {2021})}\BibitemShut {NoStop}%
\bibitem [{\citenamefont {Brahma}\ \emph {et~al.}(2022)\citenamefont {Brahma},
  \citenamefont {Berera},\ and\ \citenamefont
  {Calder{\'o}n-Figueroa}}]{Brahma:2022yxu}%
  \BibitemOpen
  \bibfield  {author} {\bibinfo {author} {\bibfnamefont {S.}~\bibnamefont
  {Brahma}}, \bibinfo {author} {\bibfnamefont {A.}~\bibnamefont {Berera}},\
  and\ \bibinfo {author} {\bibfnamefont {J.}~\bibnamefont
  {Calder{\'o}n-Figueroa}},\ }\bibfield  {title} {\bibinfo {title} {Quantum
  corrections to the primordial tensor spectrum: Open efts \& markovian
  decoupling of uv modes},\ }\href@noop {} {\bibfield  {journal} {\bibinfo
  {journal} {arXiv preprint arXiv:2206.05797}\ } (\bibinfo {year}
  {2022})}\BibitemShut {NoStop}%
\bibitem [{\citenamefont {K{\"a}ding}\ and\ \citenamefont
  {Pitschmann}(2022)}]{Kading:2022jjl}%
  \BibitemOpen
  \bibfield  {author} {\bibinfo {author} {\bibfnamefont {C.}~\bibnamefont
  {K{\"a}ding}}\ and\ \bibinfo {author} {\bibfnamefont {M.}~\bibnamefont
  {Pitschmann}},\ }\bibfield  {title} {\bibinfo {title} {A new method for
  directly computing reduced density matrices},\ }\href@noop {} {\bibfield
  {journal} {\bibinfo  {journal} {arXiv preprint arXiv:2204.08829}\ } (\bibinfo
  {year} {2022})}\BibitemShut {NoStop}%
\bibitem [{\citenamefont {Kolioni}\ and\ \citenamefont
  {Anastopoulos}(2020)}]{Charis2020nonmarkov}%
  \BibitemOpen
  \bibfield  {author} {\bibinfo {author} {\bibfnamefont {T.}~\bibnamefont
  {Kolioni}}\ and\ \bibinfo {author} {\bibfnamefont {C.}~\bibnamefont
  {Anastopoulos}},\ }\bibfield  {title} {\bibinfo {title} {Detectors
  interacting through quantum fields: Non-markovian effects, nonperturbative
  generation of correlations, and apparent noncausality},\ }\href
  {https://doi.org/10.1103/PhysRevA.102.062207} {\bibfield  {journal} {\bibinfo
   {journal} {Phys. Rev. A}\ }\textbf {\bibinfo {volume} {102}},\ \bibinfo
  {pages} {062207} (\bibinfo {year} {2020})}\BibitemShut {NoStop}%
\end{thebibliography}%

\end{document}